\documentclass[twocolumn]{aastex7}
\usepackage{enumitem} % for spacing in enumerate environments
\setlist[enumerate]{noitemsep,topsep=1ex}
\usepackage{graphicx}
\usepackage{amsmath}
\usepackage{amssymb}
\usepackage{textcomp}
\usepackage[caption=false]{subfig}
\usepackage{color}     %for \textcolor
\usepackage{gensymb}   %for \degree
\usepackage{hyperref}  %for links
\usepackage{xspace}    %for HI command
\usepackage{afterpage} %for forcing typesetting of floats
\newcommand{\HI}{\textsc{Hi}\xspace}
\newcommand{\HII}{\textsc{Hii}\xspace}
\newcommand{\msun}{$\text{M}_{\odot}$\xspace}

\newcommand{\mhk}{[M/H]$_k$\xspace}
\newcommand{\sfhjl}{\href{https://github.com/cgarling/StarFormationHistories.jl}{\texttt{StarFormationHistories.jl}}\xspace}
\newcommand{\stjl}{\href{https://github.com/cgarling/StellarTracks.jl}{\texttt{StellarTracks.jl}}\xspace}
\newcommand{\bcjl}{\href{https://github.com/cgarling/BolometricCorrections.jl}{\texttt{BolometricCorrections.jl}}\xspace}
\defcitealias{Garling2025}{Paper~I}

\shorttitle{Connecting Chemical Enrichment with Resolved Star Formation Histories}
\shortauthors{Garling et al.}
\graphicspath{{./}{figures/}}
\begin{document}
\title{A Unified Framework Connecting Chemical Enrichment to Resolved Star Formation Histories with Applications to Local Group Dwarf Irregulars}

\correspondingauthor{Christopher T. Garling}

\author[0000-0001-9061-1697]{Christopher T. Garling}
\affiliation{Department of Astronomy, University of Virginia, 530 McCormick Road, Charlottesville, VA 22904, USA}
\email{txa5ge@virginia.edu}

\author[0000-0002-8111-9884]{Alex M. Garcia}
\affiliation{Department of Astronomy, University of Virginia, 530 McCormick Road, Charlottesville, VA 22904, USA}
\email{aku7cf@virginia.edu}

\author[0009-0002-1233-2013]{Niusha Ahvazi}
\altaffiliation{Galaxy Evolution and Cosmology (GECO) Fellow}
\affiliation{Department of Astronomy, University of Virginia, 530 McCormick Road, Charlottesville, VA 22904, USA}
\email{dkk9en@virginia.edu}

\author[0000-0002-3204-1742]{Nitya Kallivayalil}
\affiliation{Department of Astronomy, University of Virginia, 530 McCormick Road, Charlottesville, VA 22904, USA}
\email{njk3r@virginia.edu}

\author[0000-0001-5538-2614]{Kristen B. W. McQuinn}
\affiliation{Space Telescope Science Institute, 3700 San Martin Drive, Baltimore, MD 21218, USA}
\affiliation{Department of Physics and Astronomy, Rutgers, the State University of New Jersey,  136 Frelinghuysen Road, Piscataway, NJ 08854, USA}
\email{kmcquinn@stsci.edu}

\author[0000-0002-1109-1919]{Robert Feldmann}
\affiliation{Department of Astrophysics, Universit\"at Z\"urich, Zurich, CH-8057, Switzerland}
\email{robert.feldmann@uzh.ch}

\author[0000-0002-2970-7435]{Roger E. Cohen}
\affiliation{Department of Physics and Astronomy, Rutgers, the State University of New Jersey,  136 Frelinghuysen Road, Piscataway, NJ 08854, USA}
\email{rc1273@physics.rutgers.edu}

\begin{abstract}

We present a new framework for modeling the chemical enrichment histories of galaxies by integrating chemical evolution with resolved star formation histories (SFHs) derived from color-magnitude diagrams. This novel approach links the time evolution of the metallicity of the star-forming ISM to the cumulative stellar mass formed in the galaxy, enabling a self-consistent description of chemical evolution. We apply this methodology to four isolated, gas-rich Local Group dwarf galaxies -- WLM, Aquarius, Leo A, and Leo P -- using deep HST and JWST imaging. For WLM, Aquarius, and Leo A, we independently validate our metallicity evolution results against ages and metallicities of individual red giant stars with spectroscopic measurements. We quantify systematic uncertainties by repeating our analysis with multiple stellar evolution and bolometric correction libraries. We compare the observed chemical enrichment histories to predictions from the TNG50 and FIREbox cosmological hydrodynamic simulations and the Galacticus semi-analytic model. Of our four galaxies, only WLM is sufficiently massive to be reliably represented in these simulations; the remaining three fall below current resolution limits. We find that the enrichment history of WLM is best reproduced by FIREbox, while TNG50 and Galacticus predict higher metallicities at early times, suggesting that differences in stellar feedback and metal recycling prescriptions drive significant variation in predicted enrichment histories. This work demonstrates the power of combining resolved SFHs with physically motivated chemical evolution models to constrain galaxy formation physics and highlights the need for further observational and theoretical studies of metal retention and recycling in low-mass dwarf galaxies.

\end{abstract}

\keywords{Galaxy chemical evolution (580); Hertzsprung Russell diagram (725); Hubble Space Telescope (761); James Webb Space Telescope (2291); Stellar photometry (1620); Stellar populations (1622)}


\section{Introduction} \label{sec:intro} 

Resolved star formation histories (SFHs) derived from deep color-magnitude diagrams (CMDs) provide the most precise measurements of the star formation rates (SFRs) for galaxies as a function of time. This precision comes from the ability to distinguish stars in different evolutionary phases, allowing for detailed reconstructions of the SFHs of galaxies. Other methods for measuring the SFHs of galaxies, like full spectrum fitting and spectral energy distribution modeling \citep[e.g.,][]{Walcher2011,Conroy2013,Johnson2021,Cappellari2023}, rely on integrated light from the entire stellar population of the galaxy. Such integrated light measurements suffer from the ``outshining" effect, where young stars contribute much of the luminosity of the galaxy, masking the contribution from older stars and limiting the ability of these methods to accurately recover the early-time star formation rates of galaxies \citep{Narayanan2024,Harvey2025,Wang2025}.

To accurately model a galaxy's CMD and disentangle age and metallicity degeneracies, resolved SFH analyses must assume or fit a model for the metallicity evolution of the galaxy as the metallicity at each epoch influences the luminosities and colors of stars \citep[e.g.,][]{Aparicio2004}. In this way, the inferred SFRs are inherently coupled to the chemical enrichment history. Most previous approaches to modeling metallicity evolution in resolved SFHs treat the age-metallicity relation (AMR) as an independent function of time, often taking simple forms with parameters fit simultaneously with the SFH \citep[non-parametric methods have also been used; see, e.g.,][]{Dolphin2016a}. However, there is no physical coupling between the metallicity evolution and the actual star formation activity in these models. This means that changes to the SFH (i.e., the amount or timing of star formation) do not affect the metallicity evolution, and vice versa. The metallicity at a given time is determined solely by the chosen AMR parameterization, not by the integrated effects of past star formation or the enrichment expected from stellar feedback.

As a result, these simple AMR models do not self-consistently evolve the metallicity in response to star formation. This can lead to non-physical solutions, such as metallicity increasing independently of the galaxy's actual star formation activity \cite[see, e.g., appendix B of][]{Savino2023}. It also makes it difficult to compare observational results to theoretical models in which the metallicity evolves self-consistently as galaxies form stars. In summary, traditional AMR-based approaches lack a physical link between the SFH and chemical enrichment, and therefore do not capture the feedback-driven, self-consistent evolution of metallicity that is expected in real galaxies.

We introduce a novel framework in which the metallicity evolution is directly coupled to the SFH. In this model, the mean metallicity of stars forming at a particular epoch is a function of the cumulative stellar mass formed prior to that epoch. As such, changes to the SFH directly propagate into changes in the metallicity evolution as is expected from galaxy evolution theory. In \S \ref{sec:sfh_model}, we describe how we integrate the chemical evolution with the resolved SFH and implement a minimal two parameter model relating them. The primary systematics in resolved SFH measurements based on well-populated CMDs are the stellar models that are assumed when making the measurement. In \S \ref{sec:systematics} we describe our method for quantifying these systematics by repeating our measurements with isochrones interpolated from three different libraries of stellar tracks and two different libraries of bolometric corrections. In order to validate our metallicity evolution results, we derive ages and metallicities for single red giant stars in WLM, Aquarius, and Leo A using the method presented in \S \ref{sec:ages}. In \S \ref{sec:obs} we present applications of this new method to deep HST/ACS and JWST/NIRCam photometry of four isolated, gas-rich Local Group galaxies -- WLM, Aquarius, Leo A, and Leo P. These galaxies range in stellar mass from $5.7 \times 10^5$ \msun \citep[Leo P;][]{McQuinn2024b} to $4.3 \times 10^7$ \msun \citep[WLM;][]{McConnachie2012}, allowing us to probe the chemical evolution of dwarf galaxies over a wide range in stellar mass. To interpret these results, we calculate chemical enrichment histories from the TNG50 and FIREbox cosmological hydrodynamic simulations and the \textsc{galacticus} semi-analytic model in \S \ref{sec:theory}. We conclude in \S \ref{sec:results} with discussion of, and comparison between, the theoretical predictions and observational measurements.

\section{Chemical Enrichment Model} \label{sec:sfh_model}
Here we describe our novel method for simultaneously modeling the SFH and chemical evolution of galaxies. We make use of the framework for hierarchical SFH modeling developed in \citet{Garling2025}, hereafter \citetalias{Garling2025}, which applied this framework to AMRs -- here we extend it to include mass-metallicity histories (MZHs) as well. We begin by outlining the basis of this framework as developed in \citetalias{Garling2025} in \S \ref{subsec:hierarchical} before introducing our new MZH model in \S \ref{subsec:mzhmodel}.

\subsection{Hierarchical Star Formation History Modeling} \label{subsec:hierarchical}

We start from the fact that the CMD of a complex stellar population, consisting of stars belonging to multiple populations with different ages and/or metallicities, can be modeled as the superposition of the CMDs of the constituent populations \citep[e.g.,][]{Hodge1989,Tosi1991,Greggio1993,Dolphin1997}. This remains true for the Hess diagrams of stellar populations, as the Hess diagram is simply a 2-D discretization of the CMD. \citetalias{Garling2025} developed methods for generating Hess diagram models for simple stellar populations (SSPs) from stellar models that can be superposed with each other to model observations of complex stellar populations. The SSP Hess diagram models (which we will refer to as \emph{templates} for brevity) are typically constructed for a regular grid of SSP ages and metallicities to cover the range of relevant stellar populations that may coexist in the observed stellar population. A weighted sum of these templates then gives the composite model Hess diagram that will be compared to the observed Hess diagram. Mathematically, the composite model Hess diagram is constructed as

\begin{equation} \label{eq:composite}
    m_i = \sum_{j,k} r_{j,k} \, c_{i,j,k} \\
\end{equation}

\noindent where $m_i$ are the counts in bin $i$ of the complex model Hess diagram, $c_{i,j,k}$ is bin $i$ of the template with age index $j$ (and corresponding age $t_j$) and metallicity index $k$ (with corresponding metallicity \mhk), and $r_{j,k}$ is the multiplicative coefficient that determines how significantly the template contributes to the complex model. Put another way, the complex Hess diagram model $m$ is a linear combination of the templates $c_{j,k}$ with amplitudes $r_{j,k}$. For the purpose of this work, we will assume the templates are normalized to uniform birth stellar mass, such that $c_{i,j,k}$ are the expected number of stars predicted to be observed in bin $i$ per unit solar mass of stars formed in the SSP and the weight $r_{j,k}$ gives the total stellar mass attributed to the SSP with age $t_j$ and metallicity \mhk.

The simplest way to model the observed Hess diagram is to simultaneously fit the $r_{j,k}$ under the sole constraint that all $r_{j,k} \ge 0$ to avoid attributing negative stellar mass to any population. However, as discussed in section 2.7 of \citetalias{Garling2025} and references therein, this approach tends to overfit the data as the grids of templates used can contain thousands of SSP models, resulting in an equal -- and very large -- number of free parameters. Additionally, such unconstrained fits can produce non-physical AMRs that undermine the interpretation of the resulting SFH. For this reason, it is common to fit a set of coefficients $R_j$ that are a function of only the age index $j$ and not the metallicity index $k$. If the templates are normalized to uniform birth stellar mass, then $R_j$ is the sum of the stellar mass formed in all SSPs with age $t_j$. In this case, a hierarchical model is used that defines how to convert the $R_j$ into the per-template coefficients $r_{j,k}$ that are needed to compute Equation \ref{eq:composite}. \citetalias{Garling2025} developed a model that used parametric AMRs to determine the mean metallicity at time $t_j$, which we define as $\mu_j \equiv \langle[\text{M}/\text{H}]\rangle(t_j)$. Given this mean value, the $R_j$ are distributed to templates with SSP ages $t_j$ and different metallicities \mhk by assuming a Gaussian metallicity distribution function (MDF) at fixed time with standard deviation $\sigma$, given by Equation 8 in \citetalias{Garling2025},

\begin{equation} \label{eq:hmodel}
  \begin{aligned}
    \mu_j &= \langle[\text{M}/\text{H}]\rangle(t_j) \\
    a_{j,k} &= \exp{ \left(-\left(\frac{[\text{M}/\text{H}]_k - \mu_j}{\sigma}\right)^2\right)} \\
    r_{j,k} &= R_j \frac{a_{j,k}}{\sum_k a_{j,k}}. \\
  \end{aligned}
\end{equation}

This class of hierarchical model is highly flexible in that $\mu_j=\langle[\text{M}/\text{H}]\rangle(t_j)$ can be any time-dependent function. \citetalias{Garling2025} used simple parametric AMRs to model the time evolution of the mean metallicity, for example $\langle[\text{M}/\text{H}]\rangle(t)=\alpha \left( T_\mathrm{max} - t \right) + \beta$ where $T_\mathrm{max}$ is the maximum lookback time for which the AMR is valid. In this model, $\alpha$ and $\beta$ are free parameters and are optimized at the same time as $R_j$, allowing the covariance between $R_j$ and the AMR parameters to be fully captured. Such parametric AMR models are attractive as they have few free parameters, mitigating concerns about overfitting the observed data, while providing sufficient freedom to enable good fits to observed Hess diagrams. Additionally, the simple forms of these AMR models allow the gradient of the likelihood function used to optimize the fitting parameters to be derived analytically with little difficulty. With an analytic gradient, finding the best-fit coefficients $R_j$ and AMR parameters (e.g., $\alpha$ and $\beta$ above) can be done efficiently and robustly.

However, these kinds of AMR models also have several undesirable characteristics. First, these AMR models are not directly linked to the SFH parameters (i.e., the $R_j$ discussed above) -- with an AMR model like $\langle[\text{M}/\text{H}]\rangle(t)=\alpha \left( T_\mathrm{max} - t \right) + \beta$, changing the SFH parameters $R_j$ does not change $\langle[\text{M}/\text{H}]\rangle(t)$. Physically, we expect the metallicity of a galaxy to rise due to stellar feedback (e.g., supernovae and AGB winds) over time -- therefore, it would be more physical for $\langle[\text{M}/\text{H}]\rangle(t_j)$ to have some link to the SFH parameters $R_j$ for times earlier than $t_j$. Additionally, these AMR models are difficult to compare between galaxies -- even if observations of two galaxies imply that they have consistent AMRs, they will have different population-averaged metallicities ($\langle[\text{M}/\text{H}]\rangle$) if their SFHs ($M_* (t)$) are different, as the average age (and, therefore, metallicity) of the two galaxies will be different.

Below we develop a class of MZH models as an alternative to the AMR models used to calculate the $\mu_j$ in Equation \ref{eq:hmodel} to address these issues. 

\subsection{The Mass-Metallicity History Model} \label{subsec:mzhmodel}

As outlined above, the goal of our new MZH model is to connect the chemical evolution of the galaxy $\mu_j = \langle[\text{M}/\text{H}]\rangle(t_j)$ to the SFH parameters $R_j$ directly, so that the $\mu_j$ change in response to changes in the $R_j$, as is physically expected since the galactic ISM is enriched primarily due to stellar feedback. Our basic approach is to convert the time dependence of $\mu_j = \langle[\text{M}/\text{H}]\rangle(t_j)$ into a dependence on the total stellar mass formed up to time $t_j$, which we write as $M_*(t_j)$, such that $\mu_j = \langle[\text{M}/\text{H}]\rangle(M_*(t_j))$ in the MZH model. This connection between the stellar mass and the metallicity of galaxies has some similarities to the mass-metallicity relation \citep[MZR; e.g.,][]{Andrews2013,Curti2020}, but is fundamentally different. MZRs describe the relation between stellar mass and gas-phase metallicity for a population of galaxies with different stellar masses at \emph{one particular epoch}. In contrast, our method describes how the \emph{historical} accumulation of stellar mass relates to an individual galaxy's chemical enrichment over time; for this reason we discuss our model as a mass-metallicity \emph{history} to differentiate it from the well-known MZR.

Given a set of proposed SFH parameters $R_j$, it is straightforward to derive the cumulative stellar mass formed in the galaxy as a function of time. Let the $R_j$ be sorted in order from earliest time $t_j$ to most recent time. The cumulative stellar mass $M_*(t_j)$ can therefore be expressed as the sum over the $R_{j^\prime}$ for which $j^\prime \le j$, such that

\begin{equation}
    M_*(t_j) = \sum_{j^\prime=0}^{j^\prime=j} R_{j^\prime}
\end{equation}

\noindent assuming that we have normalized the templates to a uniform birth stellar mass of 1 M$_\odot$ as discussed in \S \ref{subsec:hierarchical} so that the SFH parameters $R_j$ give the total stellar mass formed at each unique epoch $t_j$. With the ability to calculate $M_*(t_j)$, we can now convert the time dependence of $\mu_j = \langle[\text{M}/\text{H}]\rangle(t_j)$ into a dependence on the total stellar mass, $\mu_j = \langle[\text{M}/\text{H}]\rangle(M_*(t_j))$. This alternative form for the $\mu_j$ that enter into Equation \ref{eq:hmodel} is the basis for our new class of MZH chemical evolution models.

Given this form, there are many approaches one could take to modeling $\langle[\text{M}/\text{H}]\rangle(M_*)$. We consider a simple power-law MZH,

\begin{equation} \label{eq:mzhmodel}
  \begin{aligned}
    \langle[\text{M}/\text{H}]\rangle(M_*) &= [\text{M}/\text{H}]_0 + \text{log} \left( \left( \frac{M_*}{M_{*,0}} \right)^\alpha \right) \\
    &= [\text{M}/\text{H}]_0 + \alpha \left( \text{log} \left( M_* \right) - \text{log} \left( M_{*,0} \right) \right)
  \end{aligned}
\end{equation}

\noindent as is typically used to model the low-redshift MZR in the dwarf regime \citep[$M_* \le 10^8$ M$_\odot$; e.g.,][]{Lee2006,Kirby2013}. This model has only two free parameters, the MZH normalization $[\text{M}/\text{H}]_0$ and the power-law slope $\alpha$. The parameter $M_{*,0}$ can be chosen arbitrarily and simply defines the stellar mass at which the mean metallicity equals $[\text{M}/\text{H}]_0$. In this work we will set $M_{*,0}=10^6$ M$_\odot$, matching the convention taken by \citet{Kirby2013} who measured the relation between stellar mass and stellar metallicity for Local Group dwarf galaxies at present-day (see their Equation 4 and Figure 9).

The form of this model is very similar to that of the AMR model from \citetalias{Garling2025} mentioned previously, $\langle[\text{M}/\text{H}]\rangle(t)=\alpha \left( T_\mathrm{max} - t \right) + \beta$; we have simply swapped the time dependence in the AMR for a stellar mass dependence in the MZH. Comparing the performance of the two models on observational data should be relatively straightforward, as both models have the same number of free parameters, and the free parameters consist only of a slope and a normalization for each model.

We implement this model in \sfhjl\footnote{https://github.com/cgarling/StarFormationHistories.jl} alongside the AMR models developed in \citetalias{Garling2025}. Our implementation is general in the sense that it is easy to define new forms for $\langle[\text{M}/\text{H}]\rangle(M_*)$ that will plug into the existing infrastructure of \sfhjl to measure resolved SFHs from photometry. This generality is important because this simple power-law model will be insufficient for some types of galaxies. In particular, it is well-established observationally that the MZR saturates at $\sim10^{10.5}$ \msun \citep[e.g.,][]{Andrews2013,Curti2020} -- modeling a similar turnoff in the MZH would require a more complicated model (e.g., Equation 2 of \citealt{Curti2020}), but is irrelevant for the lower mass galaxies we study here. Additionally, it is possible that the power-law MZH may not appropriate at very low masses either -- spectroscopic studies of ultra-faint dwarf galaxies with stellar masses below $10^5$ \msun generally measure mean galaxy metallicities $\langle[\text{M}/\text{H}]\rangle \ge -3$ dex, indicating that there may be something like a metallicity floor in UFDs. In such a case, extrapolating a power-law MZH to $\langle[\text{M}/\text{H}]\rangle = -4$ dex would be inappropriate, and again a more complicated model would be required. In this work we stay between these regimes, where the power-law model is well-supported observationally -- we may define alternative models in future work as necessary.

\subsection{Interpreting the Mass-Metallicity History} \label{subsec:interpretmzr}

We note that while the form of the MZH $\langle[\text{M}/\text{H}]\rangle(M_*)$ we propose in Equation \ref{eq:mzhmodel} is motivated by observational results for low-redshift dwarf galaxies \citep[e.g.,][]{Lee2006,Kirby2013}, the observational measurements of the MZR differ in an important way from our definition of the MZH. The observational results take a population of different galaxies observed at one epoch, each of which have unique stellar masses and metallicities, and measure the correlation between their metallicities and stellar masses. Such observed MZRs are thus valid at one particular redshift and cannot be used to infer the metallicities of the observed galaxies at earlier times. In our application, we are modeling $\langle[\text{M}/\text{H}]\rangle(M_*(t))$ -- that is, we are assuming a time-independent form for the MZH and assuming that the galaxy evolves along the MZH as it accumulates stellar mass over time. Such a mass-metallicity \emph{history} is therefore not identical to the MZRs that have been measured observationally. Given that our MZH model is defined differently from the MZR, it may therefore deviate from the simple power-law form we have assumed based on the observational results for the MZR. In \S \ref{sec:theory}, we measure the MZH in several theoretical datasets to examine how well our power-law MZH model matches expectations from theory, taking into account the historical definition of our MZH model. We show that different galaxy formation models produce different predictions for the MZH, but the power-law MZH is sufficient to model their morphologies at dwarf galaxy masses.

We expect the MZH model to capture several important aspects of the relation between ISM enrichment and galactic stellar mass growth. The metallicity of the ISM is determined by the competing mechanisms of enrichment (due to stellar feedback) and dilution (due to accretion of low-metallicity gas). In equilibrium models of galaxy evolution, a relation between galactic stellar mass and metallicity arises naturally as a consequence of these processes \citep[e.g.,][]{Dave2012,Lilly2013}. However, the process of enrichment is complicated by the fact that not all metals ejected in a supernova are expected to stay in the ISM -- particularly in the shallow potential wells of dwarf galaxies, supernovae can drive winds with significant metal content into the circumgalactic medium (CGM) or out of the halo entirely \citep[e.g.,][]{Dave2012,Lilly2013,Feldmann2015,Muratov2015,Muratov2017,Christensen2016,Christensen2018,AnglesAlcazar2017,Emerick2018,Buck2019,Nelson2019,Tollet2019,Agertz2020a,Mitchell2020a,Pandya2021,Bassini2024}. These metals will not be available in the ISM to form the next generation of stars, and so these metals do not 
effectively contribute to the enrichment of the ISM. 

In the context of galactic chemical evolution models, this effect is quantified by the \emph{effective yield} which is, by definition, less than or equal to the total nucleosynthetic yield of stellar enrichment processes -- the difference between the canonical ``closed box'' and ``leaky box'' chemical evolution models is the substitution of the total nucleosynthetic yield in the former for the effective yield in the latter \citep[see, e.g., section 5.1 of][and references therein]{Kirby2013}. In high-resolution simulations, it is well established that the outflow mass-loading factor of these winds, defined as the ratio of the mass flux of the outflow to the galactic SFR $\left(\eta = \dot{M}_\text{out} / \text{SFR} \right)$, correlates with halo properties (e.g., halo circular velocity) and galaxy properties (e.g., stellar mass), though these correlations differ depending on the galaxy formation model assumed \citep[e.g.,][]{Muratov2015,Muratov2017,Christensen2016,Christensen2018,Nelson2019,Mitchell2020a,Pandya2021,Rey2025}. Generally, the fraction of created metals that is retained in the ISM is expected to be low at dwarf galaxy scales, as a significant portion of the metals created in supernovae are blown out of the ISM by their associated winds. For example, analyses using the FIRE-1 and FIRE-2 galaxy evolution models predict that nearly all metals produced in core-collapse supernovae should be ejected from the ISM in galaxies with $M_* \le 10^9 \ M_\odot$ \citep{Muratov2017,Pandya2021}. These metals persist in the circumgalactic medium of these galaxies and can be accreted into the ISM given enough time, resulting in the metallicity of accreted gas evolving as a function of time in the FIRE-2 model \citep{Bassini2024}. Other galaxy evolution models typically predict weaker metal outflows, but there is significant variation between models \citep[see, e.g.,][]{Christensen2018,Nelson2019,Mitchell2020a,Wright2024,Garcia2024a,Garcia2025}. These galaxy evolution models also differ in the redshift dependence they predict for these outflow parameters at fixed galaxy stellar mass.

In summary, star-formation-driven outflows in dwarf galaxies are expected to entrain a significant quantity of metals so that the effective metal yield (the quantity of metals ejected in stellar feedback that is retained in the ISM) is substantially less than the total nucleosynthetic metal yield. However, variations in the effective metal yields between galaxy evolution models are significant -- for example, \cite{Torrey2019} find that $\sim50\%$ of the metal mass created in galaxies with M$_* \approx 10^9$ \msun at present-day are retained in the ISM under the IllustrisTNG galaxy evolution model, while \cite{Pandya2021} find that nearly all metals created in galaxies of this mass are ejected from the ISM in the FIRE-2 galaxy evolution model. It is therefore of significant interest to design observational tests that can differentiate between these scenarios. 

It is useful to consider what relation our power-law model for $\langle[\text{M}/\text{H}]\rangle(M_*)$ has to other chemical evolution models. Most obviously, our model tracks only the \emph{relative abundance of metals and hydrogen} and does not consider the galaxy's gas mass whatsoever, while traditional one-zone analytic galactic chemical evolution models (closed box, leaky box, etc.) track the total galactic gas and metal masses separately in order to calculate the metal mass fraction $Z$ \citep[see, e.g.,][for an overview of these models]{Peimbert1994}. These models require boundary conditions for the galactic gas and metal mass fractions in the earliest and latest time bins that are often not available observationally, requiring the researcher to either adopt reasonable values or fit these boundary conditions as part of the model. These chemical evolution models can include free parameters to model metal-poor gas inflows, metal-rich outflows driven by stellar feedback, and other processes \citep[e.g.,][]{Peimbert1994,Kirby2013,Kvasova2025}, but priors on these parameters tend to be weak, increasing the number of free parameters in the MZH model and the risk of overfitting the photometric data. These models have been shown to successfully reproduce the time-integrated MDFs of observed dwarf galaxies \citep[e.g.,][]{Kirby2011,Kirby2013,Kirby2019,Kvasova2025} but are rarely used to predict the time-dependence of the metallicity and thus are mostly untested in the time domain \citep[for a recent example, see Figure 3 of][]{Kvasova2025}. A promising solution is to use external spectroscopic measurements to constrain the parameters of a one-zone chemical evolution model jointly with the photometry. As significant technical challenges must be overcome to perform such joint photometric-spectroscopic measurements \citep{Dolphin2016a}, we leave this development for future work.

Due to the modularity of our modeling framework, it is straightforward to implement new functional forms for $\langle[\text{M}/\text{H}]\rangle(M_*)$. The model can even be extended to include a time dependence in the MZH, by generalizing $\mu_j = \langle[\text{M}/\text{H}]\rangle(M_*(t_j))$ to have additional time dependence beyond that implied by $M_*(t_j)$ -- for example, by making the power-law slope $\alpha$ have a dependence on redshift. However, without priors on what form this dependence should take we would risk overfitting the data with a more complex MZH model. While higher redshift observations for higher mass dwarfs (generally M$_* \geq 10^8$ \msun) are becoming available thanks to JWST \citep[e.g.,][]{Li2023c,Nakajima2023,Curti2024,Jain2026}, we leave investigation of time dependence in the MZH parameters to future work.

\section{Systematic Uncertainties Due to Stellar Models} \label{sec:systematics}

It is well-established that the stellar models used to forward model observed CMDs contribute significantly to the overall systematic uncertainty of the measurement and can even dominate the overall error budget for nearby galaxies with well-populated CMDs and deep photometry \citep[e.g.,][]{Gallart2005,Dolphin2012,Weisz2014,Skillman2017}. It is therefore important to consider how the choice of stellar interior models, and the bolometric corrections (BCs) that transform them into observational magnitudes, affects the measured SFHs and chemical evolution parameters.

The methodology used by the \textsc{match} code to quantify systematic uncertainties due to the assumed stellar models is described by \cite{Dolphin2012}. In brief, the method shifts the fiducial isochrone models in temperature and bolometric magnitude and refits the SFH with these altered isochrones. The sizes of the shifts are randomly sampled in a Monte Carlo procedure and are drawn from distributions designed to reproduce the age and metallicity differences observed under different stellar model libraries. Credible intervals can then be measured from these Monte Carlo samples, enabling the estimation of the systematic errors.

While this approach is flexible, simple shifts in temperature and bolometric magnitude cannot reproduce morphological differences between stellar models, such as the shape of the MSTO, nor does it provide a mechanism to separate the systematic uncertainty contribution of the stellar tracks (i.e., the stellar interior models) from the contribution due to the assumed BCs (i.e., the stellar atmosphere models).

For this reason we choose to estimate systematic uncertainties in our SFH measurements and chemical evolution parameters by refitting the data directly with different combinations of stellar interior models and BC libraries. This approach \emph{can} be less robust than the method of \cite{Dolphin2012}, as often very few different stellar models are considered -- often only two stellar model libraries are compared, and typically only one BC library is used for each stellar interior library. In order for this approach to be robust, several stellar libraries should be used, and ideally several sets of BCs should be considered as well.

Our software \sfhjl is structured to work with generic isochrones, including those that can be downloaded from webforms maintained by different groups; such user-downloaded isochrones were used in \citetalias{Garling2025}. This approach is convenient when only a single stellar library is needed, but becomes cumbersome when one wishes to use several stellar interior and BC libraries. Additionally, most webforms only support a single library of BCs, and mixing different stellar tracks with different tables of BCs is generally not supported.

To support the calculation of systematic uncertainties due to stellar models, we have developed and publicly released new packages to interpolate isochrones on-the-fly. The \stjl\footnote{\url{https://github.com/cgarling/StellarTracks.jl}} package provides access to stellar track libraries from several groups and can interpolate isochrones at arbitrary metallicities and ages. This integrates with the \bcjl\footnote{\url{https://github.com/cgarling/BolometricCorrections.jl}} package that provides access to pre-computed tables of BCs derived from several different libraries of synthetic spectra to transform the stellar bolometric magnitudes into the observed magnitudes. The packages are structured so that any stellar track library can be used with any BC library, allowing arbitrary combinations. This modularity allows us to separately examine the systematics due to the stellar interiors and those due to the BCs.

\subsection{Stellar Tracks} \label{subsec:stellar_tracks}

All stellar tracks accessible through \stjl include equivalent evolutionary points (EEPs) to support accurate isochrone interpolation using the method of \citet{Dotter2016}. The package currently provides access to the PARSEC, MIST, and BaSTI stellar model grids. All have been widely used for resolved SFH studies as they have wide coverage of stellar metallicity and initial stellar masses, and such wide coverage is critical to simultaneously model populations of many ages and metallicities, as is necessary when fitting resolved SFHs. The PARSEC, MIST, and BaSTI stellar model grids are described briefly below.

We utilize the PARSEC V1.2S stellar tracks \citep{Bressan2012,Chen2014,Tang2014,Chen2015,Marigo2017} reprocessed with EEPs by \citet{Rosenfield2016}\footnote{\url{https://github.com/philrosenfield/padova_tracks}}. The tracks with EEPs were originally produced for use in the \textsc{match} code and are still in use today. This library covers a wide range of stellar parameters with scaled-solar metallicities ([M/H]) ranging from $-2.2$ to 0.5 dex and initial stellar masses ranging from 0.1 to 350 M$_\odot$, although the grid of available metallicities and stellar masses is not entirely regular. The models include evolution through the thermally-pulsating AGB phase by integrating with results from the COLIBRI code \citep{Marigo2017}. Horizontal branch evolution is also included.

The MIST stellar tracks \citep{Dotter2016, Choi2016} cover stars with scaled-solar metallicities ([M/H]) ranging from $-4$ to 0.5 dex and initial stellar masses ranging from 0.1 to 300 M$_\odot$. These models were run with the \textsc{mesa} code \citep{Paxton2013,Paxton2015} and feature continuous evolution through many phases of stellar evolution, which is unique for these types of stellar track libraries. PARSEC, for comparison, computes the core hydrogen burning, core helium burning, and thermally-pulsating AGB phases separately (see section 5.3 of \citealt{Bressan2012}). MIST isochrones are also included in \textsc{match}. MIST includes both rotating and non-rotating models, and both are available in our package.

The previous generation of BaSTI stellar tracks (which we refer to as \texttt{BaSTIv1} for short) presented in \citet{Pietrinferni2004,Pietrinferni2006,Pietrinferni2013} includes stars with metallicities ([M/H]) ranging from $-3$ to 0.5 dex and initial stellar masses ranging from 0.5 to 10 M$_\odot$. While the limited range of initial stellar masses means this grid cannot be used on \emph{all} data sets, sets of stellar models are available for a number of different modeling assumptions. For example, models are available with and without convective core overshooting (so-called non-canonical and canonical models), for different values of the Reimers mass loss parameter ($\eta=0.2$ and $0.4$), and for different relative $\alpha$-element abundances ($[\alpha/\text{Fe}] = 0.0$ and $0.4$ dex). All of these sets of tracks are made available and can be easily switched between in our package, enabling differential analyses as a function of these additional parameters. These models have been largely superceded by newer BaSTI models described below, but can still provide a useful point of comparison.

The current generation of BaSTI models (which we refer to as \texttt{BaSTIv2} for short) presented in \citet{Hidalgo2018,Pietrinferni2021,Salaris2022,Pietrinferni2024} cover stars with metallicities ([M/H]) ranging from $-3.2$ to 0.45 dex and initial stellar masses ranging from 0.1 to 15 M$_\odot$. This range of stellar initial masses is typically sufficient for modeling the CMDs of galaxies with little recent star formation \citep[e.g.,][]{Savino2023,Savino2025,Durbin2025}. Like \texttt{BaSTIv1}, this library also includes tracks for a wide range of modeling parameters -- tracks were calculated with and without convective core overshooting, atomic diffusion, and mass loss (Reimers mass loss parameters $\eta=0.0$ and $0.3$). Additionally, tracks are available with different relative $\alpha$-element abundances ($[\alpha/\text{Fe}] = -0.2$, $0.0$ and $0.4$ dex; \citealt{Pietrinferni2021,Pietrinferni2024}). He-enhanced tracks are also available for the $\alpha$-enhanced models \citep{Pietrinferni2021}. All of these tracks are included in our package and can be easily selected by users.

As mentioned previously, \stjl only considers stellar photospheric quantities (e.g., temperature, radius) and interpolates isochrones in the Hertzsprung-Russell diagram space (temperature and luminosity). To transform the isochrones into the required observational magnitudes, 
\stjl integrates with the \bcjl package, which supports interpolation of multiple libraries of BCs. A common interface is defined between the two packages to facilitate the use of isochrones from any of the above stellar model sets with any of the defined BC libraries below. This modularity allows us examine systematics due to the stellar interior models separately from those due to the assumed BCs. 

\subsection{Bolometric Corrections} \label{subsec:bcs}

\bcjl currently supports the MIST \citep{Choi2016} and YBC \citep{Chen2019a} libraries of BCs. These libraries are chosen as they are publicly available, cover wide ranges of stellar parameters, and have been calculated for many observational filter sets. However, they have been calculated using different model atmospheres and spectral synthesis codes such that differences in the BCs should reflect the systematic uncertainties in the underlying theoretical models. Brief descriptions are given below.

The MIST BCs were presented by \citet{Choi2016}. This library is composed of four sub-components that are combined to form the final BC tables that are publicly distributed.\footnote{\url{https://waps.cfa.harvard.edu/MIST/model_grids.html}} The principal component of the library are C3K synthetic spectra \citep{Conroy2018} which use hydrostatic, plane-parallel atmospheres computed with the \textsc{atlas12} code. Synthetic spectra were computed from the atmosphere models with \textsc{synthe} \citep{Kurucz2005, Kurucz2014}. The C3K synthetic spectra are supplemented with the \citet{Koester2010} spectral library for white dwarfs with H-rich atmospheres (DA type), blackbody spectra for stars hotter than 200,000 K, and new synthetic spectra for cool carbon stars with C/O = 1.05. After combining these sub-components, the final tables have extremely wide coverage of the stellar parameter space, covering scaled-solar metallicities ([M/H]) from $-4$ to 0.5 dex, effective temperature from 2,500 K to 1,000,000 K, and surface gravity\footnote{When discussing surface gravity, we always mean log $(g)$ with $g$ in cgs units.} from $-4$ to 9.5 dex. BCs are provided for V-band extinction $0 \leq A_\text{V} \leq 6$ mag with $R_\text{V} = 3.1$ assuming the \citet{Cardelli1989} reddening law.

While the large grid coverage is useful from a practical perspective, there are several phases of stellar evolution where these models are not appropriate; for example, hot O- and B-type stars are known to have radiation-driven winds that affect their spectra, and this effect is not captured in these BCs. Additionally, the MIST BCs were calculated with pre-launch filter curves for JWST/NIRCam and have not been updated to reflect the most recent measurements of the system throughputs. While this is suboptimal, comparisons of the BCs in the F090W and F150W filters between MIST and YBC (which uses the updated filter curves) for hot MS stars shows very good agreement ($\sim0.01$ mag at $T_e =$ 5000 K) as expected since both use variants of \textsc{atlas} models for these stars. By comparison, the YBC and MIST BCs differ much more for cool stars ($\sim0.09$ mag for $T_{e} =$ 2700 K) as a result of the different atmospheres they adopt at low temperatures. As such, we estimate the outdated F090W and F150W filter curves for JWST/NIRCam used in MIST constitute a systematic error of $\sim2\%$. It is worthwhile to compare SFH measurements using MIST BCs to measurements using YBC as this systematic is minor compared to the differences in the atmospheres between these libraries.

The YBC BCs were presented in \citet{Chen2019a}. These are typically associated with the PARSEC stellar models as the YBC BCs are the default choice in the web-based interface to the PARSEC stellar models. YBC is also made up of several sub-components like MIST, but does not provide tables that combine them in a predetermined way. Rather, the BCs for each sub-component are made public\footnote{\url{https://gitlab.com/parsec-group/public_repos/YBC_tables}} and user routines must then stitch them together in an appropriate way. We make use of four sub-components in our initial implementation. In all cases the transitions between the sub-components are smoothly interpolated over to prevent discontinuities in the joint BC tables. The BCs of all sub-components are provided for V-band extinction $0 \leq A_\text{V} \leq 20$ mag with $R_\text{V} = 3.1$ assuming the \citet{Cardelli1989} reddening law with the updated parameters of \citet{ODonnell1994}. The sub-components are summarized below.

\begin{enumerate}
  \item PHOENIX BT-Settl models \citep{Allard2012} are used for cool stars with temperatures 2570 -- 6000 K. The PHOENIX code assumes a hydrostatic, spherical geometry and is specially designed to model the atmospheres of cool systems, including stars and even sub-stellar objects. These models are available for metallicities between $-4$ and $0.5$ dex and surface gravities between $-0.5$ and $6$ dex. Sub-solar metallicites are $\alpha$-enhanced relative to the solar chemical composition with [$\alpha$/Fe] = $0.2$ dex at [M/H] = $-0.5$ dex and [$\alpha$/Fe] = $0.4$ dex for lower metallicities.
  \item The \textsc{atlas9} ODFNEW models of \citet{Castelli2003} are used for hot stars with temperatures 6000 -- 50,118 K. These are hydrostatic, plane-parallel atmospheres computed with the opacity distribution function method which is less accurate for cool stars with significant molecular features compared to direct opacity sampling as is used in PHOENIX and \textsc{atlas12} \citep[see, e.g.,][]{Plez2011}. These models are available for scaled-solar metallicities between $-2.5$ and $0.5$ dex and surface gravities between $0$ and $5$ dex.
  \item The white dwarf models of \citet{Tremblay2009,Koester2010} are used when the surface gravity is greater than 6 dex and the temperature is greater than 5700 K. These synthetic spectra assume H-rich atmospheres (DA type).
  \item Models for hot O- and B-type stars computed with the \texttt{WM-basic} code by \citet{Chen2019a} are used for stars with temperatures greater than 21,000 K and mass outflow rates greater than $10^{-7}$ solar masses per year. Such stars are luminous enough to drive significant outflows that can greatly affect their observed spectra \citep[e.g.,][]{Telford2024} and these models explicitly include the effects of the outflows on the resulting synthetic spectra. Generally stellar evolution codes do not provide mass outflow rates as they only evolve stellar interior quantities and so do not simultaneously resolve outflows driven from the atmosphere. If stellar tracks include mass loss rates, they are typically calculated from scaling relations that depend on the interior properties -- for example, section 2.3 of \citet{Tang2014} and \citet{Chen2015} describe the PARSEC implementation and section 3.7 of \cite{Choi2016} describes the MIST implementation. BaSTI only uses the well-known Reimers mass loss prescription for RGB stars \citep{Reimers1975}. Recently, some commonly used mass-loss prescriptions \cite[e.g.,][]{Vink2001} have been shown to overpredict the mass-loss rates of low-metallicity stars \citep[e.g.,][]{Hawcroft2024,Telford2024}, resulting in spectral models that strongly disagree with observations, particularly at short wavelengths. As most resolved SFH work focuses on populations with sub-solar metallicities, we deem it prudent to support alternative mass-loss prescriptions. We therefore include the ability to calculate mass-loss rates on-the-fly from the stellar interior quantites using different scaling relations. Our fiducial choice is the mass-loss prescription of \citet{Bjorklund2021} which shows significantly better agreement with observations of low-metallicity stars \citep{Hawcroft2024,Telford2024}.
\end{enumerate}

\subsection{Systematic Uncertainty Estimation Procedure}

Given these available libraries of stellar tracks and BCs, we estimate the model-dependent systematic uncertainties in the following straightforward manner: using the PARSEC, MIST, and \texttt{BaSTIv2} stellar track libraries and the MIST and YBC BC libraries, we repeat the CMD fitting procedure under each of their six possible combinations and quote the systematic uncertainties in quantities like the cumulative SFH and metallicity as function of time as the smallest interval that contains all six solutions. We always use non-rotating MIST models and for \texttt{BaSTIv2} we use scaled-solar models with convective overshooting, atomic diffusion, and Reimers mass-loss parameter $\eta=0.3$.

\section{Ages for Individual Stars} \label{sec:ages}

\begin{figure}
\centering
\includegraphics[width=1\linewidth]{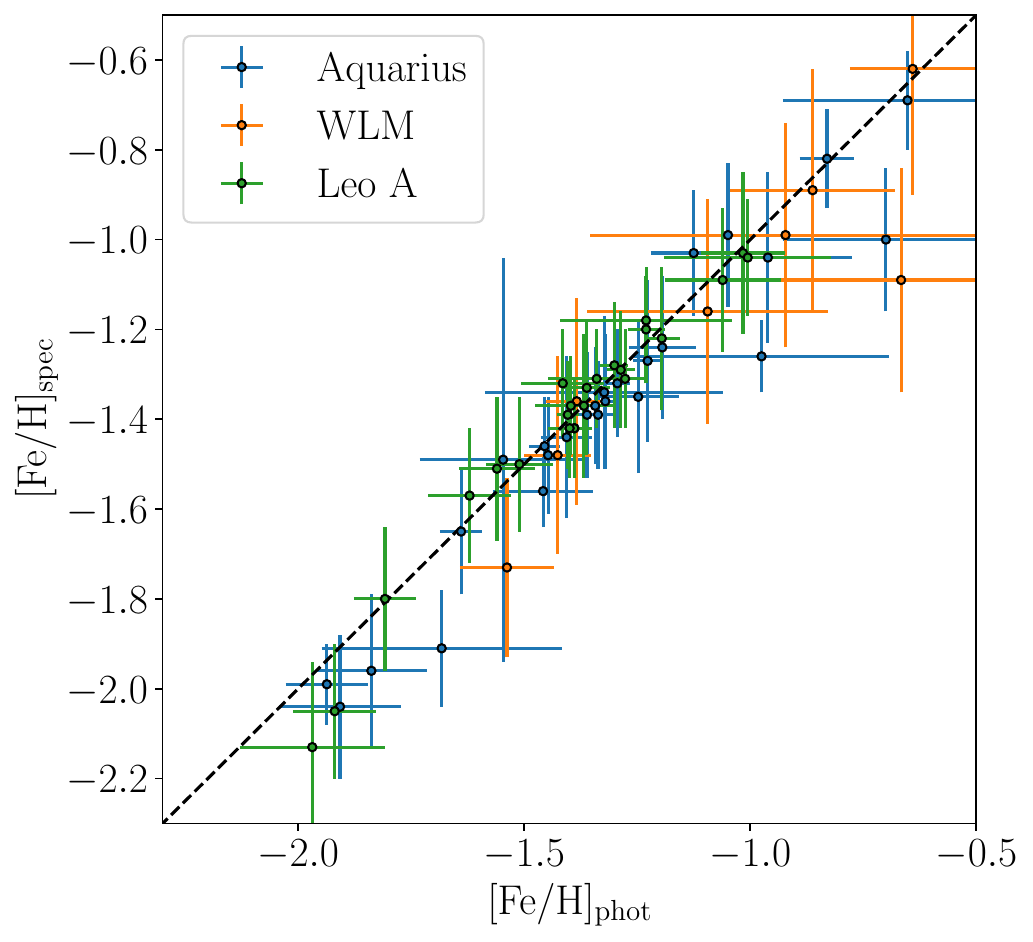}
\caption{Comparison of [Fe/H] measurements of individual red giants from Aquarius, WLM, and Leo A. The vertical axis shows the spectroscopic [Fe/H] measurements, which are used as a prior when measuring the metallicity and ages from the photometry in \S \ref{sec:ages}. The resulting photometrically-measured [Fe/H] values are shown on the horizontal axis. The line where [Fe/H]$_\text{spec} = $ [Fe/H]$_\text{phot}$ is overplotted for comparison. Agreement between the two is generally excellent, although the photometric technique prefers marginally higher metallicities for stars with [Fe/H]$_\text{spec} < -1.8$ dex.}
\label{fig:photfeh}
\end{figure}

For Leo A \citep{Kirby2017a}, Aquarius \citep{Kirby2017a,HermosaMunoz2020}, and WLM \citep{Leaman2009} there are available catalogs of spectroscopic metallicities for red giant stars. With stellar metallicities for individual stars, we can estimate their ages as well based on our high-precision photometry. With ages and metallicities for individual stars in these galaxies, we can perform an independent check of the metallicity evolutions determined from our resolved SFH fits. If a star is observed in a sufficient number of filters, multi-band photometry can be used to extract stellar masses, ages, and metallicities simultaneously \citep[e.g.,][]{Gordon2016}, but given that we have only 2--3 broad bands of photometry for the galaxies in our sample, we require metallicity priors in order to obtain robust ages. Additionally, we marginalize over the range of possible stellar masses for each star in order to increase the precision of our age estimates as the stellar masses are not of particular interest here. 

For a proposed age $t$ and metallicity $Z$ we interpolate an appropriate isochrone using the methods of \ref{subsec:stellar_tracks} and \ref{subsec:bcs}. For all calculations we assume the distance modulii and interstellar reddening values shown in Table \ref{table:summary}, which are also used for the resolved SFH fits in the next section. Letting the list of apparent magnitudes of the star be $\mathbf{m}$, the associated magnitude uncertainties be $\boldsymbol{\sigma}$\footnote{The uncertainties of the observed magnitudes that enter into Equation \ref{eq:ages} ($\boldsymbol{\sigma}$) are estimated from artificial star tests in the same way as the uncertainties are estimated when performing the full resolved SFH measurements in \S \ref{sec:results}.}, the apparent magnitude (error) in filter $i$ be $\mathbf{m}_i$ ($\boldsymbol{\sigma}_i$), and the predicted apparent magnitude from the isochrone for age $t$, stellar mass $M$, and metallicity $Z$ in filter $i$ be $I(t, Z, M)_i$, we write the likelihood of the data given the model as

\begin{equation} \label{eq:ages}
  L(\mathbf{m} | t, Z, M) = \prod_i \left( \frac{1}{\sqrt{2\pi} \, \boldsymbol{\sigma}}_i \exp \left[ -\frac{\left(\mathbf{m}_i - I(t, Z, M)_i \right)^2}{2 \, \boldsymbol{\sigma}_i^2} \right]   \right)
\end{equation}

\noindent We then apply Bayes' theorem, multiplying this likelihood with our priors to form a probability distribution that is proportional to the posterior. As is common, we neglect the Bayesian evidence $P(\mathbf{m})$ that appears in the denominator of Bayes' theorem as it is unnecessary for the BFGS optimization scheme we employ \citep[see Section 6.1 of][]{Nocedal2006}. We assume the \cite{Kroupa2001} initial mass function (IMF) as the prior on the stellar mass and a uniform prior on the stellar age from 1 -- 13.7 Gyr. % 1.6 Gyr lower limit from Leaman2009
The lower limit is a rough estimate of the youngest RGB stars that could fall within the selection region of these studies, while the upper limit is the approximate age of the Universe. We note that early AGB stars, younger than 1 Gyr, could also fall within this selection region, overlapping with the upper RGB; studies suggest 10--20\% of stars in this region may be early AGB stars \citep[e.g.,][]{Gratton2010,Harmsen2023}. The spectroscopically measured metallicity is used as the prior on the metallicity, modeled as a Gaussian distribution with standard deviation equal to the reported metallicity uncertainty. We additionally increase the weighting of the metallicity prior by multiplying it by the number of filters to better match the scale of the photometric likelihood function, which is a product over the $i$ filters. We marginalize over the stellar mass $M$ to estimate $t$ and $Z$ for each star. 

To account for systematic uncertainties due to the assumed stellar models, we repeat the above procedure for each combination of the PARSEC, MIST, and \texttt{BaSTIv2} stellar track libraries and the MIST and YBC BC libraries -- therefore, six measurements are made in total for each star. Our final $t$ and $Z$ values are taken to be the sample mean of these six measurements and we report errors as the sample standard deviation. We note that these uncertainties are several times larger than the formal random uncertainties obtained, for example, via MCMC sampling of the likelihood function as the photometric uncertainties that enter into Equation \ref{eq:ages} are small, typically $\sigma \leq 0.02$ mag. This again highlights the importance of considering the systematic impact of stellar model choice when performing comparisons against high-precision photometry.

The metallicities we find by optimizing Equation \ref{eq:ages} are generally consistent with the spectroscopic measurements -- this can be seen quantitatively in Figure \ref{fig:photfeh}, which compares the spectroscopic [Fe/H] measurements to our photometric measurements that use the spectroscopic values as a prior. These spectrophotometric metallicities and ages are shown superposed with our metallicity evolutions derived from the global resolved SFH in the next section.

% Using capital letters, e.g. “C”, “L”, or “R”, in the cols column identifier will set that specific column in math mode so that $s are unnecessary
\centerwidetable
\begin{deluxetable*}{cccccccc}
\tablecaption{Galaxy Properites \label{table:summary}}
  \tablehead{
    \colhead{Galaxy} & \colhead{R.A. (J2000)} & \colhead{Decl. (J2000)} & \colhead{$E(B-V)$} & \colhead{M$_*$} & \colhead{M$_\text{\HI}$} & \colhead{$D_\odot$} & \colhead{$\mu$} \\ 
    \colhead{} & \colhead{(hms)} & \colhead{(dms)} & \colhead{(mag)} & \colhead{(M$_\odot$)} & \colhead{(M$_\odot$)} & \colhead{(kpc)} & \colhead{(mag)}}
\startdata
WLM & 00:01:58.53 (1) & $-$15:27:27.20 (1) & 0.032 (2) & $4.3 \times 10^{7}$ (3) & $6.3 \times 10^{7}$ (4) & $968 \pm 41$ (5) & $24.93 \pm 0.09$ (5) \\
Leo A & 09:59:26.5 (3) & $+$30:44:47.0 (3) & 0.018 (2) & $1.3 \times 10^{6}$ (6) & $6.91 \times 10^{6}$ (7) & $798 \pm 44$ (8) & $24.51 \pm 0.12$ (8) \\
Aquarius & 20:46:51.8 (3) & $-$12:50:53.0 (3) & 0.043 (2) & $2.6 \times 10^6$ (3) & $4.1 \times 10^6$ (9) & $977 \pm 45$ (10) & $24.95 \pm 0.10$ (10) \\
Leo P & 10:21:42.509 (6) & $+$18:05:16.09 (6) & 0.022 (2) & $5.7 \times 10^5$ (6) & $8.1 \times 10^5$ (11, 12) & $1620 \pm 150$ (11) & $26.05 \pm 0.20$ (11) \\
\enddata
\tablerefs{(1) \cite{Cohen2025}; (2) \cite{Schlegel1998} \& \cite{Schlafly2011}; (3) \cite{McConnachie2012}; (4) \cite{Kepley2007}; (5) \cite{Albers2019}; (6) \cite{McQuinn2024b}; (7) \cite{Hunter2012}; (8) \cite{Dolphin2002}; (9) \cite{Putman2021}; (10) \cite{Cole2014}; (11) \cite{McQuinn2015}; (12) \cite{Giovanelli2013}}
\end{deluxetable*}

\section{Observational Results} \label{sec:obs}
Here we present results applying our MZH SFH model to four isolated Local Group dwarf galaxies. As environmental processes can affect the star formation of dwarf galaxies after infall to a more massive host \citep[][]{Woo2013,Zu2016,Kawinwanichakij2017,Moutard2018,Papovich2018,Davies2019,VanNest2023,Wang2024a,TorresRios2024}, we select only isolated Local Group dwarfs that do not lie within the halo of a more massive galaxy to facilitate comparisons with theory in \S \ref{sec:theory}. We apply our methodology to the WLM, Aquarius, Leo A, and Leo P dwarf galaxies in this work, as they all have high-precision archival photometry available as well as literature SFH and metallicity measurements that we can compare our results to. Properties of these galaxies relevant to our application are summarized in Table \ref{table:summary}. Summary $\tau$ statistics for our SFH fits are given in Table \ref{table:sfhsummary}, where $\tau_{x}$ is the lookback time at which a galaxy formed $x\%$ of its total stellar mass (e.g., $\tau_{10}$ is the lookback time at which the galaxy formed $10\%$ of its total stellar mass).

We use the \cite{Kroupa2001} IMF for all calculations. A metallicity spread of $\sigma = 0.1$ dex (see Equation \ref{eq:hmodel}) is used for all galaxies except WLM, where a value $\sigma = 0.2$ dex gives a better fit to the data. When making our measurements, we assume color excesses $E(B-V)$ based on the dust maps of \cite{Schlegel1998} with the updated scaling of \cite{Schlafly2011} and calculate V-band extinctions ($A_V$) assuming $R_V=3.1$. Reddening is added to the synthetic CMD models via interpolation of the BC tables (see \S \ref{subsec:bcs}) -- both the MIST and YBC tables are tabulated on a grid of $A_V$ values, taking into account the integrated effect of the reddening over each filter in a self-consistent fashion. MIST uses the \cite{Cardelli1989} reddening law and YBC uses the same functional form but with the updated coefficients from \cite{ODonnell1994}. These differ by a few percent at near-UV wavelengths, but in the optical (where our data lie) the typical difference between the two averages to $<1\%$.

\begin{deluxetable*}{LCCCCCC} % caps make data render in math mode
\tablecaption{Summary Statistics of SFH Measurements \label{table:sfhsummary}}
\tablehead{
  \colhead{$\tau$} & \colhead{PARSEC + YBC} & \colhead{PARSEC + MIST} &
  \colhead{\texttt{BaSTIv2} + YBC} & \colhead{\texttt{BaSTIv2} + MIST} & \colhead{MIST + YBC} &
  \colhead{MIST + MIST}
}
\tablecolumns{7} % Needed to center the \cutinhead
\startdata
\cutinhead{WLM}
\tau_{10} & 11.50^{+0.08}_{-0.09} & 11.69^{+0.04}_{-0.06} & 12.49^{+0.04}_{-0.05} & 12.28^{+0.05}_{-0.05} & 12.54^{+0.02}_{-0.02} & 13.21^{+0.01}_{-0.01} \\
\tau_{25} & 8.18^{+0.02}_{-0.03} & 7.41^{+0.02}_{-0.03} & 7.86^{+0.04}_{-0.05} & 7.63^{+0.03}_{-0.03} & 11.66^{+0.02}_{-0.02} & 7.88^{+0.01}_{-0.02} \\
\tau_{50} & 4.68^{+0.07}_{-0.05} & 4.68^{+0.12}_{-0.08} & 4.74^{+0.02}_{-0.02} & 4.64^{+0.02}_{-0.02} & 5.81^{+0.04}_{-0.06} & 5.39^{+0.05}_{-0.08} \\
\tau_{75} & 2.41^{+0.03}_{-0.01} & 2.37^{+0.03}_{-0.02} & 2.26^{+0.01}_{-0.01} & 2.26^{+0.01}_{-0.01} & 2.65^{+0.01}_{-0.01} & 2.54^{+0.02}_{-0.01} \\
\tau_{90} & 1.24^{+0.01}_{-0.01} & 1.21^{+0.01}_{-0.01} & 1.14^{+0.01}_{-0.01} & 1.08^{+0.01}_{-0.02} & 1.12^{+0.01}_{-0.01} & 1.09^{+0.01}_{-0.01} \\
\cutinhead{Leo A}
\tau_{10} & 7.04^{+0.17}_{-0.07} & 7.36^{+0.21}_{-0.17} & 7.52^{+0.14}_{-0.09} & 7.75^{+0.13}_{-0.06} & 7.94^{+0.43}_{-0.03} & 8.45^{+0.10}_{-0.08} \\
\tau_{25} & 5.36^{+0.05}_{-0.04} & 5.38^{+0.06}_{-0.05} & 5.69^{+0.09}_{-0.05} & 5.79^{+0.09}_{-0.06} & 5.97^{+0.07}_{-0.03} & 5.94^{+0.12}_{-0.09} \\
\tau_{50} & 3.58^{+0.03}_{-0.04} & 3.54^{+0.05}_{-0.07} & 3.89^{+0.04}_{-0.04} & 3.95^{+0.04}_{-0.04} & 4.02^{+0.06}_{-0.04} & 3.94^{+0.05}_{-0.05} \\
\tau_{75} & 2.18^{+0.01}_{-0.03} & 2.16^{+0.01}_{-0.03} & 2.32^{+0.02}_{-0.04} & 2.30^{+0.03}_{-0.04} & 2.38^{+0.03}_{-0.04} & 2.33^{+0.02}_{-0.03} \\
\tau_{90} & 1.13^{+0.01}_{-0.05} & 1.11^{+0.01}_{-0.04} & 1.04^{+0.01}_{-0.02} & 1.06^{+0.01}_{-0.02} & 1.00^{+0.01}_{-0.04} & 1.02^{+0.01}_{-0.03} \\
\cutinhead{Aquarius}
\tau_{10} & 13.23^{+0.01}_{-0.02} & 13.21^{+0.01}_{-0.02} & 13.21^{+0.01}_{-0.03} & 13.18^{+0.01}_{-0.04} & 13.09^{+0.03}_{-0.04} & 13.20^{+0.02}_{-0.05} \\
\tau_{25} & 8.87^{+0.25}_{-0.08} & 8.86^{+0.09}_{-0.07} & 12.35^{+0.09}_{-0.19} & 12.15^{+0.07}_{-0.25} & 9.79^{+0.04}_{-0.03} & 10.89^{+0.81}_{-0.90} \\
\tau_{50} & 6.70^{+0.05}_{-0.10} & 6.50^{+0.11}_{-0.12} & 7.73^{+0.05}_{-0.15} & 8.20^{+0.02}_{-0.14} & 7.74^{+1.00}_{-0.56} & 7.95^{+0.23}_{-0.65} \\
\tau_{75} & 4.26^{+0.03}_{-0.03} & 4.27^{+0.03}_{-0.04} & 4.76^{+0.04}_{-0.09} & 4.77^{+0.06}_{-0.17} & 4.75^{+0.08}_{-0.10} & 4.53^{+0.35}_{-0.09} \\
\tau_{90} & 2.36^{+0.01}_{-0.04} & 2.36^{+0.02}_{-0.04} & 2.49^{+0.06}_{-0.05} & 2.51^{+0.20}_{-0.05} & 2.67^{+0.03}_{-0.03} & 2.66^{+0.09}_{-0.05} \\
\cutinhead{Leo P}
\tau_{10} & 13.29^{+0.01}_{-0.04} & 13.28^{+0.01}_{-0.04} & 13.09^{+0.03}_{-0.08} & 13.06^{+0.03}_{-0.11} & 13.30^{+0.01}_{-0.04} & 13.32^{+0.01}_{-0.04} \\
\tau_{25} & 12.70^{+0.03}_{-0.10} & 12.67^{+0.02}_{-2.86} & 8.87^{+0.62}_{-0.21} & 8.84^{+1.26}_{-0.28} & 12.72^{+0.02}_{-0.10} & 12.78^{+0.01}_{-0.10} \\
\tau_{50} & 5.81^{+0.31}_{-0.22} & 5.84^{+0.12}_{-0.10} & 5.30^{+0.15}_{-0.08} & 5.46^{+0.38}_{-0.33} & 6.17^{+0.09}_{-0.16} & 5.96^{+0.16}_{-0.24} \\
\tau_{75} & 3.08^{+0.50}_{-0.09} & 3.13^{+0.18}_{-0.08} & 3.20^{+0.12}_{-0.46} & 2.99^{+0.37}_{-0.12} & 3.84^{+0.12}_{-0.16} & 3.31^{+0.13}_{-0.12} \\
\tau_{90} & 1.62^{+0.09}_{-0.06} & 1.62^{+0.07}_{-0.03} & 1.37^{+0.12}_{-0.05} & 1.40^{+0.18}_{-0.03} & 1.91^{+0.15}_{-0.13} & 1.45^{+0.13}_{-0.03} \\
\enddata
\tablecomments{$\tau_{x}$ values indicate the lookback time at which the galaxy formed $x$ percent of its stellar mass -- for example, $\tau_{10}$ indicates the lookback time at which the galaxy formed $10\%$ of its total stellar mass. All values are in Gyr and all uncertainties are random (statistical) uncertainties. $\tau$ values measured under our new MZH metallicity evolution model assuming each combination of stellar track library (PARSEC, \texttt{BaSTIv2}, and MIST) and bolometric correction library (YBC and MIST) are given.}
\end{deluxetable*}

\subsection{WLM} \label{subsec:wlm}

\begin{figure*}
\centering
\includegraphics[width=0.85\textwidth]{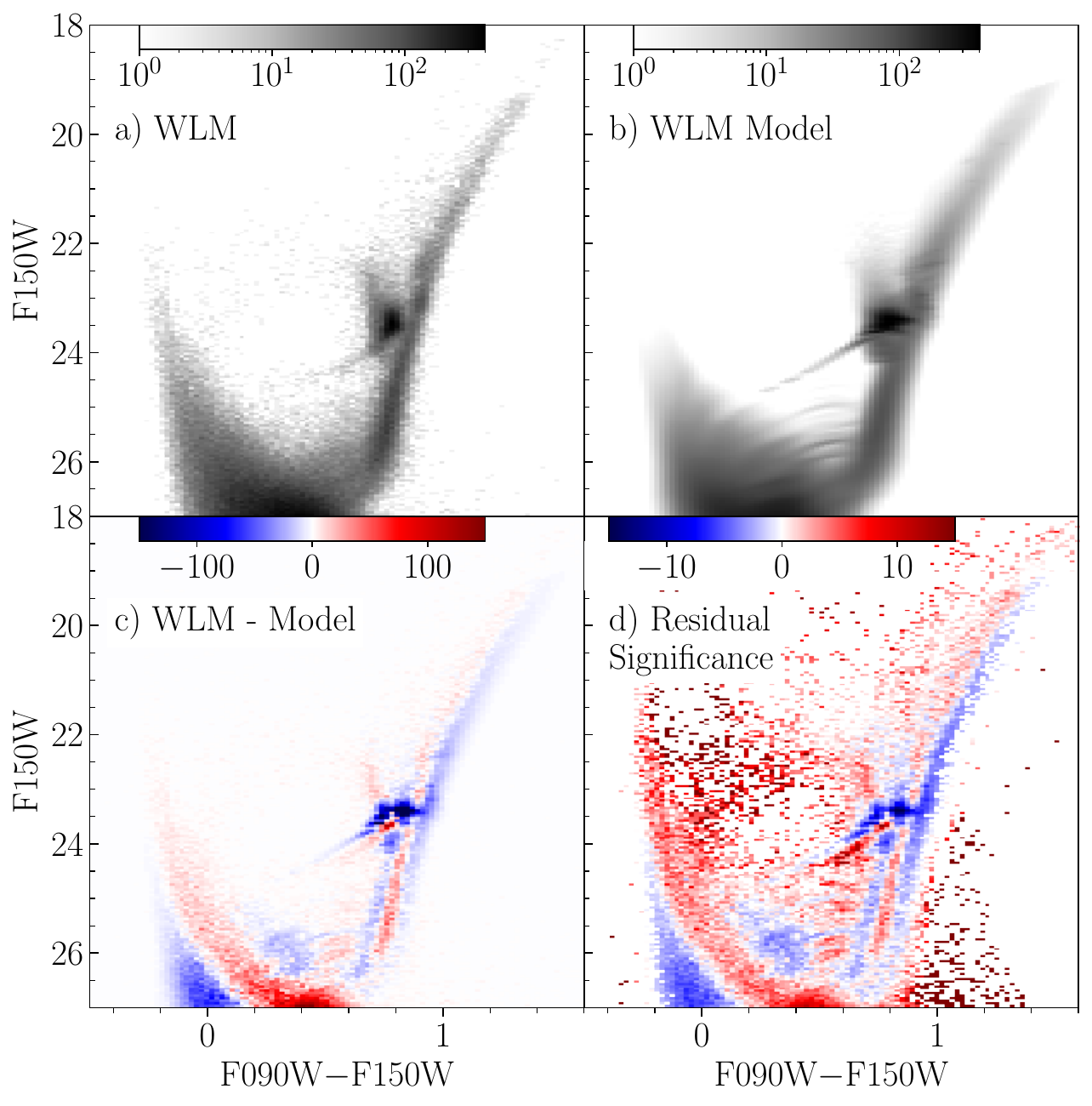} \\
\caption{Comparison of the Hess diagram of WLM with our best-fit model. (a) Hess diagram of WLM constructed from the JWST/NIRCam photometric catalog of \cite{Weisz2024}. (b) Best-fit model Hess diagram using our MZH metallicity evolution model with PARSEC stellar tracks and YBC bolometric corrections. (c) Residual between the observed Hess diagram and the best-fit model in raw star counts. (d) Residual between the observed Hess diagram and the best-fit model in units of standard deviations (i.e., the residual significance).}
\label{fig:wlm_hess}
\end{figure*}

\begin{figure*}[hpbt]
\centering
\includegraphics[width=0.85\textwidth]{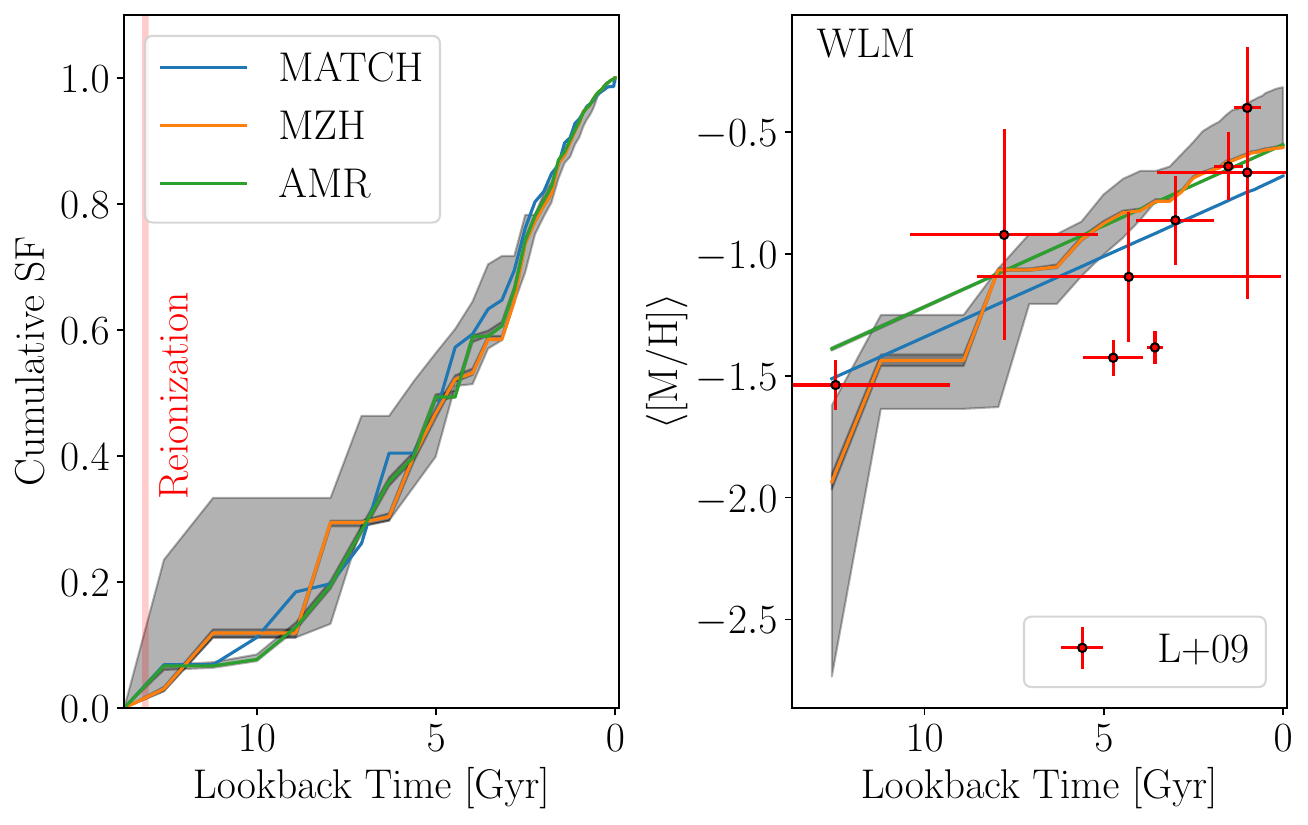} \\
\includegraphics[width=0.85\textwidth]{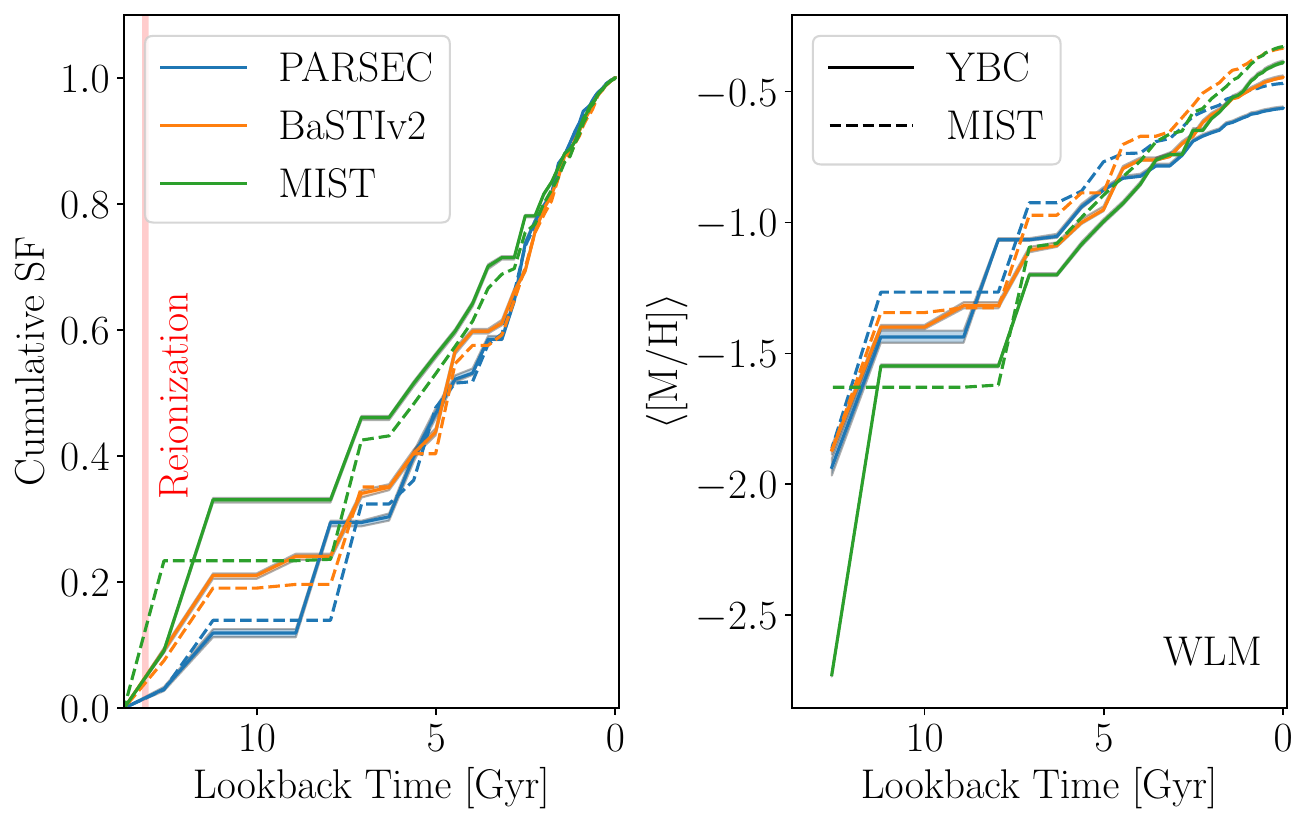}
\caption{\emph{Top row:} The cumulative SFHs (left) and AMRs (right) of WLM derived from the JWST/NIRCam imaging. Included are measurements with a linear AMR (green line) and with our hierarchical MZH model (orange line), both of which assume the PARSEC stellar models (\S \ref{subsec:stellar_tracks}) with the YBC bolometric corrections (\S \ref{subsec:bcs}). For comparison we also show the result of \cite{McQuinn2024} (blue line), who used the \textsc{match} code on the same data. The systematic uncertainty on the MZH SFH and metallicity evolution is shown as a light shaded region (see \S \ref{sec:systematics} for more details). The (much smaller) random uncertainty is shaded more darkly. Overplotted on the right panel are age and metallicity measurements for individual red giant stars obtained by combining the spectroscopic metallicities of \cite{Leaman2009} with our photometry (see \S \ref{sec:ages}). \emph{Bottom row:} The six individual measurements of the SFH of WLM under the MZH model for different combinations of stellar tracks (PARSEC, MIST, and \texttt{BaSTIv2}) and bolometric corrections (MIST, YBC) that were used to define the systematic uncertainty region in the upper row. The PARSEC and \texttt{BaSTIv2} tracks give fairly consistent results, while the MIST stellar tracks prefer more early star formation. The measurements under the MIST bolometric corrections prefer higher present-day metallicities (by $\sim0.1$ dex) compared to the solutions with the YBC bolometric corrections.}
\label{fig:wlm_sfh}
\end{figure*}

We first apply our new MZH metallicity evolution model to the WLM dwarf irregular galaxy \citep[$M_v\approx-14.2$ mag, $\text{M}_*\approx4.3\times10^7$ M$_\odot$,][]{McConnachie2012} which we previously studied with an AMR metallicity evolution model in \cite{Garling2025}. WLM is isolated and gas-rich ($M_\HI / M_* > 1$) with a distance modulus of $\mu=24.93\pm0.09$ mag \citep[$d=968_{-40}^{+41}$ kpc,][]{Albers2019} -- it is therefore near enough that deep imaging with HST/ACS and JWST/NIRCAM has enabled robust measurement of its resolved SFH \citep{Albers2019,McQuinn2024,Cohen2025,Garling2025}. These works have demonstrated that WLM has been actively star-forming over most of its history, as expected for isolated, gas-rich dwarf galaxies, with the notable exception of a period of reduced star formation in the $\sim3$ Gyr following reionization. 

We utilize the public JWST/NIRCam photometric catalogs and artificial star tests (ASTs) for WLM that were released as part of the JWST Resolved Stellar Populations Early Release Science Program \citep{Weisz2024}. We apply photometric quality cuts following the recommendations of \cite{Weisz2024} such that filtered photometric catalog and ASTs should be essentially identical to those used by \cite{Cohen2025}, who measured the global SFH of WLM and additionally examined spatially-resolved trends in SFH across the galaxy. We note that the global SFH measurement of \cite{Cohen2025} is also nearly identical to that of \cite{McQuinn2024}, who used the same measurement techniques (e.g., the \textsc{match} code) with earlier versions of the photometry and ASTs prior to the public release.

We compare the observed Hess diagram of WLM with our best-fit MZH metallicity evolution model in Figure \ref{fig:wlm_hess}. The features in the residuals are largely the same as those shown in Figure 8 of \cite{Garling2025}, which fit the SFH of WLM using the linear AMR model. As discussed in that work, the main areas of poor fit are 

\begin{enumerate}
\item An excess of stars in the model along the red side of the RGB.
\item Large residuals around the red clump.
\item An excess of observed stars on the red side of the upper main sequence (e.g., $x = 0$, $y = 24$).
\item An excess of stars in the model on the blue side of the main sequence at faint magnitudes (e.g., $x = 0$, $y = 26.5$).
\end{enumerate}

\noindent Similar residual patterns were observed by \cite{McQuinn2024} who used the \textsc{match} code to measure the resolved SFH from the same JWST/NIRCam data.

Our SFH and metallicity evolution measurements are shown in Figure \ref{fig:wlm_sfh}. We present solutions under both the traditional linear AMR model and our new MZH metallicity evolution model. For comparison we also plot the solution of \cite{McQuinn2024}, who measured the SFH of WLM with the same data using the \textsc{match} code. On the upper right panel we overplot ages and metallicities for single red giant stars measured by combining our photometry with the spectroscopic metallicities of \cite{Leaman2009} using the method presented in \S \ref{sec:ages}. 

We note first that the measurement we present here with the linear AMR model shows excellent agreement with our previous measurement in \cite{Garling2025} which used isochrones obtained from the \href{https://stev.oapd.inaf.it/cgi-bin/cmd}{PARSEC webform}. Quantitatively, the differences between the two measurements are of similar magnitude to their respective random uncertainties, which are quite low as the CMD of WLM is densely populated. Note that in Figure \ref{fig:wlm_sfh}, the light shaded region shows the systematic uncertainty region, within which there is a much smaller random uncertainty region shaded more darkly. As the PARSEC stellar tracks and YBC BCs underlying the webform are mostly the same as those included in our interpolation routines presented in \S \ref{subsec:stellar_tracks} and \S \ref{subsec:bcs}, the fact that the measurements agree so well provides a real-world demonstration that our new on-the-fly interpolation schemes perform as expected.

As in \cite{Garling2025}, our measurement under the linear AMR model agrees quite well with that of \cite{McQuinn2024}. We find a linear AMR slope that is nearly identical and an intercept that is $\sim0.1$ dex higher. Our result under the MZH model is somewhat different -- the metallicity evolution converges to nearly the same metallicity at present-day, but prefers a lower metallicity at early times which results in some deviation from the linear AMR in the earliest time bins ($t > 10$ Gyr). Agreement with both \cite{McQuinn2024} and our linear AMR model solution is still quite good for the MZH model -- as WLM has had relatively steady star formation over the last $\sim9$ Gyr, our MZH model naturally produces a fairly linear relation between age and metallicity.

The systematic uncertainties shown in Figure \ref{fig:wlm_sfh} are measured using the method described in \S \ref{sec:systematics}; in short, the MZH model is used to derive the SFH of WLM under each combination of three stellar model libraries (PARSEC, MIST, and \texttt{BaSTIv2}) and two bolometric corrections libraries (MIST, YBC). The individual solutions are shown in the bottom row of Figure \ref{fig:wlm_sfh} and show some interesting patterns. The MIST stellar tracks consistently prefer more early star formation compared to the PARSEC and \texttt{BaSTIv2} stellar tracks. Quantified by $\tau_{50}$, the lookback time at which the galaxy formed half of its stellar mass, PARSEC and \texttt{BaSTIv2} are consistent with values of $4.68^{+0.07}_{-0.05}$ Gyr and $4.74^{+0.02}_{-0.02}$ Gyr respectively, while the MIST stellar tracks prefer a value $5.81^{+0.04}_{-0.06}$ Gyr, which is about a Gyr earlier. The differences in the SFH caused by varying the stellar tracks are larger than those caused by varying the BCs. However, the bottom right panel shows that the MIST BCs prefer systematically higher metallicities compared to the YBC BCs, though the difference is minor; the mean difference between solutions using different BC libraries when fixing the stellar track library is $\sim0.1$ dex.

The upper right panel of Figure \ref{fig:wlm_sfh} also shows the ages and metallicities we measure for nine red giant stars in WLM by combining the spectroscopic metallicities of \cite{Leaman2009} with our photometry using the methodology described in \S \ref{sec:ages}. Unfortunately, these measurements are not very precise, due predominantly to the large metallicity uncertainties on the spectroscopic measurements of \cite{Leaman2009}, which are typically $\sigma \geq 0.25$ dex. While there are few stars in the sample, and they have fairly high age and metallicity uncertainties, most of the stars lie around the AMRs found via resolved SFH fitting. We note that there are two outliers at [M/H] $\sim -1.5$ dex with ages $\leq5$ Gyr. These stars are on the extreme blue edge of the RGB in our JWST/NIRCam photometry, and could be early AGBs, red supergiants, or core helium burning blue loop stars. If these are blue loop stars, then the metallicity calibration for red giants used by \cite{Leaman2009} would be inappropriate, and so our metallicity prior for these stars would be incorrect, which could account for their deviation from the rest of the sample.

\subsection{Aquarius} \label{subsec:aquarius}

To extend our MZH results to lower stellar masses, we apply our MZH model to the Aquarius dwarf irregular galaxy \citep[$M_v\approx-10.47$ mag, $\text{M}_*\approx2.6\times10^6$ M$_\odot$,][]{McConnachie2012} -- as the present-day stellar mass of Aquarius is about an order of magnitude lower than that of WLM, Aquarius probes a different regime of $\langle[\text{M}/\text{H}]\rangle(M_*)$. However, Aquarius has several properties that are similar to WLM -- both are isolated, gas-rich ($M_\HI / M_* > 1$), and about 1 Mpc away \citep[we assume $d=977 \pm 45$ kpc for Aquarius;][]{Cole2014}. At this distance, the HST/ACS imaging presented in \cite{Cole2014} is deep enough to resolve the subgiant branch and enable measurements of lifetime SFHs. The SFH measurement presented in \cite{Cole2014} has poor time resolution by modern standards, but \cite{McQuinn2024b} applied the \textsc{match} code to the same HST/ACS data to derive an independent SFH measurement with better time resolution (equal, in fact, to the time resolution of the WLM SFH measurements of \citealt{McQuinn2024,Cohen2025}). We use the same photometric catalogs and ASTs (including quality cuts) as used by \cite{McQuinn2024b} for our analysis here. 

\begin{figure*}
\centering
\includegraphics[width=0.85\textwidth]{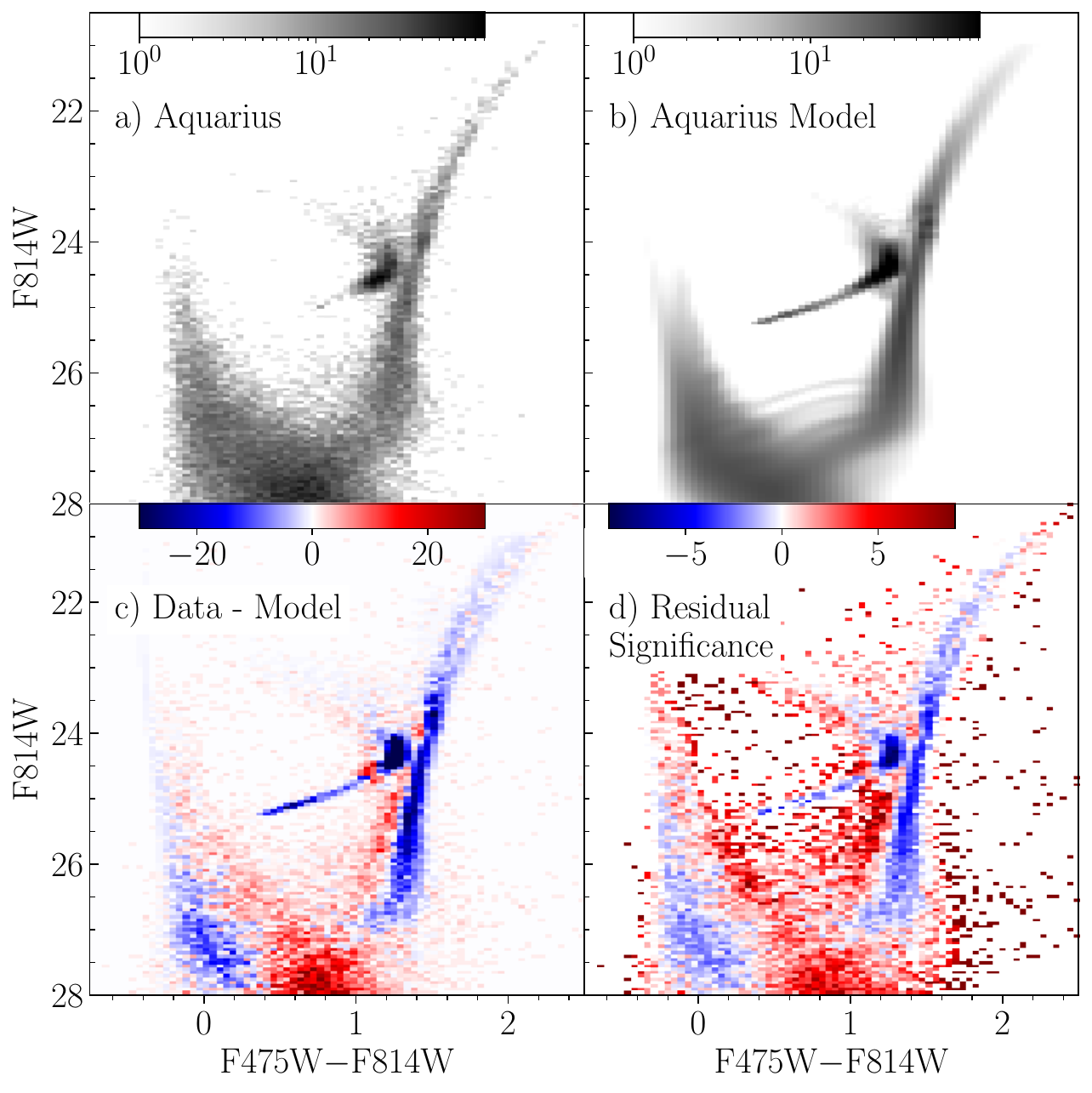} \\
\caption{Comparison of the Hess diagram of Aquarius with our best-fit model. The panels follow the same layout as in Figure \ref{fig:wlm_hess}.}
\label{fig:aquarius_hess}
\end{figure*}

We compare the observed Hess diagram of Aquarius with our best-fit MZH model in Figure \ref{fig:aquarius_hess}. In our discussion of the WLM fit (Figure \ref{fig:wlm_hess}), we noted a color offset between the model and the data along the upper RGB. In comparison, the upper RGB of Aquarius is very well fit, while the lower RGB (fainter than the red clump) in the model Hess diagram of Aquarius exhibits a minor color offset. Residuals around the red clump remain large, as was the case for WLM. We observe a less significant, but still present, excess of model stars on the blue side of the faint MS. Similar residual patterns can be seen in Figure 3 of \cite{Cole2014} showing their model Hess diagram for Aquarius.

\begin{figure*}[hpbt]
\centering
\includegraphics[width=0.85\textwidth]{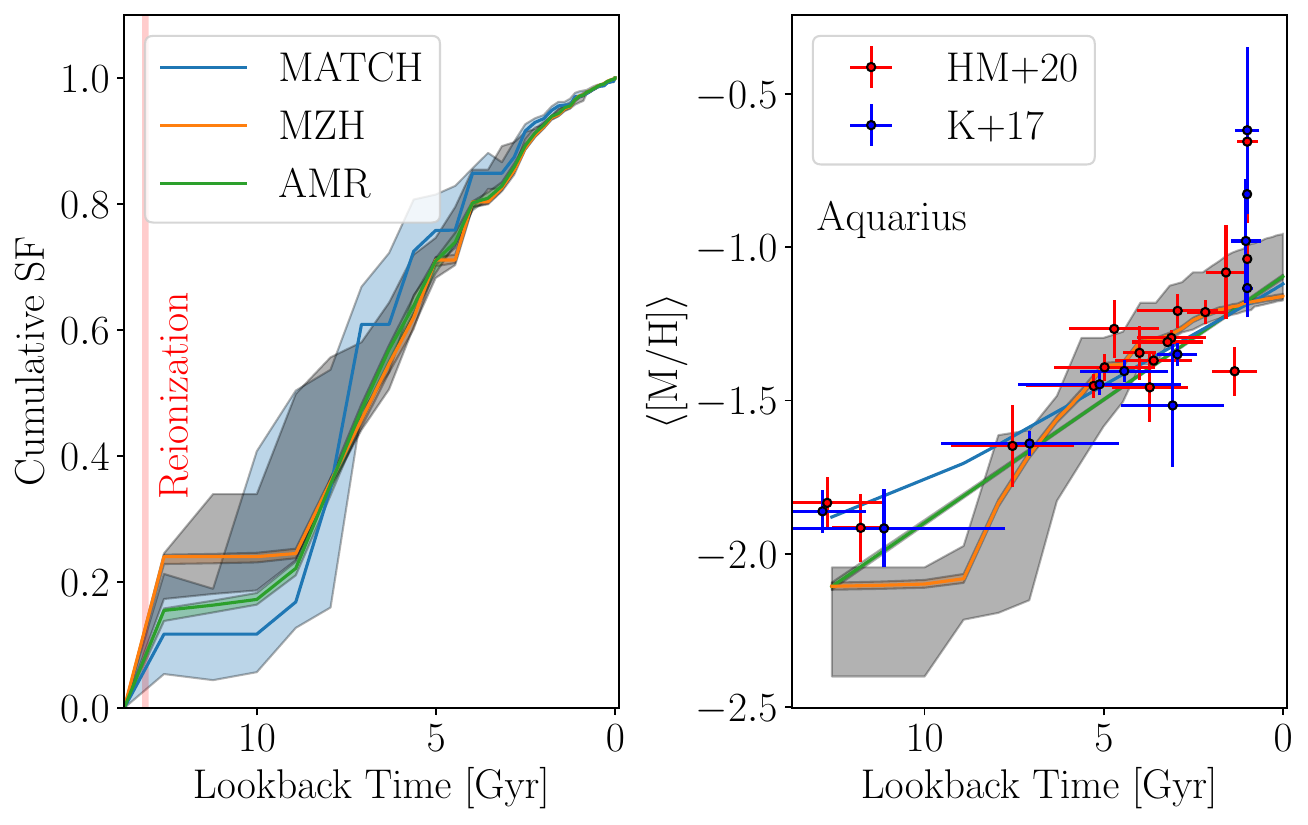} \\
\includegraphics[width=0.85\textwidth]{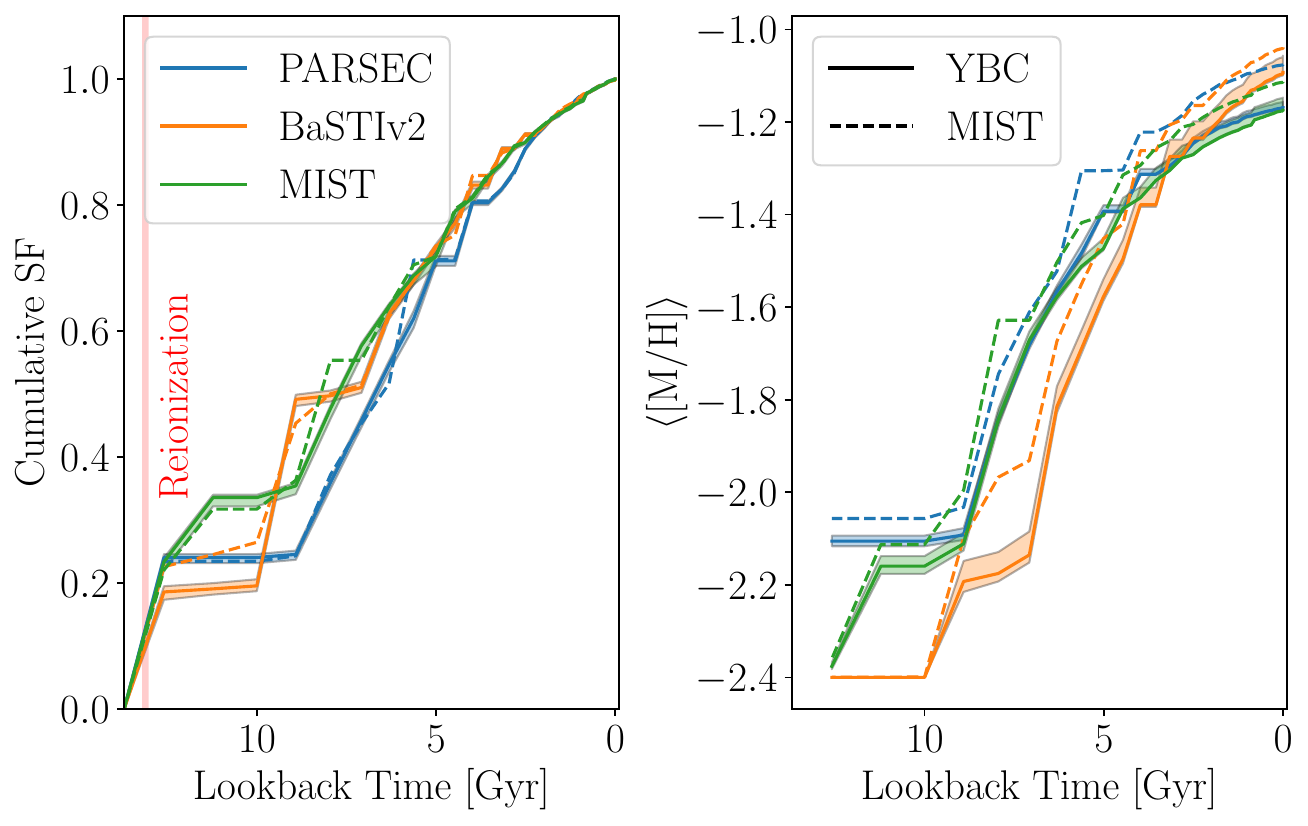}
\caption{The results of our resolved SFH and metallicity evolution fit for Aquarius based on the HST/ACS data following the same format as Figure \ref{fig:wlm_sfh}. In the top row the result of \cite{McQuinn2024b} (blue line), who used the \textsc{match} code on the same data, is shown for comparison. Overplotted on the upper right panel are age and metallicity measurements for individual red giant stars obtained by combining the spectroscopic metallicities of \cite{HermosaMunoz2020} (HM+20, red) and \cite{Kirby2017a} (K+17, blue) with our photometry (see \S \ref{sec:ages}).}
\label{fig:aquarius_sfh}
\end{figure*}

Our SFH and metallicity evolution measurements are shown in Figure \ref{fig:aquarius_sfh}. We present solutions under both the traditional linear AMR model and our new MZH metallicity evolution model. We also show the solution of \cite{McQuinn2024}, who measured the SFH of Aquarius with the same data using the \textsc{match} code. While our measurements are statistically consistent with those of \cite{McQuinn2024}, there are a few differences between our measurements that are apparent. While we reproduce the pause in star formation from roughly 10--13 Gyr ago noted by \cite{McQuinn2024b}, our measurements find larger initial bursts of star formation. In our fiducial MZH measurement, we find that Aquarius formed $\sim25\%$ of its stellar mass in the first 2 Gyr of cosmic time, while \cite{McQuinn2024b} found that only $\sim10\%$ of the stellar mass formed in this period. In our measurement, this difference is compensated by lower average SFR from 5--8 Gyr ago compared to \cite{McQuinn2024b}. Therefore, the average age of stars Aquarius is slightly older in our measurement compared to \cite{McQuinn2024b}. As older stellar populations have higher stellar mass-to-light ratios, we find a slightly higher total stellar mass formed in Aquarius of $4.35^{+0.58}_{-0.41} \times 10^6$ M$_\odot$, while \cite{McQuinn2024b} find $3.9 \times 10^6$ M$_\odot$.

We perform an independent check of the metallicity evolution found in our resolved SFH fit by comparing against the ages and metallicities of individual red giant stars following the methods in \S \ref{sec:ages}. \cite{Kirby2017a} measured the metallicities of 23 RGB stars from Keck/DEIMOS spectra and found a median [Fe/H] = $-1.49$ dex with standard deviation 0.37 dex. Later, \cite{HermosaMunoz2020} measured the metallicities of 53 RGB stars from VLT/FORS2 spectra and found a median [Fe/H] = $-1.58$ dex with standard deviation 0.29 dex. By combining these spectroscopic metallicity measurements with our high-precision photometry we are able to measure the ages of the stars that fall within the HST/ACS field of view. These single-star age and metallicity measurements are shown overplotted on the upper right panel of Figure \ref{fig:aquarius_sfh}. At older ages ($>8$ Gyr ago), the single-star measurements have slightly higher metallicities compared to the resolved SFH result ($\sim0.2$ dex).

We note that seven of the eight stars with measured ages $<2$ Gyr are found to have ages $\sim1$ Gyr, which is the youngest age allowed by our prior. If we allow younger ages for these stars we obtain better fits (statistically speaking), but this is physically unlikely given that these stars are selected to be on the upper RGB. To assess the expected fraction of stars younger than 1 Gyr that could contaminate the RGB selection region, we construct a mock CMD\footnote{We construct the mock CMD using the process described in Appendix \ref{appendix:leoA}.} of Aquarius assuming our measured SFH and data quality and measure the fraction of stars younger than 1 Gyr in the RGB selection region to be $\sim10\%$, indicating it is unlikely ($p<0.01$) that 7/24 stars in our sample would truly be younger than 1 Gyr. Overall, the young stars for which we measure ages $\approx1$ Gyr are likely not well fit, and so their metallicities and ages may not be reliable. Overall, the majority of the stars in the sample have intermediate ages (2 -- 8 Gyr ago) and show excellent agreement with the results from the resolved SFH fits.

\subsection{Leo A} \label{subsec:leoA}

Leo A is of similar stellar mass to Aquarius with $\text{M}_*\approx1.3\times10^6$ M$_\odot$ \citep{McQuinn2024b} and is similarly isolated and gas-rich \citep[$M_\HI / M_* \ge 2$;][]{Hunter2012,McConnachie2012,Putman2021}. The first global SFH for Leo A was presented by \cite{Cole2007} based on HST/ACS observations in the F475W and F814W filters. These data were recently reanalyzed by \cite{McQuinn2024b}, who used \textsc{match} to derive a new SFH with improved time resolution. We utilize the same photometric catalogs and ASTs as were used by \cite{McQuinn2024b} to facilitate comparison between our work. We take the distance of Leo A to be $798 \pm 44$ kpc ($\mu = 24.51 \pm 0.12$ mag) based on measurements of standard candle RR Lyrae stars by \cite{Dolphin2002}. This measurement is generally consistent with others from the literature measured via other methods, such as the TRGB magnitude \citep{Bernard2013} and Cepheid variable stars \citep{Tammann2011,Bernard2013}.

\begin{figure*}
\centering
\includegraphics[width=0.85\textwidth]{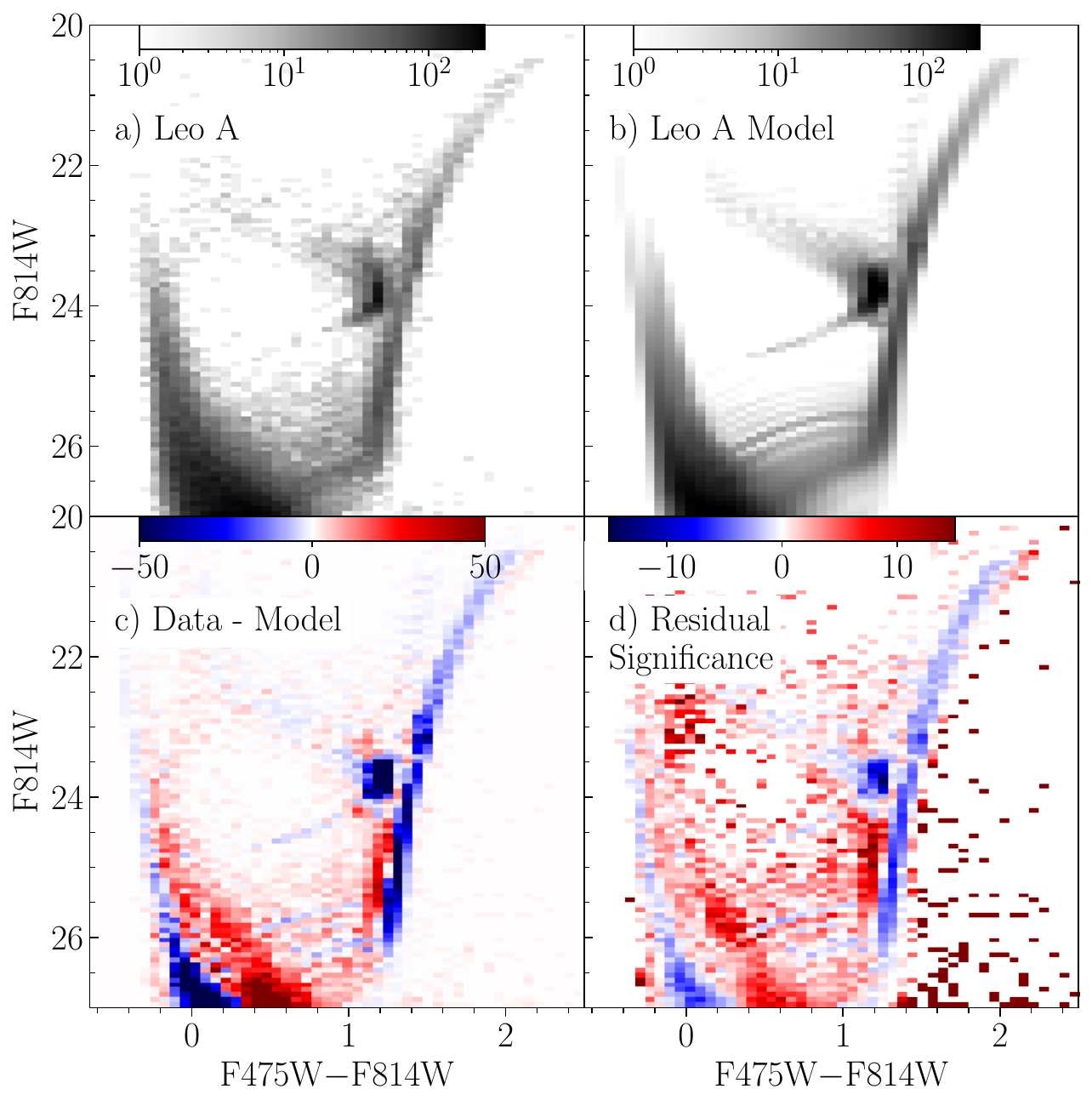} \\
\caption{Comparison of the Hess diagram of Leo A with our best-fit model. The panels follow the same layout as in Figure \ref{fig:wlm_hess}.}
\label{fig:leoA_hess}
\end{figure*}

Figure \ref{fig:leoA_hess} compares the observed Hess diagram of Leo A with our best-fit MZH metallicity evolution model. Overall, the residual patterns we observe are very similar to those we saw in Aquarius (Figure \ref{fig:aquarius_hess}). Of particular note in Leo A is the significant population of young, high-mass main sequence stars -- these are reasonably fit in our model, but the upper main sequence in the data is noticably broader than in the model. This has been noted in previous works and improved fits may be obtained by randomly adding extinction to the young stars in order to broaden and redden the main sequence \citep[e.g.,][]{Dolphin2003,Albers2019}. We experimented with this approach and found that it could reduce the statistical significance of the residuals on the upper main sequence, but it did not noticably affect the overall SFH or metallicity evolution results. Additionally, as all of the galaxies studied in this work are low metallicity and therefore likely to be dust-poor, it is unclear from a physical standpoint what model one should use for the distribution of extinction values to apply to young stars. Therefore, we do not include such age-dependent differential extinction in our models here.

From spectroscopy of young blue supergiants, the metallicity of Leo A is estimated to be [M/H] $\approx -1.35$ dex \citep{Urbaneja2023}. The young stars appear to have solar $\alpha$-element abundance patterns as the oxygen abundance is consistent with the overall metallicity; $12 +$ log(O/H) $= 7.4 \pm 0.2$ has been measured from \HII regions \citep{Zee2006,RuizEscobedo2018}, corresponding to roughly [O/H] $\approx -1.3$ dex. \cite{Kirby2017a} measured spectroscopic metallicities of 113 RGB stars and found a median [Fe/H] = $-1.59$ dex with standard deviation 0.35 dex. This is only $\sim0.2$ dex lower than the metallicities of the young supergiants, indicating that the chemical evolution of Leo A has been slow in comparison to other galaxies of similar stellar mass. A slow chemical evolution is also implied by the AMR measured by \cite{Kirby2017a} based on single-star age and metallicity estimates (see their Figure 12). This conclusion is consistent with the AMR fit by \cite{Cole2007} in their resolved SFH measurement -- they find the data can be adequately explained with a constant metallicity of [M/H] = $-1.4 \pm 0.2$ dex.

\begin{figure*}[hpbt]
\centering
\includegraphics[width=0.85\textwidth]{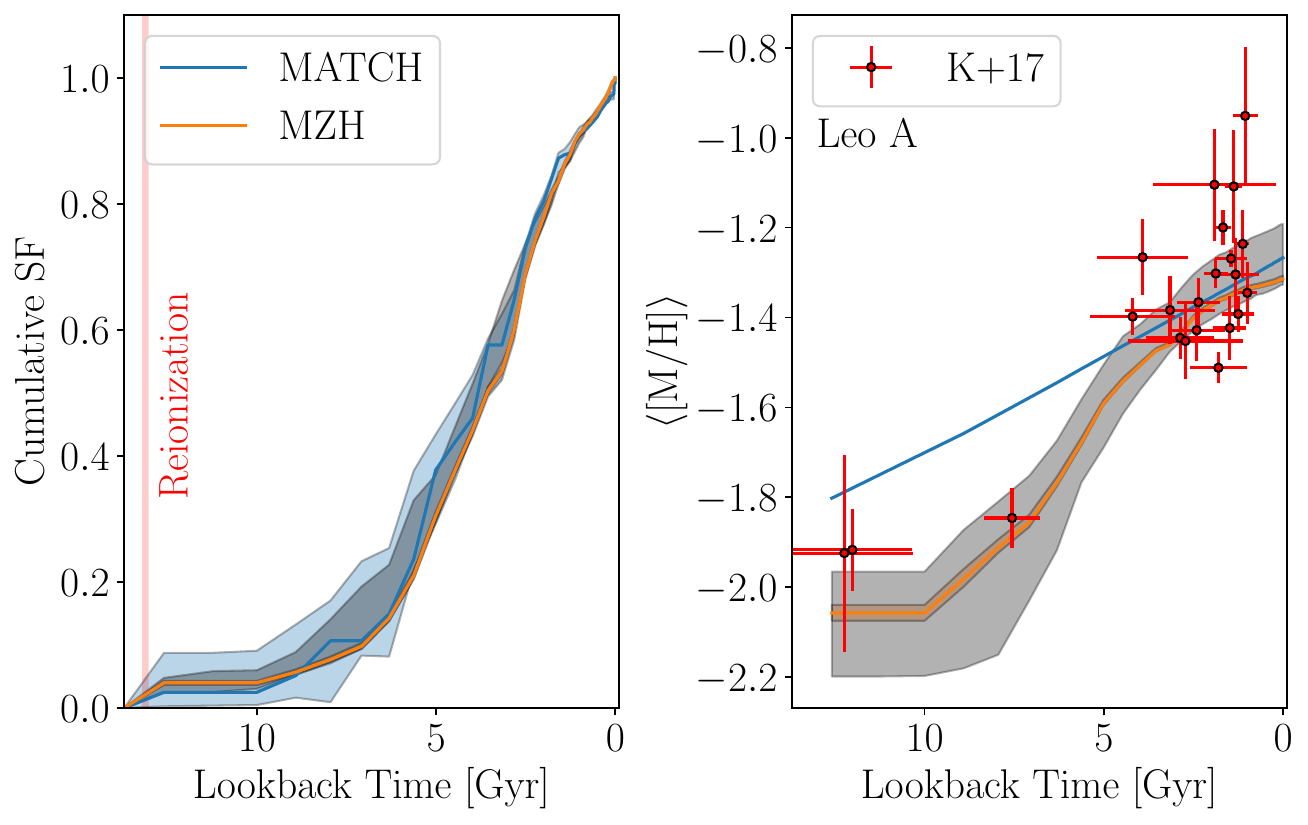} \\
\includegraphics[width=0.85\textwidth]{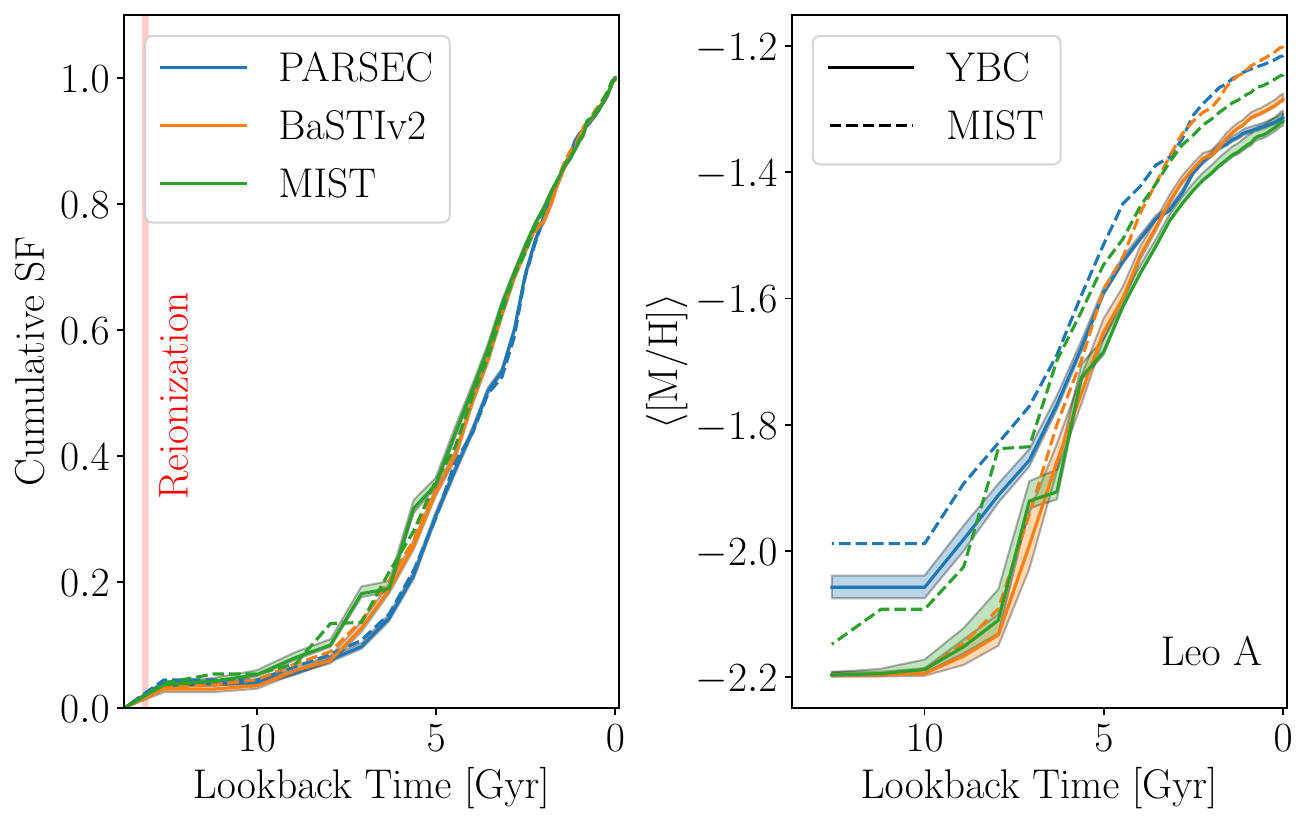}
\caption{The results of our resolved SFH and metallicity evolution fit for Leo A based on the HST/ACS data following the same format as Figure \ref{fig:wlm_sfh}. In the top row the result of \cite{McQuinn2024b} (blue line), who used the \textsc{match} code on the same data, is shown for comparison. Overplotted on the upper right panel are age and metallicity measurements for individual red giant stars obtained by combining the spectroscopic metallicities of \cite{Kirby2017a} with our photometry (see \S \ref{sec:ages}).}
\label{fig:leoA_sfh}
\end{figure*}

In contrast, our best-fit resolved SFH solution under the MZH model, shown in Figure \ref{fig:leoA_sfh}, shows significant metallicity evolution -- this measurement indicates that stars formed $>10$ Gyr ago in Leo A should have metallicities [M/H] $\sim-2.0$ dex, while stars forming in Leo A at present-day have metallicities [M/H] $\sim-1.3$ dex, in good agreement with the metallicity measurements of young supergiants by \cite{Urbaneja2023}. The AMR fit of \cite{McQuinn2024b} also prefers substantial metallicity evolution. Using the methodology of \S \ref{sec:ages} we leverage the spectroscopic metallicities of \cite{Kirby2017a} in combination with our high-precision HST/ACS photometry to measure ages for single red giant stars in Leo A in order to perform an independent check of the AMR. We note that our photometry is much more precise than the ground-based photometry that \cite{Kirby2017a} used to estimate ages for these stars -- the random errors on the HST/ACS photometry for these bright red giants is typically less than 0.02 mag. Shown overplotted in the upper right panel of Figure \ref{fig:leoA_sfh}, these single star ages illustrate that the old stars of Leo A ($>5$ Gyr old) do indeed have low metallicities ($<-1.8$ dex). Overall, the single star measurements agree very well with the metallicity evolution from the resolved SFH.

The previous results favoring little metallicity evolution may be understood as a result of Leo A showing very little early star formation. Compared to other galaxies of similar present-day stellar mass, the bulk of Leo A's stellar mass has been formed relatively recently -- about $80\%$ of its stellar mass was formed in the last 5 Gyr. As a result, there are many more young stars (ages $<5$ Gyr) on the upper RGB than old stars. We illustrate this in Appendix \ref{appendix:leoA} by generating a Monte Carlo simulation of the CMD of Leo A assuming our best-fit SFH. We then examine the age distribution of stars that lie on the upper RGB, where the spectroscopic sample of \cite{Kirby2017a} lies. Our mock CMD predicts that roughly $19\%$ of stars on the upper RGB should have ages $>5$ Gyr; this is in reasonable agreement with the observations, as only $14\%$ of the stars for which we measure ages are older than 5 Gyr. Given the paucity of bright, old stars in Leo A, it is understandable that its metallicity evolution has been difficult to measure accurately.

\subsection{Leo P} \label{subsec:leoP}

Leo P is less massive ($M_* \approx 5.7 \times 10^5$ M$_\odot$; \citealt{McQuinn2024b}) and more metal-poor ($Z \approx 0.03 \, Z_\odot$; \citealt{Skillman2013}) than the other galaxies in this section, extending our census of the metallicity evolution of dwarf galaxies. Leo P is isolated and gas-rich ($M_\HI/ M_* \approx 2$; \citealt{BernsteinCooper2014}) but is more distant than the other galaxies in our sample with $d \approx 1.62 \pm 0.15$ Mpc \citep[$\mu = 26.05 \pm 0.20$ mag;][]{McQuinn2015}. 

The resolved SFH of Leo P was measured with HST/ACS data by \cite{McQuinn2015} and remeasured with new JWST/NIRCam data by \cite{McQuinn2024b}. While the SFHs of the two measurements are fairly consistent, the metallicity models differ -- both solutions find a flat (or nearly flat) AMR, but the HST/ACS data prefer a metallicity [M/H] between $-1.6$ and $-1.7$ dex (see Figure 8 of \citealt{McQuinn2015}) while the JWST/NIRCam data prefer a higher metallicity between $-1.4$ and $-1.2$ dex (see Figure 7 of \citealt{McQuinn2024b}). Presently the best estimate of the metallicity of Leo P comes from the oxygen abundance of its single known \HI region, $12 +$ log(O/H) $= 7.17 \pm 0.04$ \citep{Skillman2013}, corresponding to roughly [O/H] $\approx -1.5$ dex. Spectroscopy targeting candidate massive stars in Leo P was presented by \cite{Evans2019}, but these data were insufficient to measure stellar metallicities. Spectroscopy of a bright O star in Leo P was presented by \cite{Telford2024} who found that the star's metallicity is consistent with the nebular oxygen abundance and that no internal extinction due to dust in Leo P is required to fit the star's spectral energy distribution.

\begin{figure*}
\centering
\includegraphics[width=0.85\textwidth]{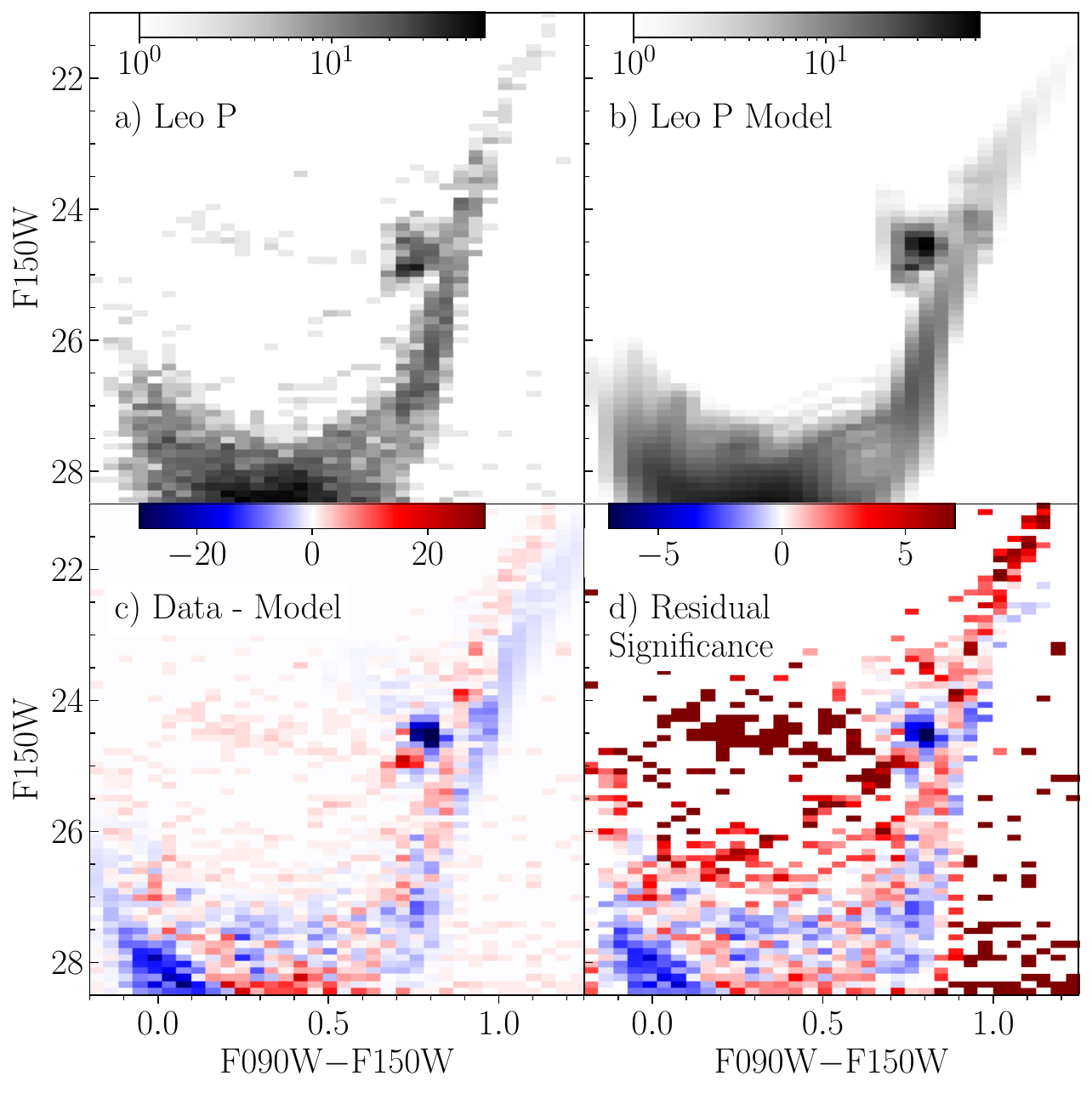} \\
\caption{Comparison of the Hess diagram of Leo P with our best-fit model. The panels follow the same layout as in Figure \ref{fig:wlm_hess}.}
\label{fig:leoP_hess}
\end{figure*}

In Figure \ref{fig:leoP_hess} we compare the observed Hess diagram of Leo P with our best-fit MZH metallicity evolution model. The model fits the data reasonably well, with the main area of disagreement being the color of the upper RGB -- in the model Hess diagram, the upper RGB (brighter than the red clump) is noticably redder than in the data. This mismatch is also present in the models presented by \cite{McQuinn2024b} (see their Figure 6). 

\begin{figure*}[hpbt]
\centering
\includegraphics[width=0.85\textwidth]{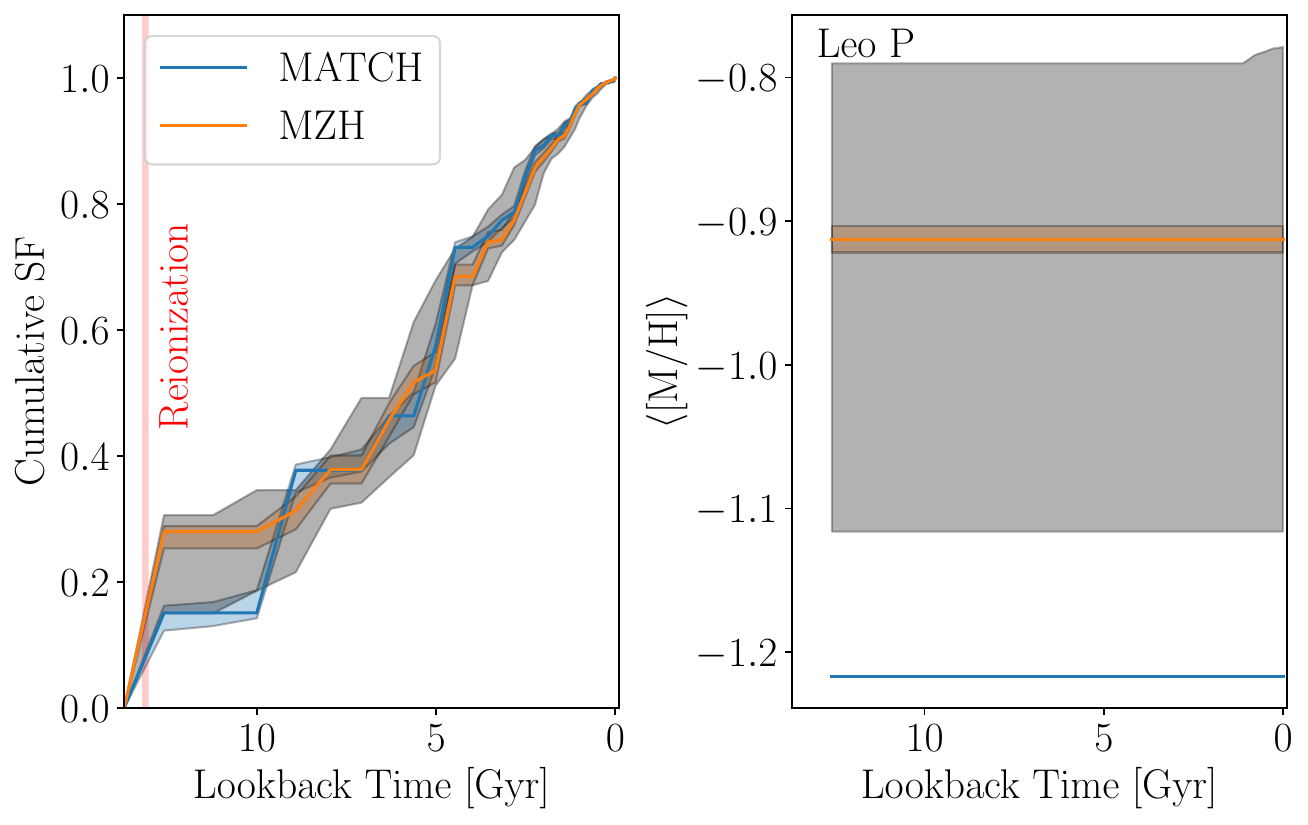} \\
\includegraphics[width=0.85\textwidth]{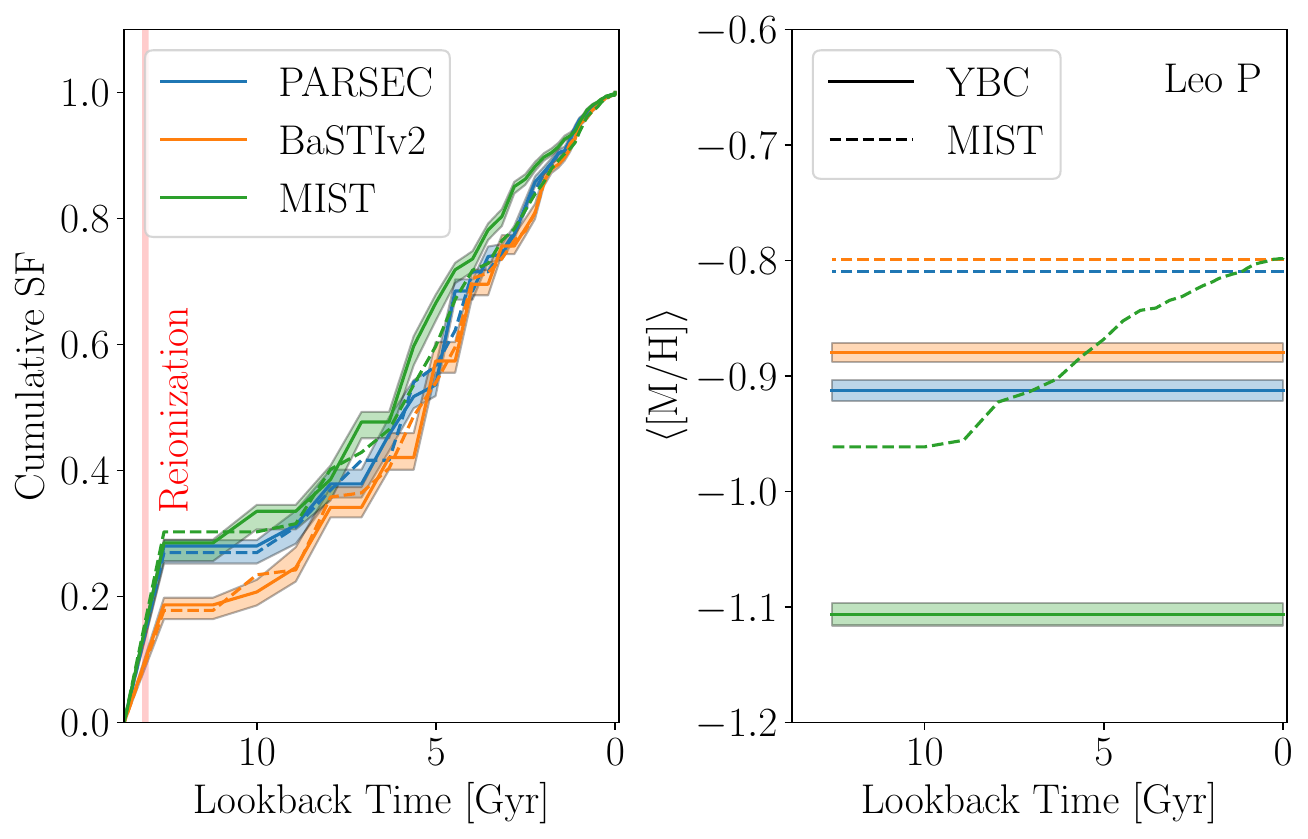}
\caption{The results of our resolved SFH and metallicity evolution fit for Leo P based on the JWST/NIRCam data following the same format as Figure \ref{fig:wlm_sfh}. In the top row the result of \cite{McQuinn2024b} (blue line), who used the \textsc{match} code on the same data, is shown for comparison.}
\label{fig:leoP_sfh}
\end{figure*}

As Leo P is fainter and more distant than the other galaxies in our sample, the random uncertainties in the SFH solution are accordingly larger. This is illustrated in Figure \ref{fig:leoP_sfh}, where we present our SFH and metallicity evolution measurements using our new MZH model. Our fiducial solution agrees well with that of \cite{McQuinn2024b}, though our solution prefers a larger initial burst of star formation and does not show the same burst in star formation 10 Gyr ago as found by \cite{McQuinn2024b}. We find negligible metallicity evolution under all combinations of stellar tracks and BCs except for the combination of MIST stellar tracks and MIST BCs. Interestingly, our fit prefers a higher metallicity of roughly $-0.9^{+0.1}_{-0.2}$ dex -- the SFH measurement of \cite{McQuinn2024b} also found negligible metallicity evolution, but they measured [M/H] between $-1.4$ and $-1.2$ dex, although they only included isochrones in their model with metallicities [M/H] $\leq -0.9$ dex. Both measurements prefer a higher metallicity than implied by the oxygen abundance of the single \HII region in Leo P of [O/H] $\approx -1.5$ dex \citep{Skillman2013}. To explore this discrepancy, we performed experiments fixing the variables describing the metallicity evolution to better match that measured from the \HII region with results shown in Appendix \ref{appendix:leoP}. Assuming [M/H] $\approx -1.5$ dex gives a better fit to the upper RGB, but results in a poorer fit on the lower RGB and the upper main sequence. As there are many more stars in these areas than on the upper RGB, the overall fit quality is worse when assuming [M/H] $\approx -1.5$ dex than in our fiducial model. Overall, the metallicity of Leo P's stellar population is poorly constrained as there are no spectroscopic measurements available for red giants in Leo P as there are for the other galaxies in our sample. Future work probing the stellar metallicities of Leo P is needed to understand this discrepancy.

\section{Historical Mass-Metallicity Histories from Theory} \label{sec:theory}
The MZH is an important point of comparison between observations and theory as stellar feedback processes regulate the enrichment of the ISM. These stellar feedback models have significant uncertainties and the strength of stellar feedback varies considerably between different galaxy evolution models, leading to differences in ISM metal abundances of a factor of two or more for massive galaxies \citep[M$_* \ge 10^9$ \msun; see, e.g., Figure 10 of][]{RocaFabrega2024}. We expect these differences to be even more significant for low-mass dwarf galaxies as their dark matter halos have shallower potential wells, allowing for significant metal-loaded outflows in models with strong stellar feedback \citep[e.g.,][]{Muratov2015,Muratov2017,Christensen2016,Christensen2018,Nelson2019,Mitchell2020a,Pandya2020,Pandya2021,Rey2025}. Therefore, observationally probing the MZH at dwarf scales using our SFH-informed methodology can constrain stellar feedback models, but only if the observations and theoretical models are compared in a consistent way.

Observationally, traditional MZRs are typically measured from statistical samples of galaxies with different stellar masses in small redshift bins \citep[e.g.,][]{Andrews2013,Blanc2019,Curti2020} -- in the limit of infinitesimally small redshift bins, this amounts to measuring metallicity as a function of stellar mass at fixed time. The MZRs measured from theoretical sources (e.g., cosmological hydrodynamic simulations and semi-analytic models) mirror this design \citep[e.g.,][]{Ma2016,Dave2019,Torrey2019,Bassini2024,RocaFabrega2024,Marszewski2024,Marszewski2025}. In contrast, our MZH model is historical -- it models the metallicity of stars forming at time $t$ as a function of the total stellar mass formed prior to $t$. As such, the only time-dependent parameters of our model are the SFRs. We must therefore analyze the theoretical data in the same historical fashion to derive MZHs that mirror the definition we use in our model.

To provide context and comparisons to our observational results, we analyze the TNG50 \citep{Nelson2019,Pillepich2019} and FIREbox \citep{Feldmann2023} cosmological hydrodynamic simulations, as well as results from the \textsc{galacticus} \citep{Benson2012,Ahvazi2024} semi-analytic model. While TNG50 uses a galaxy evolution model whose stellar feedback is relatively smooth, the stellar feedback model used in FIREbox is comparatively bursty and drives stronger mass- and metal-loaded outflows in dwarf galaxies \citep{Muratov2015,Nelson2019,Pandya2021}. Given the differences in the models, it is reasonable to expect their MZHs may differ at dwarf scales. We additionally consider \textsc{galacticus} to contrast the predictions of a state-of-the-art semi-analytic model with the aforementioned hydrodynamical simulations. \textsc{galacticus} is somewhat unique amongst semi-analytic models in how much attention has been given to improving its agreement with observations at dwarf scales \citep[e.g.,][]{Weerasooriya2023,Weerasooriya2024,Ahvazi2024} making it particularly interesting for our application.

In this section we will describe our methodologies for measuring the MZHs from the theoretical data in order to mirror the definition of the MZH used in our observational work. We then present the measured MZHs in \S \ref{sec:results} and discuss how the results compare amongst the theoretical models and against the observational results.

\subsection{Analog Selection} \label{subsec:selection}
As neither TNG50 nor FIREbox have sufficient baryonic resolution to resolve low-mass dwarf galaxies with $M_* \le 10^6$ \msun, we are limited to considering higher mass dwarf galaxies. These simulations do have sufficient resolution to study analogs of the WLM dwarf irregular, which has $M_* \approx 4.3 \times 10^7$ \msun -- we therefore choose to focus our theoretical analysis on analogs of WLM to compare against the MZH we measured for WLM from JWST/NIRCAM photometry in \S \ref{sec:obs}. 

Since it is known that the classical MZR exhibits secondary correlations with other galaxy properties in several galaxy evolution models (e.g., specific star formation rate and ISM gas mass; see, for example, \citealt{Lilly2013,Dave2019,Torrey2019,Garcia2024,Garcia2025a}), we first select galaxies from the theoretical data that resemble WLM in several important ways. All selection criteria are imposed at present-day. We select only central galaxies in the theoretical data since WLM is not a satellite of a more massive host galaxy. We further select only galaxies with stellar masses in the range $7 \le \text{log}\left(\text{M}_* \left[\text{M}_\odot\right]\right) \le 9$ to ensure that the selected galaxies are of similar stellar mass to WLM. As WLM is actively forming stars \citep[e.g.,][]{Hodge1995,Albers2019,McQuinn2024,Garling2025}, we further select only galaxies with specific star formation rates (i.e., star formation rate per unit stellar mass) greater than $10^{-11}$ $\text{yr}^{-1}$. Like most star-forming dwarf galaxies, WLM is relatively gas-rich with a ratio of gas mass to stellar mass of order unity \citep{Kepley2007,Ianjamasimanana2020}. We therefore select only galaxies with $\text{M}_\text{gas} / \text{M}_* \ge 0.75$ to ensure the simulated galaxies are similarly gas-rich.

\subsection{TNG50}
The IllustrisTNG galaxy evolution model \citep{Pillepich2018} was purpose-built to enable simulation of large cosmological volumes with magnetohydrodynamics. As such, it uses a subgrid star formation model that forms stars stochastically in gas denser than $n_\mathrm{H} \simeq 0.1$ cm$^{-3}$ following an empirical Kennicutt-Schmidt relation \citep{Springel2003}. Generally speaking, the IllustrisTNG model is designed to operate at resolutions insufficient to resolve the dense molecular clouds where most star formation is thought to occur ($n_\mathrm{H} \geq 100$ cm$^{-3}$). The computational efficiency of these methods has allowed IllustrisTNG to be run in cosmological volumes with resolutions sufficient to resolve dwarf galaxies.

The TNG50 simulation \citep{Nelson2019,Pillepich2019} applies the IllustrisTNG galaxy formation and evolution model \citep{Weinberger2017,Pillepich2018} to a cosmological volume of (51.7 Mpc)\textsuperscript{3} with a baryon particle mass of $8.5 \times 10^4$ \msun which is sufficient to resolve isolated dwarf galaxies like WLM with $\sim10^3$ star particles. The MZR was studied in the larger volume, lower resolution TNG100 simulation by \citet{Torrey2019} who found good agreement with observations for galaxies with M$_* \ge 10^9$ \msun (see their Figure 6). The MZR and the fundamental metallicity relation in the IllustrisTNG model were compared to results from other galaxy evolution models in \citet{Garcia2024,Garcia2025a}. Below we describe the procedure used to measure the MZH (metallicity as a function of cumulative stellar mass \emph{regardless of time}) for comparison to our observational results.

We first perform the analog selection described in \S \ref{subsec:selection} to identify galaxies in TNG50 that resemble WLM. We identify 6261 galaxies in the simulation that pass our selection criteria. Our subsequent procedure is similar to that developed in section 4.2 of \citet{Escala2018} to measure dispersions in stellar metallicities in FIRE-2 simulations. Each star particle in TNG50 is tagged at birth with its formation time and metallicity. The metallicity of the star particle is inherited from its progenitor gas particle and therefore reflects the metallicity of the star-forming ISM at the time it was formed. Given the population of star particles in a galaxy at present-day, one can fully reconstruct the stellar mass of the galaxy over time $\left(\text{M}_* \left(t\right)\right)$ under the assumption of in-situ star formation. As such, if star particle $i$ was formed at time $t_i$ with metallicity $Z_i$, we can correlate time $t_i$ with a total stellar mass of $\text{M}_* \left(t_i\right)$, allowing us to form the MZH that we require $\left(Z\left(\text{M}_*\right)\right)$. While there is likely to be some contamination due to star particles formed ex-situ (i.e., in other galaxies that were later accreted), \citet{Escala2018} found that $\ge98\%$ of star particles in their sample of isolated dwarf galaxies had formed in-situ, indicating that contamination from ex-situ stars should be minimal.

Considering the above, we proceed as follows. For each of the 6261 galaxies that pass our selection criteria, we use the \textsc{subfind} \cite{Springel2001} halo catalog to identify star particles bound to the galaxy at present-day and extract their formation times and metallicities. We then form $\text{M}_* \left(t\right)$ and $Z\left(t\right)$ by discretizing the formation times of the star particles onto a regular grid. For each time bin, the total stellar mass formed in the time bin is simply the sum of the birth stellar masses of all particles formed in the time bin. Calculating the birth mass of a star particle is non-trivial as star particles in TNG50 and FIREbox have non-uniform birth stellar masses and the mass decreases over time following models of the mass loss from evolved massive stars. In Appendix \ref{appendix:birthmass}, we describe how we calculate an approximate correction for this stellar mass loss to infer the birth stellar masses of star particles based on their present-day masses in TNG50 and FIREbox. We find that the birth stellar masses can be reasonably approximated from the present-day stellar mass as $\text{M}_{*,\text{born}} = \text{M}_*(z=0) \div (1 - R)$ assuming a population-averaged recycled fraction of $R=0.25$ at present-day ($R=0.30$ for FIREbox). This approximation adds no computational complexity and is correct to better than $10\%$ precision for star particles older than 1 Gyr. A $10\%$ uncertainty in the birth masses of the star particles makes no significant difference to our results. Once we know how much stellar mass was formed in each time bin, the cumulative stellar mass $\text{M}_* \left(t\right)$ can then be found by taking the cumulative sum from the first snapshot to the present-day. Under this definition, $\text{M}_* \left(t\right)$ is the total stellar mass formed prior to time $t$ -- we have explicitly removed the effect of stellar mass loss due to the evolution of massive stars. This is the easiest way to compare to the observational SFHs, as the observational results give the SFR as a function of time -- forming $\text{M}_* \left(t\right)$ from the observational results therefore only requires integrating the SFRs over time. The metallicity in the time bin $Z\left(t\right)$ is measured as the median metallicity of all star particles formed in the time bin.

Once $\text{M}_* \left(t\right)$ and $Z\left(t\right)$ have been measured, we form the MZH by combining the time series data from all the selected galaxies to form a single dataset consisting of total stellar masses and corresponding metallicities. We bin these data logarithmically in stellar mass and calculate the median and $68\%$ confidence interval to form our final MZH measurements for TNG50, shown in Figure \ref{fig:mzh_result}.

\subsection{FIREbox}
When it was first introduced, the FIRE galaxy evolution model \citep{Hopkins2014} was unique in its approach to resolving the multiphase ISM in simulations of galaxy evolution. FIRE was particularly notable for its treatment of the cold, dense ISM -- resolving temperatures $\sim 10$ K and densities $n_\text{H} \geq 100$ cm$^{-3}$, FIRE formed stars in gas that was colder and denser than other contemporary models. Star formation in these cold, dense regions of the ISM could be highly efficient, and after a short period of forming stars, these regions would be destroyed due to significant stellar feedback. This cycling between the formation of dense, star forming clouds and violent dissolution due to stellar feedback resulted in star formation that was more ``bursty" (i.e., more time variable and spatially inhomogenous) in FIRE simulations compared to other models like IllustrisTNG. Over the years, the FIRE model been updated with new and revised physics in FIRE-2 \citep{Hopkins2018} and FIRE-3 \citep{Hopkins2023b}. Until recently, the FIRE models had only been applied to zoom-in simulations due to computational constraints \citep[see section 2.3 of][]{Wetzel2023}. With the completion of FIREbox \citep{Feldmann2023}, the first application of FIRE-2 to a cosmological volume, we now have a large statistical sample of galaxies across a range of environments that we can compare to other volume simulations like TNG50.

The FIREbox simulation \citep{Feldmann2023} applies the FIRE-2 galaxy evolution model \citep{Hopkins2018} to a cosmological volume of (22.1 Mpc)\textsuperscript{3} with a baryonic particle mass of $6.26 \times 10^4$ \msun. The baryonic particle mass is of the same order of magnitude as TNG50 and the volume of FIREbox is roughly $8\%$ that of TNG50. Given that we identified about 6000 galaxies matching our selection criteria in TNG50, we expected to find of order 500 galaxies matching our criteria in FIREbox given the differences in the volumes of the simulations. Applying the selection criteria described in \S \ref{subsec:selection} to FIREbox left us with 574 galaxies, well in line with our expectations. We select star particles associated with these galaxies at present-day from halo catalogs built with AHF \citep{Gill2004,Knollmann2009}.

Conveniently, the star particles in FIREbox have formation time and metallicity tags that are analogous to those we used when measuring the MZH in TNG50, enabling us to use the same procedure to measure the MZH in FIREbox. As our analysis procedure is essentially identical for both TNG50 and FIREbox, any differences in their MZHs should be attributable solely to their different galaxy evolution models. 

\subsection{Galacticus}
We make predictions using the \textsc{galacticus} semi-analytic model \citep{Benson2012} to compare against the hydrodynamic simulations described above. As a semi-analytic model, \textsc{galacticus} requires merger trees to define the halos within which galaxies will form and evolve. We generate random merger trees using the built-in mechanisms based on the extended Press-Schecter formalism \citep{Press1974,Bond1991,Cole2000,Cole2008,Benson2005,Zentner2007,Parkinson2008,Zhang2008,Jiang2014,Benson2017,Nadler2023a}. In this study, we track the evolution of 340 galaxy merger trees with present-day halo masses in the range $8\times10^9$ to $8\times10^{10}$ M$_{\odot}$ and apply the selection criteria described in \S \ref{subsec:selection} to the galaxy population at present-day, leaving us with a sample of 78 galaxies that we use to derive the MZH.

The modeling of baryonic processes in \textsc{galacticus} includes prescriptions for gas cooling, star formation, feedback, gas re-incorporation and metal enrichment, and many other processes important in galaxy evolution (see \citealt{Benson2012} for an overview of the physical processes modeled). Gas cooling is implemented based on \citet{White1991} to calculate the cooling radius, with the cooling time set by the halo's dynamical time. Cooling rates are metallicity-dependent and computed using CLOUDY \citep[v23.01,][]{Gunasekera2023} assuming collisional ionization equilibrium. Star formation in the disk is modeled using a hydrostatic pressure-based formulation following \citet{Blitz2006}. The star formation rate in the spheroid component of galaxies depends on the gas mass and the system's dynamical time. In \textsc{galacticus}, star formation drives gaseous outflows whose strength scales with the circular velocity of the halo and the rate of energy input due to the stellar populations (Equation A2 in \citealt{Ahvazi2024}). Gas expelled from galaxies by winds is retained in an outflow reservoir, from which it gradually returns to the hot halo on a timescale proportional to the halo's dynamical time. Metal enrichment follows an instantaneous recycling approximation, assuming a fixed recycled fraction and metal yield, with metals fully mixed within all gas phases to track their transport between different reservoirs (for more details, see section 2.2 of \citealt{Knebe2018} and Appendix A of \citealt{Ahvazi2024}).

As \textsc{galacticus} can save output at any user-requested times, we output results on the same redshift grid as that of TNG50 for consistency. We can measure $\text{M}_* \left(t\right)$ for each galaxy straightforwardly from the stellar mass of the galaxy along the main progenitor branch of the merger tree. Once we have $\text{M}_* \left(t\right)$ and $Z\left(t\right)$ for each galaxy, we form the time-independent MZH the same way as we did for TNG50 and FIREbox. 

\section{Results and Discussion} \label{sec:results}
\begin{figure*}
\centering
\includegraphics[width=0.85\textwidth]{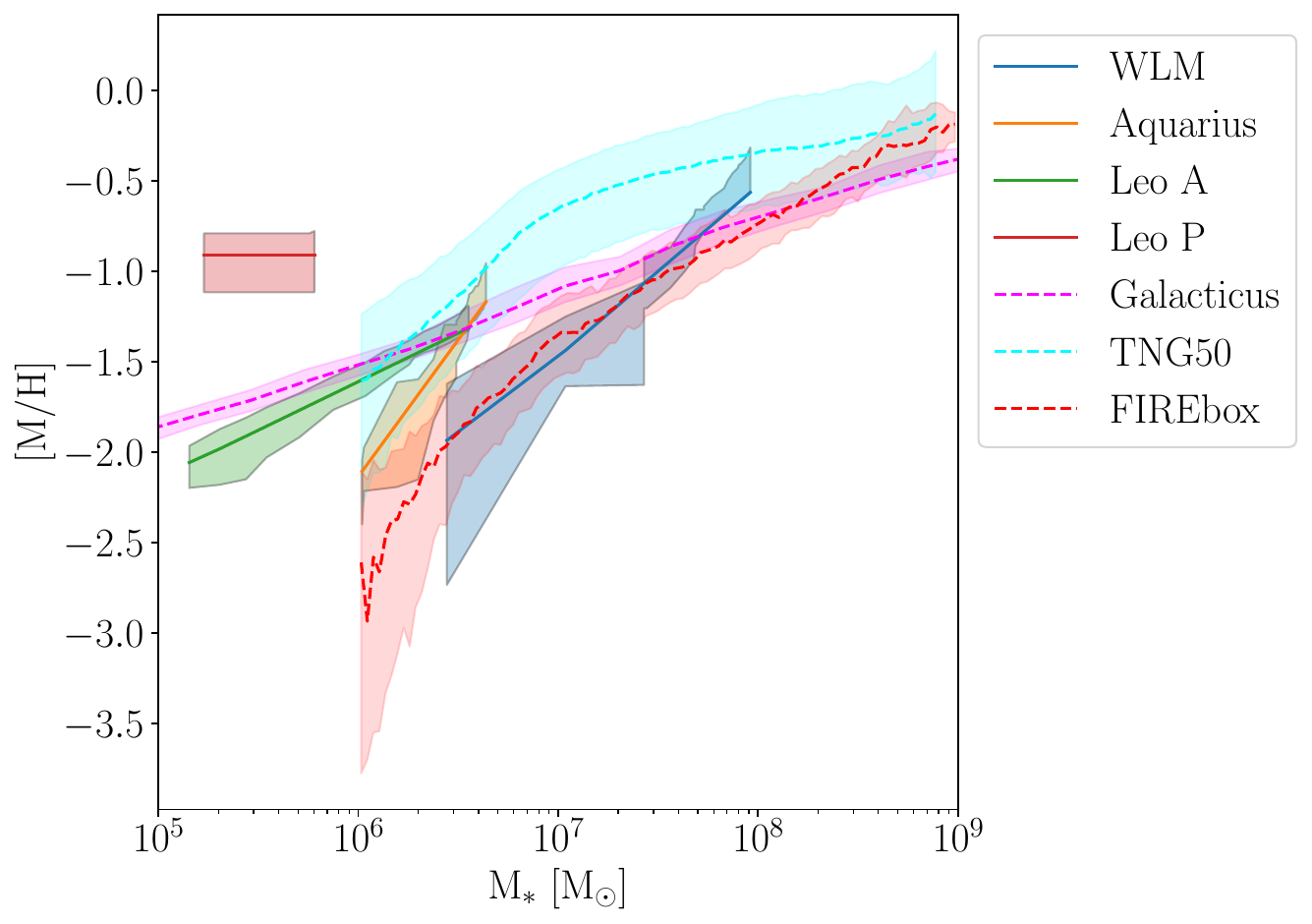} \\
\caption{MZHs for WLM-analog galaxies measured from TNG50, FIREbox, and \textsc{galacticus} (dashed lines) compared to our observational results for nearby isolated dwarfs (solid lines). For the simulation measurements, the shaded regions represent the $68\%$ confidence interval on the median metallicity in each stellar mass bin. The solid lines show our fiducial observational measurements under our new MZH model, with shaded regions representing the systematic uncertainty region (measured as discussed in \S \ref{sec:systematics}).}
\label{fig:mzh_result}
\end{figure*}

In Figure \ref{fig:mzh_result} we present the MZHs measured from TNG50, FIREbox, and \textsc{galacticus} compared to our observational results for nearby isolated dwarfs. As discussed above, the selection of analogs from the theoretical sources was designed to select galaxies similar to WLM due to the limited resolution of TNG50 and FIREbox; Leo A, Leo P, and Aquarius are all an order of magnitude less massive than WLM, so we expect they may not be well-represented in this comparison, but we include them in Figure \ref{fig:mzh_result} for completeness. The stellar masses on the horizontal axis represent the cumulative stellar mass formed neglecting stellar mass loss due to massive star evolution. The field of view of our imaging covers the majority of Leo A, Leo P, and Aquarius, so we take the total stellar mass formed directly from our SFH measurements; no additional corrections are performed. 

For WLM, the JWST/NIRCam data we use to measure the SFH are insufficient to cover the entire galaxy, so we must correct the normalization of our SFRs to reflect the true total stellar mass of WLM. We do this by first assuming the present-day stellar mass of WLM to be $4.3 \times 10^7$ M$_\odot$ \citep{McConnachie2012}. The present-day stellar mass is typically converted into the cumulative stellar mass formed by assuming a recycling fraction of order 30--$50\%$ \citep{Vincenzo2016a,McQuinn2024b}, with the specific value depending on the assumed IMF and SFH of the population. However, calculating the recycling fraction requires making an assumption of the mass of stellar remnants (e.g., white dwarfs, neutron stars, and black holes) as a function of initial stellar mass and metallicity which adds uncertainty to the calculation of the present-day stellar mass \cite[see, e.g., Equation 3 of][]{Vincenzo2016a}. It is more straightforward to calculate the fraction of the total stellar mass formed that remains in stars using the IMF and stellar lifetimes directly. We use stellar lifetimes from the same PARSEC models used to measure our fiducial SFH to determine, for each stellar population of unique age $i$ and metallicity $j$, the initial stellar mass of the most massive star that is still alive at present-day, $u_{i,j}$. Using our fiducial SFH, we then calculate the surviving stellar mass fraction as

\begin{equation}
f_{\text{surv}} = \frac{\sum_{i,j} c_{i,j} \int_l^{u_{i,j}} m \, \phi(m) \, dm}{\langle m \rangle \sum_{i,j} c_{i,j}}
\end{equation}

\noindent where $c_{i,j}$ is the total stellar mass formed in the stellar population with age $i$ and metallicity $j$, $\phi(m)$ is the IMF, $l$ is the lower mass limit of the IMF, and $\langle m \rangle$ is the mean stellar mass of the IMF. Assuming $l=0.08$ M$_\odot$, we find $f_{\text{surv}} = 0.47$. We therefore estimate the total stellar mass formed in WLM to be $\frac{4.3}{0.47} \times 10^7 \approx 9.1 \times 10^7$ M$_\odot$. We use this value to renormalize our SFRs when plotting the MZH for WLM in Figure \ref{fig:mzh_result}. 

\subsection{The Mass-Metallicity History of WLM} \label{subsec:wlm_discussion}

We find that the best-fit power law slope for WLM is $\alpha \approx 1$ (see Equation \ref{eq:mzhmodel}). \textsc{galacticus} shows a nearly constant power law slope of $\alpha \approx 0.4$ across the entire stellar mass range, while the MZH in TNG50 shows scaling more similar to a broken power law. At early times, when the stellar mass of the simulated galaxies was M$_* \leq 10^7$ \msun, the power law scaling in TNG50 is $\alpha \approx 1.0$ and becomes $\alpha \approx 0.2$ above this mass. As such, the low-mass slope of the TNG50 MZH actually agrees with the slope we find for WLM, but the metallicity normalization is much higher -- at M$_* = 10^7$ M$_\odot$, the TNG50 prediction is about $0.9$ dex higher than what we measure for WLM. In contrast to TNG50 and \textsc{galacticus}, FIREbox does not exhibit simple power law, or even broken power law, scaling. Across nearly the entire range of the predictions from $10^6 \leq \text{M}_* \leq 10^9$, the slope of the MZH in FIREbox is larger than in \textsc{galacticus} and TNG50. While FIREbox agrees fairly well with the other simulations at higher masses, it predicts much lower metallicities at low stellar masses compared to the other simulations ($\text{[M/H]} \leq -1.5$ dex for M$_* \leq 10^7$ M$_\odot$). This ends up putting the FIREbox prediction in quite good agreement with our result for WLM. 

Comparing the theoretical results, they are all roughly consistent at the high-mass end (M$_* \geq 10^8$ \msun), which we expect as they all reasonably reproduce the traditional MZR at these masses \citep[e.g.,][]{Torrey2019,Ahvazi2024,Bassini2024,Garcia2024,Garcia2025a}. As a result, the MZH cannot meaningfully distinguish between the models in the stellar mass range $10^8 \leq \text{M}_* \leq 10^9$ \msun. However, at lower masses the models predict significantly different metallicities. The TNG50 MZH predicts the highest metallicities between $10^7 \leq \text{M}_* \leq 10^8$ \msun, while FIREbox predicts the lowest. At M$_*=10^7$ \msun, TNG50 predicts [M/H] $\approx -0.6$ dex for our sample of WLM analogs, while \textsc{galacticus} predicts [M/H] $\approx -1.1$ dex and FIREbox predicts [M/H] $\approx -1.3$ dex. At this stellar mass, the observational result for WLM shows [M/H] $\approx -1.5$ dex, consistent with the FIREbox result given the uncertainties. Observationally, we are able to measure WLM's MZH starting from when it had formed $\sim2 \times 10^6$ M$_\odot$ of stars up to the present-day when it has formed a total of $\sim9 \times 10^7$ M$_\odot$ of stars -- across this full range, the results agree best with the predictions of FIREbox.

A possible explanation for the difference between the predictions of TNG50 and FIREbox are the supernova metal yields. However, the differences between the metal yields in TNG50 and FIREbox are too minor to explain the $\sim0.7$ dex difference in their predicted metallicities for M$_* \leq 10^7$ \msun \citep{Pillepich2018,Hopkins2018}. For our purposes we will discuss the total metal yield as none of our analysis depends on specific elemental abundances. For SN-Ia, where no compact remnant is produced, both simulations eject a Chandrasekhar mass ($\sim1.4$ M$_\odot$) of metals into the ISM. For core-collapse supernova (CCSN), FIRE-2 assumes the yield tables of \cite{Nomoto2006} but renormalizes the total ejecta mass to match a set of \textsc{starburst99} \citep{Leitherer1999} models. FIRE-2 assumes the \cite{Kroupa2001} IMF, resulting in an IMF-averaged metal ejecta mass of about 2 M$_\odot$ per CCSN. The only element with a metallicity-dependent yield in FIRE-2 is N such that the average metal ejecta mass per CCSN is nearly constant as a function of initial stellar metallicity. In comparison, the IllustrisTNG model uses the \cite{Kobayashi2006} yield table and extends the mass coverage with the \cite{Portinari1998} yield table \citep{Pillepich2018}. The joint table is renormalized so that the yield ratios between species match that of the \cite{Kobayashi2006} yield table and the total ejected metal mass matches the \cite{Portinari1998} yield table. In the IllustrisTNG model, the metallicity dependence of the CCSN yields follows that of the yield tables; no other alterations are made. Using the \cite{Chabrier2003} IMF assumed by the IllustrisTNG model, we estimate an IMF-averaged metal ejecta mass of about 2.14 M$_\odot$ of metals per CCSN at [M/H] = $-1.5$ dex. This is only $\sim7\%$ higher than the IMF-averaged metal ejecta mass in FIRE-2 and is insufficient to explain the $\sim0.7$ dex difference in the MZH predictions between TNG50 and FIREbox at M$_*=10^7$ M$_\odot$. The occurrence rates of SN-Ia and CCSN (in terms of the number of supernovae per solar mass of star formation) are also comparable between FIRE-2 and TNG, such that differences between their MZHs cannot be due to differences in the supernova rates either.

We hypothesize that differences between the MZHs of TNG50 and FIREbox are likely attributable to their different implementations of stellar feedback. While both stellar feedback models drive galaxy-scale outflows at dwarf masses \citep[M$_* \leq 10^7$ M$_\odot$;][]{Muratov2015,Nelson2019,Pandya2021} the mass-loading factors ($\eta = \frac{\dot{M}_\text{outflow}}{\text{SFR}}$) in FIRE-2 can be an order of magnitude larger than those in TNG50 at dwarf scales \citep{Muratov2015,Nelson2019,Pandya2021}. While these outflows can push a significant amount of gas out of the ISM, such outflows are typically more metal-rich than the star-forming ISM as they retain a significant fraction of the metals ejected in the supernova events \citep{Chisholm2018}. The degree to which the outflows are metal-enhanced depends on simulation details such as the adopted metal diffusion model \citep[e.g.,][]{Escala2018,Hafen2019,Garcia2025,Steinwandel2025}, but it is generally accepted that outflows driven by supernovae push a significant fraction of the metals formed in dwarf galaxies out of the ISM. These theoretical results are reinforced by observational studies of the traditional MZR at dwarf scales that show that dwarf galaxy metallicities are lower than would be expected if they retained all of the metals they create, necessitating significant metal-loaded outflows to reproduce the observed MZR \citep[e.g.,][]{Kirby2011a,Kirby2013,KadoFong2024}.

If the differences in the MZHs of TNG50 and FIREbox are indeed due to differences in their supernova-driven outflows, this should manifest differences in the CGM properties they predict at dwarf scales as well. Observationally, the CGM metallicity can be probed via metal-line absorption studies. In \cite{Li2021b}, CGM metal-line absorption predictions for galaxies with $8.1 \leq \text{log} (\text{M}_*) \leq 10.4$ simulated with FIRE-2 were compared against observations, with results indicating reasonably good agreement across several different metal ions probing different phases of the CGM (e.g., C \textsc{iv} and O \textsc{vi}). These galaxies are more massive than WLM, so extending this work to dwarf galaxies of lower mass in the future would provide a better comparison. Interestingly, \cite{Hafen2019} find that low-mass FIRE-2 galaxies host significantly more metals in their CGM than in their ISM, highlighting the importance of the CGM as a metal reservoir (see their Figure 3). Recycling of metals from the CGM back into the ISM can then be a critical mechanism by which the ISM metallicity is regulated -- all accreted baryons, including pristine gas from the intergalactic medium, must pass through the CGM to reach the ISM, providing an opportunity for the metal-enriched CGM to mix with the accreting material \citep[e.g.,][]{Tung2025}. In a study examining the time evolution of the MZR for galaxies with $8 \leq \text{log} (\text{M}_*) \leq 11$ in FIREbox, \cite{Bassini2024} find that the time dependence of the MZR is due to the average metallicity of accreted gas increasing over time as the metal-enriched CGM mixes with gas inflows from other sources.

In contrast, the analysis of \cite{Torrey2019} ascribes the cause of the redshift evolution of the MZR to a different mechanism in the IllustrisTNG model. Studying galaxies with $9 \leq \text{log} (\text{M}_*) \leq 10.5$, they find that redshift evolution in the gas fraction of galaxies is the primary driver of the redshift evolution of the MZR. As the metallicity of the ISM is a relative quantity (the relative abundance of metals to hydrogen), increasing the gas mass in the ISM at fixed ISM metal mass decreases the overall metallicity of the ISM (and vice versa). \cite{Bassini2024} do not find the same mechanism to be effective in the FIRE-2 galaxy evolution model. This is interesting as both models have been shown to agree well with measurements of the MZR for M$_* \geq 10^9$ M$_\odot$ \citep{Torrey2019,Bassini2024,Jain2026}. However, these analyses looked at galaxies at least an order of magnitude more massive than WLM, so these results cannot be directly compared to our measurements. Additionally, most of the work done looking at the CGM in the IllustrisTNG model has focused on higher mass galaxies like the MW \citep[e.g.,][]{Nelson2020,Ramesh2023,Ramesh2023a}, so it is unclear how significant a role metal recycling from the CGM back into the ISM plays in regulating the ISM metallicities of dwarf galaxies in the IllustrisTNG model. Further work examining the early-time CGM and ISM properties of dwarf galaxies in the IllustrisTNG model is needed to determine if the higher metallicities it predicts at M$_* \leq 10^7$ M$_\odot$ in our MZHs can be explained by the impact of supernova-driven outflows and metal recycling from the CGM.

The stellar feedback model in \textsc{galacticus} is also capable of driving significant outflows in dwarf galaxies. These outflows are captured in a CGM-like reservoir from which gas can re-accrete onto the galaxy. As shown by \cite{Ahvazi2024}, increasing the metallicity of accreting gas over time is critical to match the metallicities of MW dwarf satellites with M$_* \leq 10^5$ M$_\odot$; this mechanism is similar to that found to be responsible for the time dependence of the MZR in FIRE-2 by \cite{Bassini2024}. \textsc{galacticus} uses a model for supernova-driven outflows that scales as a power-law with the ratio of outflow velocity (with fiducial value $V_\text{out} = 200$ km/s) to the galaxy's circular velocity ($\dot{M}_\text{out} \propto \left( \frac{V_\text{out}}{V_\text{circ}} \right)^{\alpha}$); we believe this is the primary reason \textsc{galacticus} predicts a MZH that is nearly a power law in Figure \ref{fig:mzh_result} with $\langle \text{[M/H]}(M_*) \rangle \approx -1.5 + 0.38 \, \left( \text{log} \left(M_*\right) - 6 \right)$. In contrast, the hydrodynamic simulations both show deviations from a constant power law starting at M$_* \approx 10^7$ M$_\odot$. In hydrodynamic simulations, supernova ejecta must expand into an ISM which provides mechanical resistance to the expanding shock; this effect is not modeled in \textsc{galacticus} and could explain why the hydrodynamic simulations do not exhibit simple power-law scaling in the MZH. 

At M$_* = 10^7$ M$_\odot$, the \textsc{galacticus} MZH lies between the predictions of TNG50 and FIREbox, but the hydrodynamic simulations show a steepening of the MZH below this mass while \textsc{galacticus} does not. As a result, \textsc{galacticus} predicts the highest metallicity of the three models at M$_* = 10^6$ M$_\odot$, which is the lowest mass for which we measure the MZHs in the hydrodynamic simulations due to resolution concerns. The sample variance in the MZH in \textsc{galacticus} is also lower than in the hydrodynamic simulations. The standard deviation in [M/H] at fixed M$_*$ is typically $\sigma \lesssim 0.1$ dex, indicating that the metallicity of the ISM correlates very strongly with the total stellar mass formed   and secondary variables like the SFH and accretion history impact the MZH less in \textsc{galacticus} than in the hydrodynamic simulations. This is, again, likely a result of the simple power law scaling between the supernova-driven outflow rate and the galactic circular velocity in \textsc{galacticus} introducing little variance into the MZH.

\subsection{Aquarius, Leo A, and Leo P}
Our simulation results are limited to galaxies that are of similar mass to WLM at present-day due to resolution concerns in TNG50 and FIREbox. Therefore, we would not necessarily expect these simulation results to agree well with the results for Aquarius, Leo A, and Leo P, which have formed an order of magnitude less stellar mass over their histories compared to WLM. Additional work making similar measurements with higher resolution simulations is warranted to better interpret these measurements, but we offer some general comments here.

We note first that comparing our measured MZHs for WLM, Leo A, and Aquarius indicates that the MZH is not universal. If the MZH was universal, we would expect that the MZHs for these galaxies would form a continuous power law in Figure \ref{fig:mzh_result}, but they do not. In the limited stellar mass range where they overlap (2--4 $\times 10^6$ M$_\odot$) our measurement for WLM shows a significantly lower metallicity ([M/H] $\sim -2$ dex) compared to Aquarius and Leo A ([M/H] $\gtrsim -1.5$ dex).

The most likely explanation for this is that the MZH has an additional time dependence -- WLM formed a total stellar mass of $2 \times 10^6$ M$_\odot$ over 12 Gyr ago, while Aquarius and Leo A reached this stellar mass only a few Gyr ago. Hydrodynamic simulations predict that the MZR should have such a time dependence, although the quantitative scaling of the time dependence and the underlying causal mechanism that creates the time dependence differs between galaxy evolution models \citep[e.g.,][see also the discussion in \S \ref{subsec:wlm_discussion}]{Torrey2019,Bassini2024,Garcia2025a}. Observations also indicate the existence of a time-dependent MZR, though these studies target galaxies of greater masses than we consider here \citep{Li2023c,Nakajima2023,Curti2024,Jain2026}.

Such a time dependence could be added to our MZH model by modifying Equation \ref{eq:mzhmodel}. In studies of the MZR, time dependence is often modeled by making the power law slope $\alpha$ and/or the normalization [M/H]$_0$ functions of time. This could be added to our model without much difficulty, but constraining the time dependence requires forming a statistical sample of measurements for galaxies that probe $\langle \text{[M/H]} \rangle (\text{M}_*)$ for the same range of M$_*$ at different times to constrain $\alpha(t)$ and [M/H]$_0(t)$. Our present sample is too small for such an analysis to be robust, but future extensions of this work measuring the MZHs of additional galaxies in the Local Universe will enable us to constrain the time dependence of the MZH.

Even in comparing just Leo A and Aquarius we can see indications of a time dependence in the MZH. While both galaxies have comparable stellar mass and metallicity at present-day, our measurements show that they have different MZHs. We fit power law slopes of $\alpha \approx 1.5$ for Aquarius and $\alpha \approx 0.5$ for Leo A; even considering the range of $\alpha$ values fit under different stellar tracks and BCs, the two galaxies show inconsistent power law slopes. The most conspicuous difference between these galaxies that could explain these results are their SFHs (i.e, \emph{when} they formed their stars). Relative to other Local Group dwarfs, Leo A formed its stellar mass quite late -- our measurement indicates that Leo A formed $10\%$ of its total stellar mass only $\sim7$ Gyr ago ($\tau_{10} \approx 7$; see Table \ref{table:sfhsummary}), while Aquarius has $\tau_{10} > 13$ Gyr. Even their $\tau_{50}$ values are significantly different, as $\tau_{50} \approx 3.6$ Gyr for Leo A and $\tau_{50} \approx 6.7$ Gyr for Aquarius. In the presence of a time-dependent MZH, it would be expected that these galaxies would show different metallicities at fixed total stellar mass formed as the \emph{time} at which they reached that stellar mass would be different. 

As discussed in \S \ref{subsec:leoP}, the metallicity we fit for Leo P is $\sim0.6$ dex higher than is implied by the oxygen abundance of its single \HII region \cite{Skillman2013} and no other metallicity constraints are available to constrain the metallicity of Leo P's stellar population. If Leo P is currently forming stars at [M/H] $\approx -1.5$ dex as implied by the \HII region oxygen abundance, then it would be roughly consistent with the \textsc{galacticus} prediction. As Leo P is nearly 2 dex lower in stellar mass compared to the mean stellar mass of the simulated galaxies, we refrain from commenting on it further -- additional work with higher resolution simulations, probing an appropriate range of stellar mass, is needed to better interpret the result for Leo P.

\section{Conclusion}

We have developed a new framework that directly links the chemical enrichment history of galaxies to their resolved SFHs by modeling metallicity evolution as a function of cumulative stellar mass formed. We apply this methodology to four isolated, gas-rich Local Group dwarf galaxies—WLM, Aquarius, Leo A, and Leo P—and present new measurements for their global SFHs and chemical evolution history. We validate our chemical evolution measurements against independent measurements of the ages and metallicities of individual red giant stars in WLM, Leo A, and Aquarius, finding good agreement with our results. We measure systematic uncertainties by repeating our measurements using multiple stellar evolution and bolometric correction libraries, demonstrating the effect these modeling assumptions have on our results.

We examine the chemical enrichment history of WLM-analog galaxies in the TNG50 and FIREbox cosmological hydrodynamic simulations and the \textsc{galacticus} semi-analytic model in order to examine how our observations can help to constrain galaxy evolution physics. We find that all three models predict similar metallicities at higher masses (M$_* \geq 10^8$ \msun) but deviate significantly at lower masses, highlighting the importance of studying low-mass dwarf galaxies to constrain galaxy evolution physics. The chemical evolution history we measure for WLM is most similar to the predictions of FIREbox, with TNG50 and \textsc{galacticus} predicting higher metallicities at early times. We ascribe the differences between the model predictions to their stellar feedback predictions that regulate the metallicity of the ISM by driving metal-enriched outflows out of the ISM. We suggest that studying the outflow properties of dwarf galaxies at early times may further reveal the mechanisms by which the chemical enrichment of dwarf galaxies is regulated.

Our results underscore the power of combining resolved SFHs with physically motivated chemical evolution models to constrain the physics of galaxy formation, especially in the low-mass regime. They also emphasize the need for further observational and theoretical work to better understand stellar feedback, metal retention, and the cycling of metals between the ISM and the CGM. This framework opens new avenues for testing galaxy evolution models using observations of nearby galaxies.

%% Please use the acknowledgment and contribution environments. This will 
%% be anonomyized when the "anonymous" style option is used. 
\begin{acknowledgments}
We thank Leo Girardi for helpful discussions on stellar tracks and bolometric corrections. We thank James Johnson for helpful discussions on galactic chemical evolution models. The authors acknowledge \href{https://rc.virginia.edu}{Research Computing at The University of Virginia} for providing computational resources and technical support that have contributed to the results reported within this publication. We acknowledge useful discussions during Galaxy Evolution and Cosmology (GECO) lunches at the University of Virginia. Support for this work was provided by the Owens Family Foundation and by NASA through grant HST-AR-17560 from the Space Telescope Science Institute, which is operated by AURA, Inc., under NASA contract NAS 5-26555. This work is based on observations made with the NASA/ESA/CSA James Webb Space Telescope. The data were obtained from the Mikulski Archive for Space Telescopes at the Space Telescope Science Institute, which is operated by the Association of Universities for Research in Astronomy, Inc., under NASA contract NAS 5-03127 for JWST. These observations are associated with program DD-ERS-1334. This research is also based in part on observations made with the NASA/ESA Hubble Space Telescope obtained from the Space Telescope Science Institute, which is operated by the Association of Universities for Research in Astronomy, Inc., under NASA contract NAS 5–26555. These observations are associated with programs HST GO-12925 and GO-10590. This research has made use of NASA Astrophysics Data System Bibliographic Services and the NASA/IPAC Extragalactic Database (NED), which is operated by the Jet Propulsion Laboratory, California Institute of Technology, under contract with the National Aeronautics and Space Administration.
\end{acknowledgments}

\begin{contribution}
%%This section gives authors the space to recognize author contributions. The text inside this environment is NOT counted towards the total word quanta. At a minimum, manuscripts are expected to include this text:
  %% All authors contributed equally to the Terra Mater collaboration.

CTG is the primary author responsible for planning, executing, and presenting the work. AMG performed the selection of WLM analogs from the TNG50 simulation. NA ran the \textsc{galacticus} simulations presented in this work and assisted with analyzing the results. NK helped plan, organize, and direct the work. KBWM provided the photometric data products for Aquarius, Leo A, and Leo P and the \textsc{match} SFH measurements from \cite{McQuinn2024b} to facilitate comparison to the new measurements. RF provided access to the FIREbox simulations and assisted CTG in working with the data products. RC assisted with editing of the manuscript.

%% But authors are expected to provide more specific details, e.g. 
%%
%%SC was responsible for writing and submitting the manuscript.
%%WWM came up with the initial research concept and edited the manuscript.
%%OTS obtained the funding and edited the manuscript.
%%EBF provided the formal analysis and validation. He also edited the manuscript.
%%GEH Supervised the undergraduates, wrote the software and administers the project github and Zenodo repositories.
%%
%% Authors can use the Contributor Role Taxonomy (CRediT) at
%% for ideas on how write a good statement tailored to their needs.

\end{contribution}

\facilities{JWST (NIRCAM), HST(ACS)}

\software{This research made use of the following software packages: 
\begin{enumerate}
    \item \sfhjl \citep{StarFormationHistories.jl} v1.3.1
    \item \href{https://github.com/cgarling/InitialMassFunctions.jl}{\texttt{InitialMassFunctions.jl}} v0.1.6
    \item \bcjl v0.2.0
    \item \stjl v1.1.0
    \item The Julia programming language \citep{Julia} v1.12.6
    \item \texttt{Matplotlib} \citep{Matplotlib} v3.10.8
    \item Distributions.jl \citep{Besancon2021,Distributions.jl} v0.25.125
    \item DynamicHMC.jl \citep{DynamicHMC.jl} v3.6.0
    \item Optim.jl \citep{Optim.jl} v2.0.1
\end{enumerate}
}


\appendix

\section{The Age Distribution on the Red Giant Branch of Leo A} \label{appendix:leoA}
\begin{figure*}
\centering
\includegraphics[width=0.45\textwidth]{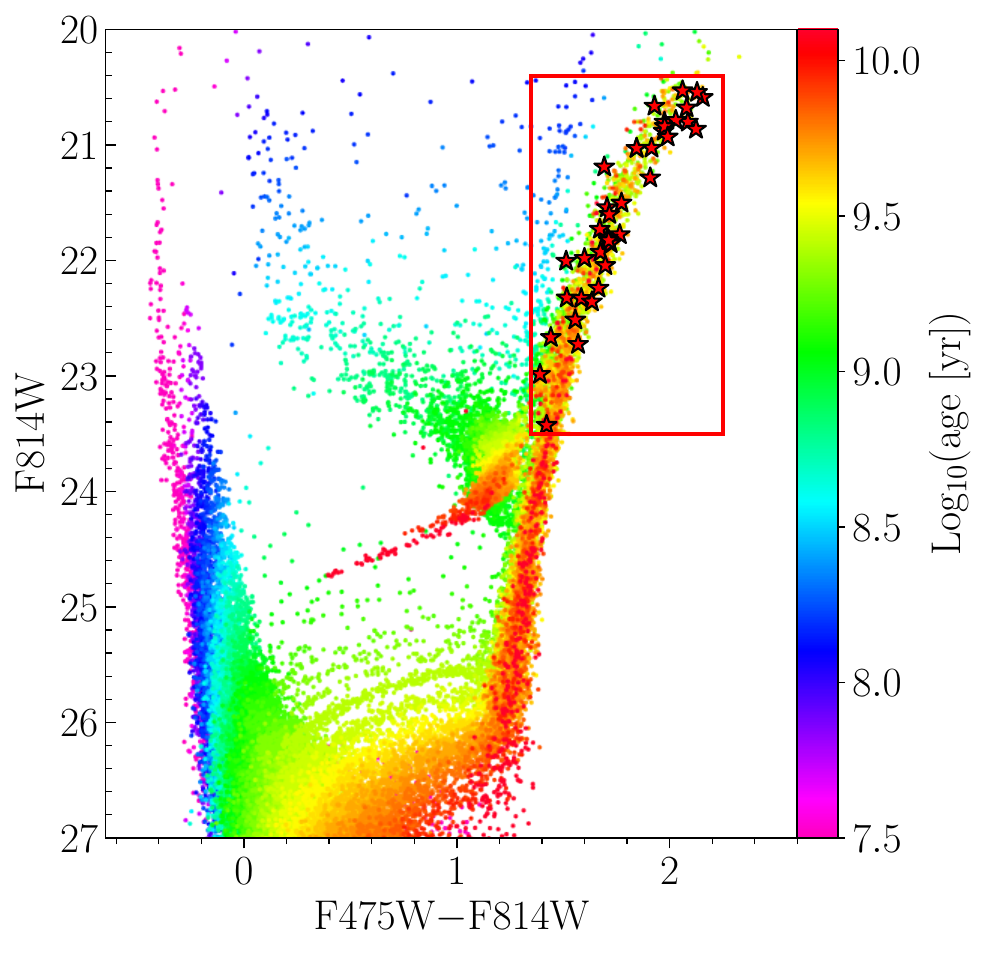}
\includegraphics[width=0.45\textwidth]{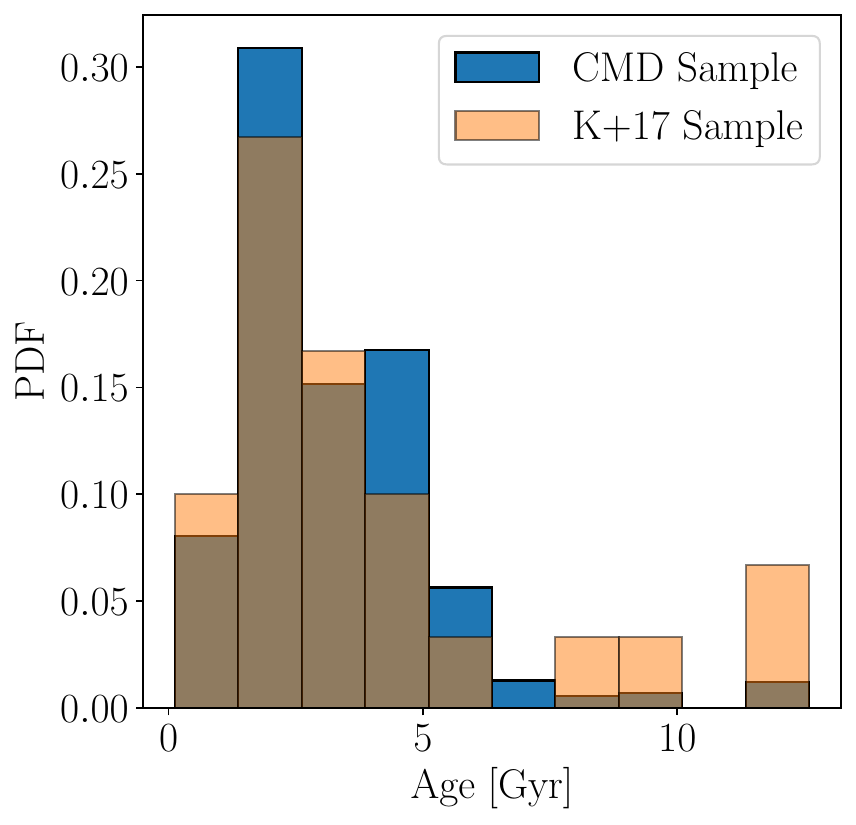}
\caption{\emph{Left:} A mock CMD simulated assuming the best-fit SFH of Leo A. Stars are colored according to the logarithms of their ages. Stars with spectroscopic metallicities from \cite{Kirby2017a} are overplotted as red stars. The red box contains the sample of \cite{Kirby2017a}. \emph{Right:} The distribution of ages for all randomly sampled stars in the red box region from the left panel (blue) compared to the age distribution of observed red giant stars with spectroscopic metallicities from \cite{Kirby2017a} (orange) illustrating that about $80\%$ of the stars in this region are expected to be younger than 5 Gyr.}
\label{fig:leoA_ages}
\end{figure*}

Here we examine the age distribution of stars on the upper RGB of Leo A in order to understand the difficulty inherent in measuring its metallicity evolution. Using the methods described in appendix A of \cite{Garling2025}, we use PARSEC stellar tracks and YBC BCs to sample a random realization of Leo A's CMD assuming our best-fit SFH. Other relevant parameters (e.g., distance, reddening, IMF) are the same as were used in the SFH fit. We add photometric error and account for incompleteness using the same methods as are used when measuring the SFH. In the left panel of Figure \ref{fig:leoA_ages} we show the mock CMD; stars with spectroscopic metallicities from \cite{Kirby2017a} for which we also have HST/ACS photometry are shown overplotted as red star symbols. The red boxed region encloses the sample of \cite{Kirby2017a}. We plot the age distribution of stars that fall within this region in the right panel. This distribution demonstrates a clear and significant skew to young ages ($<5$ Gyr), with only $19\%$ of stars in the region being older than 5 Gyr. This is in reasonable agreement with the observations, as $14\%$ of the stars for which we measure ages are older than 5 Gyr. We hypothesize that the relative paucity of bright, old stars in Leo A has made it difficult for previous work to accurately measure the historical metallicity evolution in Leo A, as younger stars with higher metallicities are far more numerous on the upper RGB.

\section{The Star Formation History of Leo P with Fixed Metallicity} \label{appendix:leoP}
\begin{figure*}
\centering
\includegraphics[width=0.85\textwidth]{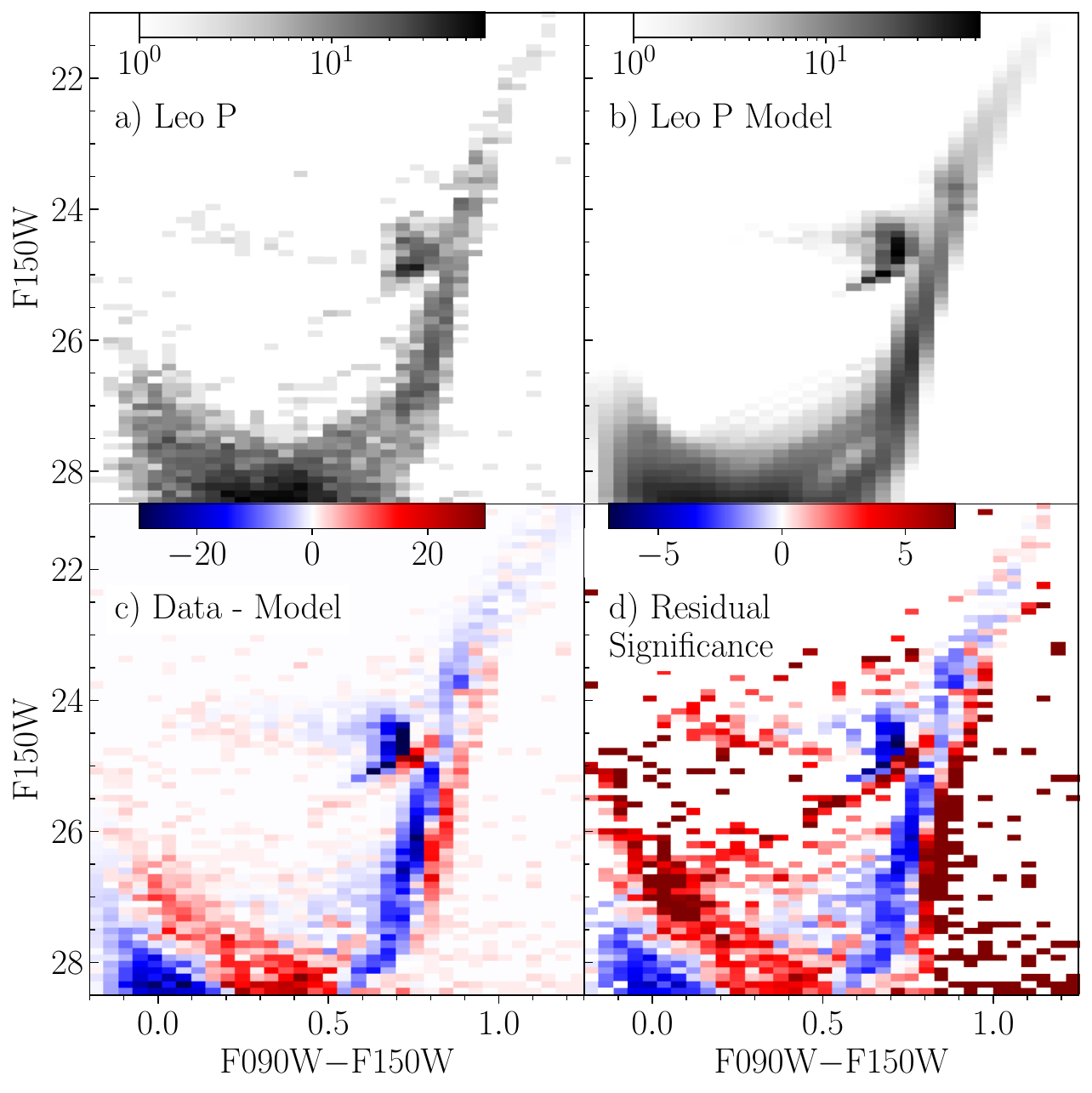} \\
\caption{Comparison of the JWST/NIRCam Hess diagram of Leo P with our best-fit model assuming [M/H] = $-1.5$ dex at present-day. The panels follow the same layout as in Figure \ref{fig:wlm_hess}. While the upper RGB in the model is a better match to the data than in our fiducial fit (Figure \ref{fig:leoP_hess}), the lower RGB and upper main sequence show significant residuals and the fit quality is worse overall.}
\label{fig:leoP_hess_appendix}
\end{figure*}

\begin{figure*}[hpbt]
\centering
\includegraphics[width=0.85\textwidth]{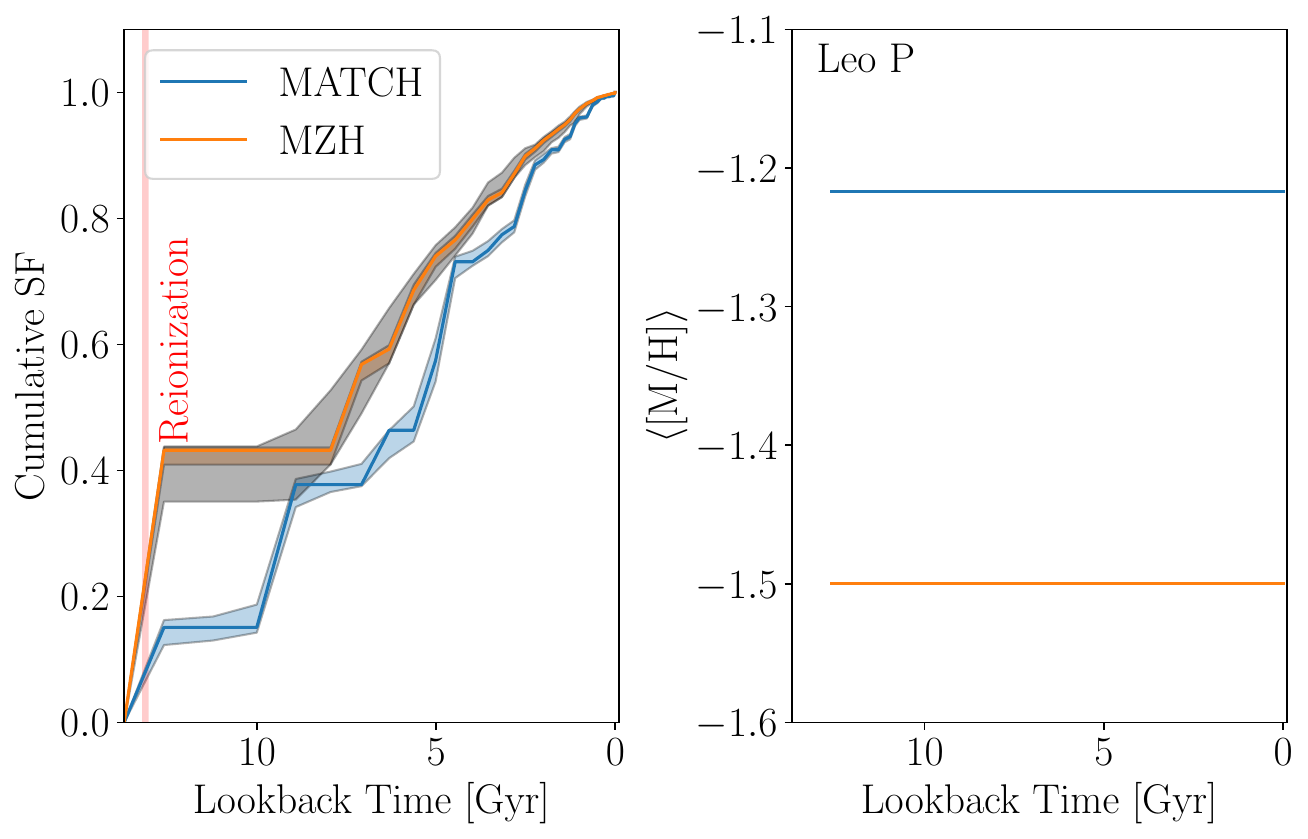} \\
\includegraphics[width=0.85\textwidth]{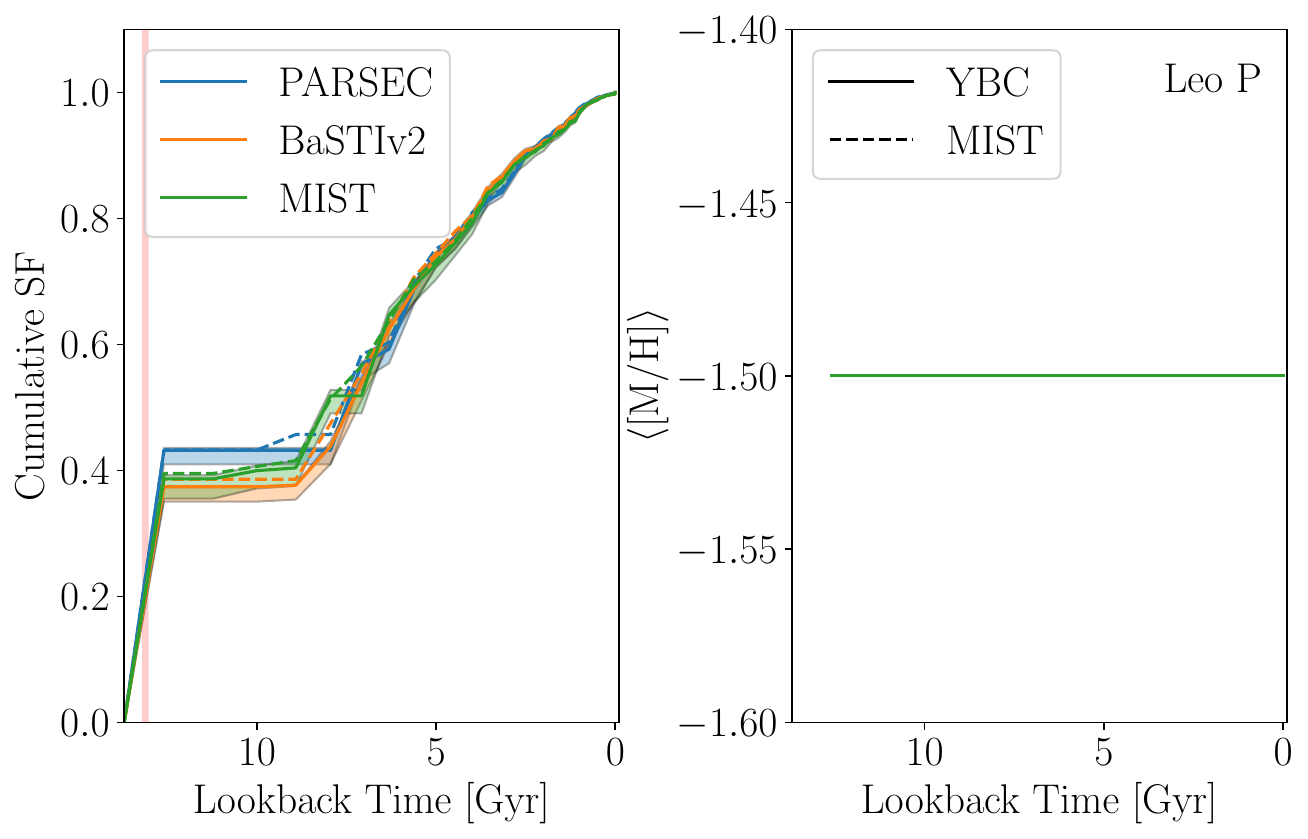}
\caption{The results of our resolved SFH and metallicity evolution fit for Leo P assuming [M/H] = $-1.5$ dex at present-day. The panels follow the same format as Figure \ref{fig:wlm_sfh}. In the top row the result of \cite{McQuinn2024b} (blue line), who used the \textsc{match} code on the same data, is shown for comparison. Assuming a lower metallicity than our fiducial fit results in a very large initial burst of star formation in which Leo P forms roughly $40\%$ of its total stellar mass. Overall, Leo P formed more stellar mass earlier in this model than in our fiducial fit.}
\label{fig:leoP_sfh_appendix}
\end{figure*}

Here we present the results of an experiment fitting the SFH of Leo P assuming a present-day metallicity of [M/H]$=-1.5$ dex, consistent with the oxygen abundance measured from the single \HII region in Leo P \citep{Skillman2013}. The MZH slope parameter ($\alpha$ in Equation \ref{eq:mzhmodel}) is left free to vary, but is found to be 0 for all solutions. We performed other experiments with different fixed metallicities, but these results are representative of the general trends we observed. Our best-fit model Hess diagram is compared to the data in Figure \ref{fig:leoP_hess_appendix} and the resulting SFH is shown in Figure \ref{fig:leoP_sfh_appendix}.

Assuming this lower metallicity results in a better fit to the upper RGB than in our fiducial fit (Figure \ref{fig:leoP_hess}), but the lower RGB and upper main sequence show significant residuals that are not present in our fiducial fit. As these areas of the CMD contain many more stars than the upper RGB, the overall fit quality is worse than in our fiducial fit. The resulting SFH shows a very large initial burst of star formation in which Leo P forms roughly $40\%$ of its total stellar mass. Generally this solution indicates that Leo P formed more stellar mass earlier than in our fiducial fit. Future work constraining the metallicity of Leo P's stellar population is needed to improve the modeling of this low mass dwarf galaxy.

\section{Inferring Birth Masses of Star Particles} \label{appendix:birthmass}
In both the IllustrisTNG and FIRE-2 galaxy evolution models, star particles are born with irregular masses as the initial mass of each star particle is inherited from its parent gas resolution element (the gas resolution elements are cells in IllustrisTNG simulations and particles in FIRE-2). The masses of the gas resolution elements change over time in order to conserve mass in the presence of stellar winds, supernovae, and other processes that transfer mass between gas resolution elements. 
In both models, star particles are massive enough that they are modeled as SSPs and the evolution of their properties is tracked accordingly. 
In particular, star particle masses decrease over time following models of stellar mass loss due to massive star evolution. Stellar mass lost due to massive star winds and supernovae is returned to the ISM, and the so the fraction of stellar mass lost is also known as the recycled fraction ($R$). The recycled fraction as a function of the SSP age can be calculated given models for stellar lifetimes, an assumed IMF, and appropriate yield tables for each relevant type of massive star. The recycled fraction will be highest for the oldest star particles which have $R\approx0.3$ in both IllustrisTNG and FIRE-2 (see, e.g., Figure 1 of \citealt{Pillepich2018} and Figure 4 of \citealt{Hopkins2023b}). 

\begin{figure*}[hpbt]
\centering
\includegraphics[width=0.75\textwidth]{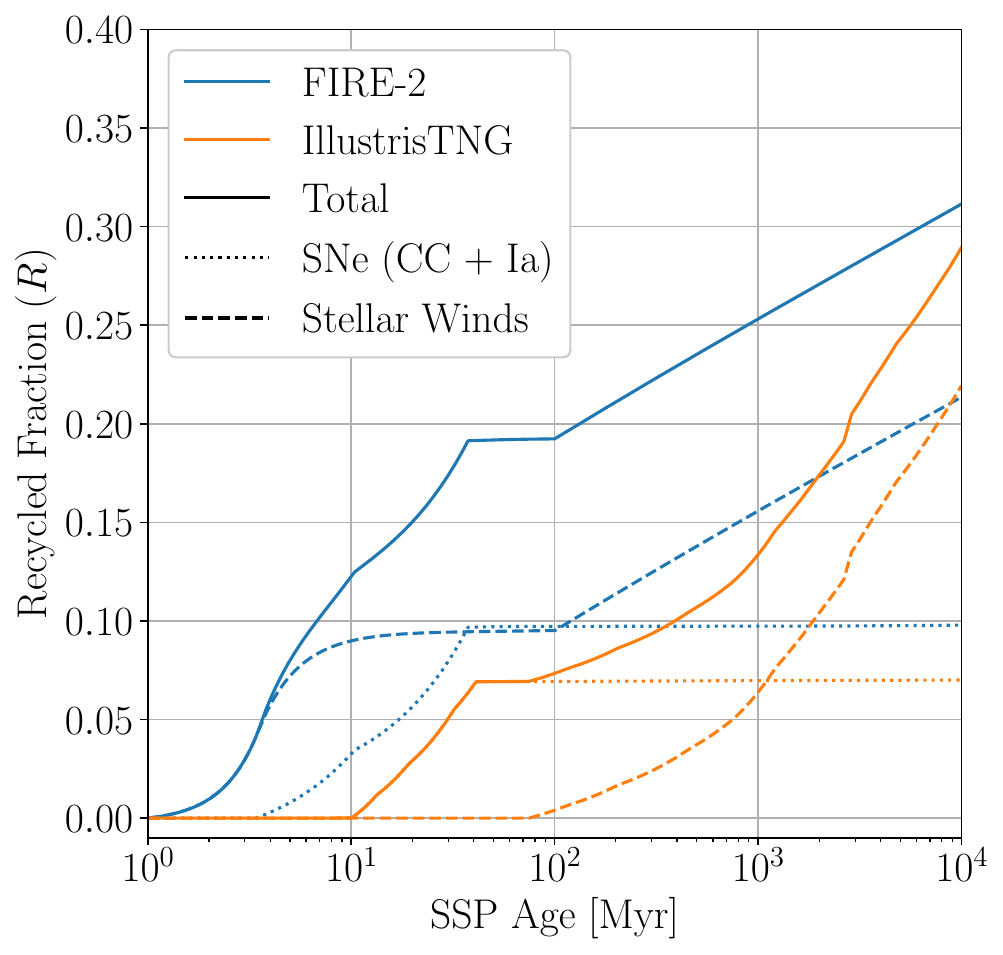}
\caption{The recycled fraction (i.e., the fraction of stellar mass returned to the ISM) as a function of SSP age under the FIRE-2 (blue) and IllustrisTNG (orange) models. Shown are the total fractional mass loss (solid lines), as well as the fractional mass loss separated by source (dotted lines show fractional mass loss due to supernovae, dashed lines show fractional mass loss due to stellar winds).}
\label{fig:massloss}
\end{figure*}

Conveniently, appendix A of \cite{Hopkins2018} gives IMF-averaged yields and fitting functions for the occurrence rates of SN-Ia and CCSN in FIRE-2, as well as stellar mass-loss rates from winds driven by massive O and B-type stars as well as evolved AGB stars. We use this information to calculate the recycled fraction as a function of SSP age shown in Figure \ref{fig:massloss} (blue). We show the fraction of stellar mass lost due to supernovae (dotted), stars (dashed), and the total recycled fraction (solid). This simple model based on fitting functions reproduces Figure 4 of \cite{Hopkins2023b} well, which compares the stellar mass fraction lost as a function of SSP age for the FIRE-2 and FIRE-3 galaxy evolution models. 

No similar fitting functions are available for IllustrisTNG, so we calculate the recycled fraction from first principles. The assumed stellar evolution physics for IllustrisTNG is discussed in section 2.3.3 of \cite{Pillepich2018} and is summarized in their Table 2 \citep[see also section 2.1 of][]{Torrey2019}. Following the above references, we adopt the IMF from \cite{Chabrier2003}, total CCSN ejecta masses and stellar lifetimes from \cite{Portinari1998}, SN-Ia rates from \cite{Maoz2012}, and total AGB wind masses from \cite{Karakas2010}. We note that the abundance ratios for CCSN in IllustrisTNG follow \cite{Kobayashi2006}, but the total ejecta masses are normalized to match \cite{Portinari1998}, which is the relevant quantity for computing the fraction of stellar mass lost from an SSP as a function of age. Additionally, IllustrisTNG integrates yield tables from other sources for AGB stars with initial masses between 6--7 \msun but we omit these for simplicity, using only the \cite{Karakas2010} yields for AGB stars with initial masses 1--6 \msun which are sufficient to capture the majority of the mass loss due to AGB star winds. Using these as inputs, we calculate the recycled fraction as a function of SSP age under the IllustrisTNG model, shown in Figure \ref{fig:massloss} (orange). The fractions of stellar mass lost after 10 Gyr ($\sim21\%$ due to AGB stars and $\sim7\%$ due to supernovae, for $\sim28\%$ total) agree well with the values shown in Figure 1 of \cite{Pillepich2018}.

Given these full models, it is possible to infer the birth mass of each star particle based on its age and present-day mass by interpolating the recycled fraction as a function of SSP age. However, this would be somewhat excessive as $>99\%$ of the star particles are older than 40 Myr at present-day, meaning that there will be no further recycling from CCSN. Additionally, the total mass recycling from SN-Ia is $<1\%$ over a Hubble time \citep[see also section 2.2 of][]{Segers2016}, making AGB star winds the only significant contributor to the evolution of the recycled fraction for SSPs older than $40$ Myr. Additionally, the fraction of mass lost due to AGB star winds only changes by 10--$15\%$ for SSP ages older than 1 Gyr in both models, indicating that the recycled fraction changes slowly over most of cosmic time. As a result, the birth stellar masses of the star particles in our WLM-analog galaxies can be well approximated by simply assuming a population-averaged recycled fraction (e.g., $R$ evaluated at $\tau_{50} = 4.68$ Gyr; see Table \ref{table:sfhsummary}) and computing the birth stellar mass based on the present-day stellar mass as $\text{M}_*(z=0) \div (1 - R)$. We find the appropriate average recycling fractions to be $R=0.25$ for the IllustrisTNG model and $R=0.3$ for FIRE-2. Neglecting the age dependence results in a birth mass uncertainty of $\sim10\%$, which does not meaningfully impact our MZH results shown in Figure \ref{fig:mzh_result} as our measurements span several orders of magnitude in stellar mass.

As noted in the main text, \textsc{galacticus} uses the instantaneous recycling approximation for returning stellar material back into the ISM with a fixed recycled fraction \citep[$R=0.46$; section 2.2 of][]{Knebe2018} so that inferring the amount of stellar mass born as a function of time is trivial.

\bibliographystyle{aasjournal}
% run with full bib file, then use JabRef -> Tools -> New sublibrary based on AUX file
% put in generated AUX file, and save new sublibrary to use here
% \bibliography{/mnt/c/Users/cgarling/Dropbox/JabRef/data/library.bib}
\bibliography{library}

@Article{Weerasooriya2023,
  author     = {Weerasooriya, Sachi and Bovill, Mia Sauda and Benson, Andrew and Musick, Alexi M. and Ricotti, Massimo},
  title      = {Devouring the {Milky} {Way} {Satellites}: {Modeling} {Dwarf} {Galaxies} with {Galacticus}},
  doi        = {10.3847/1538-4357/acc32b},
  issn       = {0004-637X},
  pages      = {87},
  url        = {https://ui.adsabs.harvard.edu/abs/2023ApJ...948...87W},
  urldate    = {2024-03-24},
  volume     = {948},
  journal    = {ApJ},
  month      = may,
  ranking    = {rank3},
  shorttitle = {Devouring the {Milky} {Way} {Satellites}},
  year       = {2023},
}

@Article{Hodge1989,
  author   = {Hodge, Paul},
  journal  = {\araa},
  title    = {Populations in Local Group galaxies.},
  year     = {1989},
  month    = jan,
  pages    = {139-198},
  volume   = {27},
  doi      = {10.1146/annurev.aa.27.090189.001035},
  url      = {https://ui.adsabs.harvard.edu/abs/1989ARA&A..27..139H},
}

@Article{Tosi1991,
  author   = {Tosi, M. and Greggio, L. and Marconi, G. and Focardi, P.},
  journal  = {AJ},
  title    = {Star Formation in Dwarf Irregular Galaxies: Sextans B},
  year     = {1991},
  month    = sep,
  pages    = {951},
  volume   = {102},
  doi      = {10.1086/115925},
  url      = {https://ui.adsabs.harvard.edu/abs/1991AJ....102..951T},
}

@Article{Greggio1993,
  author   = {Greggio, L. and Marconi, G. and Tosi, M. and Focardi, P.},
  journal  = {AJ},
  title    = {Star Formation in Dwarf Irregular Galaxies: DDO 210 and NGC 3109},
  year     = {1993},
  month    = mar,
  pages    = {894},
  volume   = {105},
  doi      = {10.1086/116481},
  url      = {https://ui.adsabs.harvard.edu/abs/1993AJ....105..894G},
}

@Article{Hodge1995,
  author   = {Hodge, Paul and Miller, Bryan W.},
  journal  = {ApJ},
  title    = {H II Regions in Four Galaxies in and near the Local Group},
  year     = {1995},
  month    = sep,
  pages    = {176},
  volume   = {451},
  doi      = {10.1086/176209},
  url      = {https://ui.adsabs.harvard.edu/abs/1995ApJ...451..176H},
}

@Article{Gratton2010,
  author        = {Gratton, R.~G. and D'Orazi, V. and Bragaglia, A. and Carretta, E. and Lucatello, S.},
  journal       = {\aap},
  title         = {The connection between missing AGB stars and extended horizontal branches},
  year          = {2010},
  month         = nov,
  pages         = {A77},
  volume        = {522},
  archiveprefix = {arXiv},
  doi           = {10.1051/0004-6361/201015405},
  eid           = {A77},
  eprint        = {1010.5913},
  primaryclass  = {astro-ph.SR},
  url           = {https://ui.adsabs.harvard.edu/abs/2010A&A...522A..77G},
}

@Article{Harmsen2023,
  author        = {Harmsen, Benjamin and Bell, Eric F. and D'Souza, Richard and Monachesi, Antonela and de Jong, Roelof S. and Smercina, Adam and Jang, In Sung and Holwerda, Benne W.},
  journal       = {MNRAS},
  title         = {Constraining the assembly time of the stellar haloes of nearby Milky Way-mass galaxies through AGB populations},
  year          = {2023},
  month         = nov,
  number        = {3},
  pages         = {4497-4514},
  volume        = {525},
  archiveprefix = {arXiv},
  doi           = {10.1093/mnras/stad2480},
  eprint        = {2308.11499},
  primaryclass  = {astro-ph.GA},
  url           = {https://ui.adsabs.harvard.edu/abs/2023MNRAS.525.4497H},
}

@Article{Koester2010,
  author   = {Koester, D.},
  title    = {White dwarf spectra and atmosphere models},
  pages    = {921-931},
  url      = {https://ui.adsabs.harvard.edu/abs/2010MmSAI..81..921K},
  volume   = {81},
  journal  = {\memsai},
  month    = jan,
  year     = {2010},
}

@Article{Wang2024a,
  author    = {Wang, Yunchong and Nadler, Ethan O. and Mao, Yao-Yuan and Wechsler, Risa H. and Abel, Tom and Behroozi, Peter and Geha, Marla and Asali, Yasmeen and de los Reyes, Mithi A. C. and Kado-Fong, Erin and Kallivayalil, Nitya and Tollerud, Erik J. and Weiner, Benjamin and Wu, John F.},
  title     = {The {SAGA} {Survey}. {V}. {Modeling} {Satellite} {Systems} around {Milky} {Way}-mass {Galaxies} with {Updated} {UniverseMachine}},
  doi       = {10.3847/1538-4357/ad7f4c},
  issn      = {1538-4357},
  note      = {ADS Bibcode: 2024arXiv240414500W Type: article},
  number    = {1},
  pages     = {119},
  url       = {https://ui.adsabs.harvard.edu/abs/2024arXiv240414500W},
  urldate   = {2024-10-10},
  volume    = {976},
  journal   = {ApJ},
  month     = nov,
  publisher = {American Astronomical Society},
  year      = {2024},
}

@Article{Weerasooriya2024,
  author     = {Weerasooriya, Sachi and Bovill, Mia Sauda and Taylor, Matthew A. and Benson, Andrew J. and Leahy, Cameron},
  title      = {Devouring the {Centaurus} {A} {Satellites}: {Modeling} {Dwarf} {Galaxies} with {Galacticus}},
  doi        = {10.3847/1538-4357/ad3924},
  issn       = {0004-637X},
  pages      = {78},
  url        = {https://ui.adsabs.harvard.edu/abs/2024ApJ...968...78W},
  urldate    = {2025-03-16},
  volume     = {968},
  journal    = {ApJ},
  month      = jun,
  publisher  = {IOP},
  shorttitle = {Devouring the {Centaurus} {A} {Satellites}},
  year       = {2024},
}

@Misc{StarFormationHistories.jl,
  author    = {Chris Garling},
  title     = {StarFormationHistories.jl},
  doi       = {10.5281/zenodo.14963317},
  copyright = {MIT License},
  publisher = {Zenodo},
  year      = {2025},
}

@Article{Bressan2012,
  author   = {Bressan, Alessandro and Marigo, Paola and Girardi, L{\'{e}}o and Salasnich, Bernardo and {Dal Cero}, Claudia and Rubele, Stefano and Nanni, Ambra},
  title    = {{PARSEC : stellar tracks and isochrones with the PAdova and TRieste Stellar Evolution Code}},
  doi      = {10.1111/j.1365-2966.2012.21948.x},
  issn     = {00358711},
  number   = {1},
  pages    = {127--145},
  url      = {https://ui.adsabs.harvard.edu/abs/2012MNRAS.427..127B/abstract},
  volume   = {427},
  journal  = {MNRAS},
  ranking  = {rank4},
  year     = {2012},
}

@Article{Cardelli1989,
  author        = {Cardelli, Jason A. and Clayton, Geoffrey C. and Mathis, John S.},
  title         = {{The relationship between infrared, optical, and ultraviolet extinction}},
  doi           = {10.1086/167900},
  eprint        = {0706.1275},
  issn          = {0004-637X},
  number        = {1},
  pages         = {245},
  url           = {http://adsabs.harvard.edu/doi/10.1086/167900},
  volume        = {345},
  archiveprefix = {arXiv},
  arxivid       = {0706.1275},
  isbn          = {doi:10.1086/167900},
  journal       = {ApJ},
  pmid          = {25246403},
  year          = {1989},
}

@Article{White1991,
  author    = {White, Simon D. M. and Frenk, Carlos S.},
  title     = {Galaxy {Formation} through {Hierarchical} {Clustering}},
  doi       = {10.1086/170483},
  issn      = {0004-637X},
  note      = {ADS Bibcode: 1991ApJ...379...52W},
  pages     = {52},
  url       = {https://ui.adsabs.harvard.edu/abs/1991ApJ...379...52W},
  urldate   = {2025-04-02},
  volume    = {379},
  journal   = {ApJ},
  month     = sep,
  publisher = {IOP},
  year      = {1991},
}

@Article{Woo2013,
  author    = {Joanna Woo and Avishai Dekel and S. M. Faber and Kai Noeske and David C. Koo and Brian F. Gerke and Michael C. Cooper and Samir Salim and Aaron A. Dutton and Jeffrey Newman and Benjamin J. Weiner and Kevin Bundy and Christopher N. A. Willmer and Marc Davis and Renbin Yan},
  title     = {Dependence of galaxy quenching on halo mass and distance from its centre},
  doi       = {10.1093/mnras/sts274},
  number    = {4},
  pages     = {3306--3326},
  volume    = {428},
  journal   = {MNRAS},
  month     = feb,
  publisher = {Oxford University Press ({OUP})},
  ranking   = {rank3},
  year      = {2013},
}

@Article{Besancon2021,
  author    = {Besançon, Mathieu and Papamarkou, Theodore and Anthoff, David and Arslan, Alex and Byrne, Simon and Lin, Dahua and Pearson, John},
  title     = {Distributions.jl: Definition and Modeling of Probability Distributions in the JuliaStats Ecosystem},
  doi       = {10.18637/jss.v098.i16},
  issn      = {1548-7660},
  number    = {16},
  volume    = {98},
  journal   = {Journal of Statistical Software},
  publisher = {Foundation for Open Access Statistic},
  year      = {2021},
}

@Article{Garling2025,
  author   = {Garling, Christopher T. and Kallivayalil, Nitya and McQuinn, Kristen B. W. and Warfield, Jack T. and Gennaro, Mario and Cohen, Roger E.},
  title    = {Measuring {Resolved} {Star} {Formation} {Histories} from {High}-{Precision} {Color}-{Magnitude} {Diagrams} with {StarFormationHistories}.jl},
  doi      = {10.3847/1538-4365/adbb64},
  eprint   = {2407.19534},
  note     = {ADS Bibcode: 2024arXiv240719534G Type: article},
  number   = {61},
  pages    = {25},
  url      = {https://ui.adsabs.harvard.edu/abs/2025ApJS..277...61G/abstract},
  volume   = {277},
  journal  = {ApJS},
  month    = apr,
  year     = {2025},
}

@Article{ODonnell1994,
  author   = {O'Donnell, James E.},
  title    = {R v-dependent Optical and Near-Ultraviolet Extinction},
  doi      = {10.1086/173713},
  pages    = {158},
  url      = {https://ui.adsabs.harvard.edu/abs/1994ApJ...422..158O},
  volume   = {422},
  journal  = {ApJ},
  month    = feb,
  year     = {1994},
}

@Article{Paxton2013,
  author        = {Paxton, Bill and Cantiello, Matteo and Arras, Phil and Bildsten, Lars and Brown, Edward F. and Dotter, Aaron and Mankovich, Christopher and Montgomery, M.~H. and Stello, Dennis and Timmes, F.~X. and Townsend, Richard},
  title         = {Modules for Experiments in Stellar Astrophysics (MESA): Planets, Oscillations, Rotation, and Massive Stars},
  doi           = {10.1088/0067-0049/208/1/4},
  eid           = {4},
  eprint        = {1301.0319},
  number        = {1},
  pages         = {4},
  url           = {https://ui.adsabs.harvard.edu/abs/2013ApJS..208....4P},
  volume        = {208},
  archiveprefix = {arXiv},
  journal       = {\apjs},
  month         = sep,
  primaryclass  = {astro-ph.SR},
  year          = {2013},
}

@Article{Paxton2015,
  author        = {Paxton, Bill and Marchant, Pablo and Schwab, Josiah and Bauer, Evan B. and Bildsten, Lars and Cantiello, Matteo and Dessart, Luc and Farmer, R. and Hu, H. and Langer, N. and Townsend, R.~H.~D. and Townsley, Dean M. and Timmes, F.~X.},
  title         = {Modules for Experiments in Stellar Astrophysics (MESA): Binaries, Pulsations, and Explosions},
  doi           = {10.1088/0067-0049/220/1/15},
  eid           = {15},
  eprint        = {1506.03146},
  number        = {1},
  pages         = {15},
  url           = {https://ui.adsabs.harvard.edu/abs/2015ApJS..220...15P},
  volume        = {220},
  archiveprefix = {arXiv},
  comment       = {See erratum https://ui.adsabs.harvard.edu/abs/2016ApJS..223...18P/abstract},
  journal       = {\apjs},
  month         = sep,
  primaryclass  = {astro-ph.SR},
  year          = {2015},
}

@Book{Nocedal2006,
  author    = {Nocedal, Jorge},
  title     = {Numerical optimization},
  edition   = {Second edition},
  editor    = {Stephen J. Wright},
  isbn      = {9780387400655},
  pagetotal = {1664},
  publisher = {Springer},
  series    = {Springer series in operations research and financial engineering},
  address   = {New York, NY},
  booktitle = {Numerical optimization},
  ppn_gvk   = {173524662X},
  year      = {2006},
}

@Article{Reimers1975,
  author   = {Reimers, D.},
  title    = {Circumstellar absorption lines and mass loss from red giants.},
  pages    = {369-382},
  url      = {https://ui.adsabs.harvard.edu/abs/1975MSRSL...8..369R},
  volume   = {8},
  journal  = {Memoires of the Societe Royale des Sciences de Liege},
  month    = jan,
  year     = {1975},
}

@Article{Nomoto2006,
  author        = {Nomoto, Ken'ichi and Tominaga, Nozomu and Umeda, Hideyuki and Kobayashi, Chiaki and Maeda, Keiichi},
  title         = {Nucleosynthesis yields of core-collapse supernovae and hypernovae, and galactic chemical evolution},
  doi           = {10.1016/j.nuclphysa.2006.05.008},
  eprint        = {astro-ph/0605725},
  pages         = {424-458},
  url           = {https://ui.adsabs.harvard.edu/abs/2006NuPhA.777..424N},
  volume        = {777},
  archiveprefix = {arXiv},
  journal       = {\nphysa},
  month         = oct,
  primaryclass  = {astro-ph},
  year          = {2006},
}

@Article{Chen2015,
  author   = {Chen, Yang and Bressan, Alessandro and Girardi, Léo and Marigo, Paola and Kong, Xu and Lanza, Antonio},
  title    = {{PARSEC} evolutionary tracks of massive stars up to 350 {M}⊙ at metallicities 0.0001 ≤ {Z} ≤ 0.04},
  doi      = {10.1093/mnras/stv1281},
  issn     = {0035-8711},
  note     = {ADS Bibcode: 2015MNRAS.452.1068C},
  pages    = {1068--1080},
  url      = {https://ui.adsabs.harvard.edu/abs/2015MNRAS.452.1068C},
  urldate  = {2024-03-24},
  volume   = {452},
  journal  = {MNRAS},
  month    = sep,
  year     = {2015},
}

@Article{Hawcroft2024,
  author        = {Hawcroft, C. and Mahy, L. and Sana, H. and Sundqvist, J.~O. and Abdul-Masih, M. and Brands, S.~A. and Decin, L. and de Koter, A. and Puls, J.},
  title         = {Empirical mass-loss rates and clumping properties of O-type stars in the Large Magellanic Cloud},
  doi           = {10.1051/0004-6361/202348478},
  eid           = {A126},
  eprint        = {2407.06775},
  pages         = {A126},
  url           = {https://ui.adsabs.harvard.edu/abs/2024A&A...690A.126H},
  volume        = {690},
  archiveprefix = {arXiv},
  journal       = {\aap},
  month         = oct,
  primaryclass  = {astro-ph.SR},
  year          = {2024},
}

@Article{Skillman2017,
  author   = {Skillman, Evan D. and Monelli, Matteo and Weisz, Daniel R. and Hidalgo, Sebastian L. and Aparicio, Antonio and Bernard, Edouard J. and Boylan-Kolchin, Michael and Cassisi, Santi and Cole, Andrew A. and Dolphin, Andrew E. and Ferguson, Henry C. and Gallart, Carme and Irwin, Mike J. and Martin, Nicolas F. and Martínez-Vázquez, Clara E. and Mayer, Lucio and McConnachie, Alan W. and McQuinn, Kristen B. W. and Navarro, Julio F. and Stetson, Peter B.},
  title    = {The {ISLAndS} {Project}. {II}. {The} {Lifetime} {Star} {Formation} {Histories} of {Six} {Andomeda} {dSphS}},
  doi      = {10.3847/1538-4357/aa60c5},
  issn     = {0004-637X},
  note     = {ADS Bibcode: 2017ApJ...837..102S},
  pages    = {102},
  url      = {https://ui.adsabs.harvard.edu/abs/2017ApJ...837..102S},
  urldate  = {2024-02-03},
  volume   = {837},
  journal  = {ApJ},
  month    = mar,
  year     = {2017},
}

@Article{Kvasova2025,
  author    = {Kvasova, Kateryna A. and Kirby, Evan N.},
  title     = {An Analytical Galactic Chemical Evolution Model with Gas Inflow and a Terminal Wind},
  doi       = {10.3847/1538-4357/adb729},
  issn      = {1538-4357},
  note      = {ADS Bibcode: 2025arXiv250307836K Type: article},
  number    = {2},
  pages     = {92},
  url       = {https://ui.adsabs.harvard.edu/abs/2025arXiv250307836K},
  urldate   = {2025-03-12},
  volume    = {982},
  comment   = {I don't really follow what they're doing to model gas inflows (their F); it looks like it is parameterized by Xi (Table 1) which scales with beta (star formation efficiency) and eta (mass-loading factor). Then in section 4 they say Xi is a "normalization factor" and I think they set it to a fixed value (Figure 1 also shows results for Xi = 10). It looks like their mass-loading factors are also fixed (eta=0.5 from the caption of Figure 1). Maybe F is a "normalization factor" because increasing F just dilutes the ISM and moves the whole MDF to lower metallicity (and vice versa)? Not sure, might be worth talking to evan at some point},
  journal   = {ApJ},
  month     = mar,
  publisher = {American Astronomical Society},
  year      = {2025},
}

@Article{Skillman2013,
  author        = {Skillman, Evan D. and Salzer, John J. and Berg, Danielle A. and Pogge, Richard W. and Haurberg, Nathalie C. and Cannon, John M. and Aver, Erik and Olive, Keith A. and Giovanelli, Riccardo and Haynes, Martha P. and Adams, Elizabeth A.~K. and McQuinn, Kristen B.~W. and Rhode, Katherine L.},
  title         = {ALFALFA Discovery of the nearby Gas-rich Dwarf Galaxy Leo P. III. An Extremely Metal Deficient Galaxy},
  doi           = {10.1088/0004-6256/146/1/3},
  eid           = {3},
  eprint        = {1305.0277},
  number        = {1},
  pages         = {3},
  url           = {https://ui.adsabs.harvard.edu/abs/2013AJ....146....3S},
  volume        = {146},
  archiveprefix = {arXiv},
  journal       = {AJ},
  month         = jul,
  primaryclass  = {astro-ph.CO},
  year          = {2013},
}

@Article{Chen2014,
  author   = {Chen, Y. and Girardi, L. and Bressan, a. and Marigo, P. and Barbieri, M. and Kong, X.},
  title    = {{Improving PARSEC models for very low mass stars}},
  doi      = {10.1093/mnras/stu1605},
  issn     = {0035-8711},
  number   = {3},
  pages    = {2525--2543},
  url      = {https://ui.adsabs.harvard.edu/abs/2014MNRAS.444.2525C/abstract},
  volume   = {444},
  journal  = {MNRAS},
  year     = {2014},
}

@Article{Choi2016,
  author    = {Jieun Choi and Aaron Dotter and Charlie Conroy and Matteo Cantiello and Bill Paxton and Benjamin D. Johnson},
  title     = {{MESA} {ISOCHRONES} {AND} {STELLAR} {TRACKS} ({MIST}). I. {SOLAR}-{SCALED} {MODELS}},
  doi       = {10.3847/0004-637x/823/2/102},
  number    = {2},
  pages     = {102},
  url       = {https://ui.adsabs.harvard.edu/abs/2016ApJ...823..102C/abstract},
  volume    = {823},
  comment   = {Stars less massive than 0.6--0.7 solar masses are too blue? See pages 28--29.},
  journal   = {ApJ},
  month     = {may},
  publisher = {American Astronomical Society},
  year      = {2016},
}

@Article{Evans2019,
  author    = {C. J. Evans and N. Castro and O. A. Gonzalez and M. Garcia and N. Bastian and M.-R. L. Cioni and J. S. Clark and B. Davies and A. M. N. Ferguson and S. Kamann and D. J. Lennon and L. R. Patrick and J. S. Vink and D. R. Weisz},
  title     = {First stellar spectroscopy in Leo P},
  doi       = {10.1051/0004-6361/201834145},
  pages     = {A129},
  url       = {https://ui.adsabs.harvard.edu/abs/2019A&A...622A.129E/abstract},
  volume    = {622},
  journal   = {A{\&}A},
  month     = {feb},
  publisher = {{EDP} Sciences},
  year      = {2019},
}

@Article{Ma2016,
  author   = {Ma, Xiangcheng and Hopkins, Philip F. and Faucher-Giguère, Claude-André and Zolman, Nick and Muratov, Alexander L. and Kereš, Dušan and Quataert, Eliot},
  title    = {The origin and evolution of the galaxy mass–metallicity relation},
  doi      = {10.1093/mnras/stv2659},
  issn     = {0035-8711},
  number   = {2},
  pages    = {2140--2156},
  url      = {https://ui.adsabs.harvard.edu/abs/2016MNRAS.456.2140M/abstract},
  urldate  = {2024-12-05},
  volume   = {456},
  journal  = {MNRAS},
  month    = feb,
  year     = {2016},
}

@Article{Hopkins2023b,
  author     = {Hopkins, Philip F. and Wetzel, Andrew and Wheeler, Coral and Sanderson, Robyn and Grudić, Michael Y. and Sameie, Omid and Boylan-Kolchin, Michael and Orr, Matthew and Ma, Xiangcheng and Faucher-Giguère, Claude-André and Kereš, Dušan and Quataert, Eliot and Su, Kung-Yi and Moreno, Jorge and Feldmann, Robert and Bullock, James S. and Loebman, Sarah R. and Anglés-Alcázar, Daniel and Stern, Jonathan and Necib, Lina and Choban, Caleb R. and Hayward, Christopher C.},
  title      = {{FIRE}-3: updated stellar evolution models, yields, and microphysics and fitting functions for applications in galaxy simulations},
  doi        = {10.1093/mnras/stac3489},
  issn       = {0035-8711},
  note       = {ADS Bibcode: 2023MNRAS.519.3154H},
  pages      = {3154--3181},
  url        = {https://ui.adsabs.harvard.edu/abs/2023MNRAS.519.3154H},
  urldate    = {2025-01-30},
  volume     = {519},
  journal    = {MNRAS},
  month      = feb,
  publisher  = {OUP},
  ranking    = {rank5},
  shorttitle = {{FIRE}-3},
  year       = {2023},
}

@Article{Weisz2014,
  author    = {Daniel R. Weisz and Evan D. Skillman and Sebastian L. Hidalgo and Matteo Monelli and Andrew E. Dolphin and Alan McConnachie and Edouard J. Bernard and Carme Gallart and Antonio Aparicio and Michael Boylan-Kolchin and Santi Cassisi and Andrew A. Cole and Henry C. Ferguson and Mike Irwin and Nicolas F. Martin and Lucio Mayer and Kristen B. W. McQuinn and Julio F. Navarro and Peter B. Stetson},
  title     = {Comparing M31 and Milky Way Satellites: The Extended Star Formation Histories of Andromeda II and Andromeda XVI},
  doi       = {10.1088/0004-637x/789/1/24},
  number    = {1},
  pages     = {24},
  volume    = {789},
  journal   = {ApJ},
  month     = {jun},
  publisher = {{IOP} Publishing},
  year      = {2014},
}

@Article{Parkinson2008,
  author    = {Hannah Parkinson and Shaun Cole and John Helly},
  title     = {Generating dark matter halo merger trees},
  doi       = {10.1111/j.1365-2966.2007.12517.x},
  number    = {2},
  pages     = {557--564},
  url       = {https://ui.adsabs.harvard.edu/abs/2008MNRAS.383..557P/abstract},
  volume    = {383},
  journal   = {MNRAS},
  month     = {dec},
  publisher = {Oxford University Press ({OUP})},
  year      = {2008},
}

@Article{Christensen2016,
  author        = {Christensen, C. R. and Dav{\'e}, R. and Governato, F. and Pontzen, A. and Brooks, A. and Munshi, F. and Quinn, T. and Wadsley, J.},
  title         = {In-N-Out: The Gas Cycle from Dwarfs to Spiral Galaxies},
  doi           = {10.3847/0004-637X/824/1/57},
  eid           = {57},
  eprint        = {1508.00007},
  pages         = {57},
  url           = {http://adsabs.harvard.edu/abs/2016ApJ...824...57C},
  volume        = {824},
  archiveprefix = {arXiv},
  journal       = {ApJ},
  month         = jun,
  ranking       = {rank4},
  year          = {2016},
}

@Article{Tremblay2009,
  author        = {Tremblay, P. E. and Bergeron, P.},
  title         = {Spectroscopic Analysis of DA White Dwarfs: Stark Broadening of Hydrogen Lines Including Nonideal Effects},
  doi           = {10.1088/0004-637X/696/2/1755},
  eprint        = {0902.4182},
  number        = {2},
  pages         = {1755-1770},
  url           = {https://ui.adsabs.harvard.edu/abs/2009ApJ...696.1755T},
  volume        = {696},
  archiveprefix = {arXiv},
  journal       = {\apj},
  month         = may,
  primaryclass  = {astro-ph.SR},
  year          = {2009},
}

@Article{Christensen2018,
  author   = {Christensen, Charlotte R. and Davé, Romeel and Brooks, Alyson and Quinn, Thomas and Shen, Sijing},
  title    = {Tracing {Outflowing} {Metals} in {Simulations} of {Dwarf} and {Spiral} {Galaxies}},
  doi      = {10.3847/1538-4357/aae374},
  issn     = {0004-637X},
  note     = {ADS Bibcode: 2018ApJ...867..142C},
  pages    = {142},
  url      = {https://ui.adsabs.harvard.edu/abs/2018ApJ...867..142C},
  urldate  = {2024-03-04},
  volume   = {867},
  journal  = {ApJ},
  month    = nov,
  ranking  = {rank4},
  year     = {2018},
}

@Article{Rey2025,
  author     = {Rey, Martin P. and Taylor, Ethan and Gray, Emily I. and Kim, Stacy Y. and Andersson, Eric P. and Pontzen, Andrew and Agertz, Oscar and Read, Justin I. and Cadiou, Corentin and Yates, Robert M. and Orkney, Matthew D. A. and Scholte, Dirk and Saintonge, Amélie and Breneman, Joseph and McQuinn, Kristen B. W. and Muni, Claudia and Das, Payel},
  journal    = {MNRAS},
  title      = {{EDGE}: {The} emergence of dwarf galaxy scaling relations from cosmological radiation-hydrodynamics simulations},
  year       = {2025},
  issn       = {1365-2966},
  month      = June,
  note       = {ADS Bibcode: 2025arXiv250303813R Type: article},
  number     = {2},
  pages      = {1195--1217},
  volume     = {541},
  doi        = {10.1093/mnras/staf1058},
  publisher  = {Oxford University Press (OUP)},
  shorttitle = {{EDGE}},
  url        = {https://ui.adsabs.harvard.edu/abs/2025MNRAS.541.1195R/abstract},
  urldate    = {2025-04-01},
}

@Article{Cappellari2023,
  author     = {Cappellari, Michele},
  title      = {Full spectrum fitting with photometry in {PPXF}: stellar population versus dynamical masses, non-parametric star formation history and metallicity for 3200 {LEGA}-{C} galaxies at redshift z ≈ 0.8},
  doi        = {10.1093/mnras/stad2597},
  issn       = {0035-8711},
  note       = {ADS Bibcode: 2023MNRAS.526.3273C},
  pages      = {3273--3300},
  url        = {https://ui.adsabs.harvard.edu/abs/2023MNRAS.526.3273C},
  urldate    = {2024-04-22},
  volume     = {526},
  journal    = {MNRAS},
  month      = dec,
  publisher  = {OUP},
  shorttitle = {Full spectrum fitting with photometry in {PPXF}},
  year       = {2023},
}

@Article{Emerick2018,
  author        = {Emerick, A. and Bryan, G. L. and Mac Low, M.-M.},
  title         = {Stellar Radiation Is Critical for Regulating Star Formation and Driving Outflows in Low-mass Dwarf Galaxies},
  doi           = {10.3847/2041-8213/aae315},
  eid           = {L22},
  eprint        = {1808.00468},
  pages         = {L22},
  url           = {http://adsabs.harvard.edu/abs/2018ApJ...865L..22E},
  volume        = {865},
  archiveprefix = {arXiv},
  journal       = {ApJL},
  month         = oct,
  year          = {2018},
}

@Article{Tang2014,
  author     = {Tang, Jing and Bressan, Alessandro and Rosenfield, Philip and Slemer, Alessandra and Marigo, Paola and Girardi, Léo and Bianchi, Luciana},
  title      = {New {PARSEC} evolutionary tracks of massive stars at low metallicity: testing canonical stellar evolution in nearby star-forming dwarf galaxies},
  doi        = {10.1093/mnras/stu2029},
  issn       = {0035-8711},
  note       = {ADS Bibcode: 2014MNRAS.445.4287T},
  pages      = {4287--4305},
  url        = {https://ui.adsabs.harvard.edu/abs/2014MNRAS.445.4287T},
  urldate    = {2024-03-24},
  volume     = {445},
  journal    = {MNRAS},
  month      = dec,
  shorttitle = {New {PARSEC} evolutionary tracks of massive stars at low metallicity},
  year       = {2014},
}

@Article{Weisz2024,
  author    = {Weisz, Daniel R. and Dolphin, Andrew E. and Savino, Alessandro and McQuinn, Kristen B. W. and Newman, Max J. B. and Williams, Benjamin F. and Kallivayalil, Nitya and Anderson, Jay and Boyer, Martha L. and Correnti, Matteo and Geha, Marla C. and Sandstrom, Karin M. and Cole, Andrew A. and Warfield, Jack T. and Skillman, Evan D. and Cohen, Roger E. and Beaton, Rachael and Bressan, Alessandro and Bolatto, Alberto and Boylan-Kolchin, Michael and Brooks, Alyson M. and Bullock, James S. and Conroy, Charlie and Cooper, Michael C. and Dalcanton, Julianne J. and Dotter, Aaron L. and Fritz, Tobias K. and Garling, Christopher T. and Gennaro, Mario and Gilbert, Karoline M. and Girardi, Leo and Johnson, Benjamin D. and Johnson, L. Clifton and Kalirai, Jason and Kirby, Evan N. and Lang, Dustin and Marigo, Paola and Richstein, Hannah and Schlafly, Edward F. and Tollerud, Erik J. and Wetzel, Andrew},
  title     = {The {JWST} {Resolved} {Stellar} {Populations} {Early} {Release} {Science} {Program}. {V}. {DOLPHOT} {Stellar} {Photometry} for {NIRCam} and {NIRISS}},
  doi       = {10.3847/1538-4365/ad2600},
  issn      = {0067-0049},
  note      = {ADS Bibcode: 2024ApJS..271...47W},
  pages     = {47},
  url       = {https://ui.adsabs.harvard.edu/abs/2024ApJS..271...47W},
  urldate   = {2024-04-27},
  volume    = {271},
  journal   = {\apjs},
  month     = apr,
  publisher = {IOP},
  year      = {2024},
}

@Article{Bernard2013,
  author        = {Bernard, Edouard J. and Monelli, Matteo and Gallart, Carme and Fiorentino, Giuliana and Cassisi, Santi and Aparicio, Antonio and Cole, Andrew A. and Drozdovsky, Igor and Hidalgo, Sebastian L. and Skillman, Evan D. and Stetson, Peter B. and Tolstoy, Eline},
  title         = {The ACS LCID Project - VIII. The short-period Cepheids of Leo A},
  doi           = {10.1093/mnras/stt655},
  eprint        = {1304.5435},
  number        = {4},
  pages         = {3047-3061},
  url           = {https://ui.adsabs.harvard.edu/abs/2013MNRAS.432.3047B},
  volume        = {432},
  archiveprefix = {arXiv},
  journal       = {MNRAS},
  month         = jul,
  primaryclass  = {astro-ph.GA},
  year          = {2013},
}

@Article{Albers2019,
  author    = {Saundra M Albers and Daniel R Weisz and Andrew A Cole and Andrew E Dolphin and Evan D Skillman and Benjamin F Williams and Michael Boylan-Kolchin and James S Bullock and Julianne J Dalcanton and Philip F Hopkins and Ryan Leaman and Alan W McConnachie and Mark Vogelsberger and Andrew Wetzel},
  title     = {Star formation at the edge of the Local Group: a rising star formation history in the isolated galaxy {WLM}},
  doi       = {10.1093/mnras/stz2903},
  number    = {4},
  pages     = {5538--5550},
  url       = {https://ui.adsabs.harvard.edu/abs/2019MNRAS.490.5538A/abstract},
  volume    = {490},
  comment   = {WFIRST, SFH},
  journal   = {MNRAS},
  month     = {oct},
  publisher = {Oxford University Press ({OUP})},
  ranking   = {rank2},
  year      = {2019},
}

@Article{Hafen2019,
  author    = {Zachary Hafen and Claude-Andr{\'{e}} Faucher-Gigu{\`{e}}re and Daniel Angl{\'{e}}s-Alc{\'{a}}zar and Jonathan Stern and Du{\v{s}}an Kere{\v{s}} and Cameron Hummels and Clarke Esmerian and Shea Garrison-Kimmel and Kareem El-Badry and Andrew Wetzel and T K Chan and Philip F Hopkins and Norman Murray},
  title     = {The origins of the circumgalactic medium in the {FIRE} simulations},
  doi       = {10.1093/mnras/stz1773},
  number    = {1},
  pages     = {1248--1272},
  volume    = {488},
  journal   = {MNRAS},
  month     = {jun},
  publisher = {Oxford University Press ({OUP})},
  year      = {2019},
}

@Article{McConnachie2012,
  author        = {McConnachie, Alan W.},
  title         = {{The Observed Properties of Dwarf Galaxies in and Around the Local Group}},
  doi           = {10.1088/0004-6256/144/1/4},
  eprint        = {arXiv:1204.1562v2},
  issn          = {0004-6256},
  pages         = {4},
  volume        = {144},
  archiveprefix = {arXiv},
  arxivid       = {arXiv:1204.1562v2},
  journal       = {AJ},
  ranking       = {rank4},
  year          = {2012},
}

@Article{Kroupa2001,
  author   = {Kroupa, P.},
  title    = {On the variation of the initial mass function},
  doi      = {10.1046/j.1365-8711.2001.04022.x},
  eprint   = {astro-ph/0009005},
  pages    = {231-246},
  url      = {http://adsabs.harvard.edu/abs/2001MNRAS.322..231K},
  volume   = {322},
  journal  = {MNRAS},
  month    = apr,
  year     = {2001},
}

@Article{Putman2021,
  author    = {Mary E. Putman and Yong Zheng and Adrian M. Price-Whelan and Jana Grcevich and Amalya C. Johnson and Erik Tollerud and Joshua E. G. Peek},
  title     = {The Gas Content and Stripping of Local Group Dwarf Galaxies},
  doi       = {10.3847/1538-4357/abe391},
  number    = {1},
  pages     = {53},
  volume    = {913},
  journal   = {ApJ},
  month     = {may},
  publisher = {American Astronomical Society},
  ranking   = {rank3},
  year      = {2021},
}

@Article{Vincenzo2016a,
  author        = {Vincenzo, F. and Matteucci, F. and Belfiore, F. and Maiolino, R.},
  title         = {Modern yields per stellar generation: the effect of the IMF},
  doi           = {10.1093/mnras/stv2598},
  eprint        = {1503.08300},
  number        = {4},
  pages         = {4183-4190},
  url           = {https://ui.adsabs.harvard.edu/abs/2016MNRAS.455.4183V},
  volume        = {455},
  archiveprefix = {arXiv},
  journal       = {MNRAS},
  month         = feb,
  primaryclass  = {astro-ph.GA},
  year          = {2016},
}

@Article{Kirby2011,
  author   = {Kirby, Evan N. and Lanfranchi, Gustavo a. and Simon, Joshua D. and Cohen, Judith G. and Guhathakurta, Puragra},
  title    = {{MULTI-ELEMENT ABUNDANCE MEASUREMENTS FROM MEDIUM-RESOLUTION SPECTRA. III. METALLICITY DISTRIBUTIONS OF MILKY WAY DWARF SATELLITE GALAXIES}},
  doi      = {10.1088/0004-637X/750/2/173},
  issn     = {0004-637X},
  pages    = {78},
  volume   = {727},
  journal  = {ApJ},
  year     = {2011},
}

@Article{Kawinwanichakij2017,
  author        = {Kawinwanichakij, L. and Papovich, C. and Quadri, R. F. and Glazebrook, K. and Kacprzak, G. G. and Allen, R. J. and Bell, E. F. and Croton, D. J. and Dekel, A. and Ferguson, H. C. and Forrest, B. and Grogin, N. A. and Guo, Y. and Kocevski, D. D. and Koekemoer, A. M. and Labb{\'e}, I. and Lucas, R. A. and Nanayakkara, T. and Spitler, L. R. and Straatman, C. M. S. and Tran, K.-V. H. and Tomczak, A. and van Dokkum, P.},
  title         = {Effect of Local Environment and Stellar Mass on Galaxy Quenching and Morphology at 0.5 < z < 2.0},
  doi           = {10.3847/1538-4357/aa8b75},
  eid           = {134},
  eprint        = {1706.03780},
  pages         = {134},
  url           = {https://ui.adsabs.harvard.edu/abs/2017ApJ...847..134K},
  volume        = {847},
  archiveprefix = {arXiv},
  journal       = {ApJ},
  month         = oct,
  ranking       = {rank3},
  year          = {2017},
}

@Article{Kirby2013,
  author   = {Kirby, Evan N and Cohen, Judith G and Guhathakurta, Puragra and Cheng, Lucy and Bullock, James S and Gallazzi, Anna},
  title    = {{The Universal Stellar Mass – Stellar Metallicity Relation for Dwarf Galaxies}},
  doi      = {10.1088/0004-637X/779/2/102},
  pages    = {102},
  url      = {https://ui.adsabs.harvard.edu/abs/2013ApJ...779..102K/abstract},
  volume   = {779},
  journal  = {ApJ},
  ranking  = {rank3},
  year     = {2013},
}

@Article{Ramesh2023,
  author        = {Ramesh, Rahul and Nelson, Dylan and Pillepich, Annalisa},
  date          = {2022-10-31},
  title         = {The Circumgalactic Medium of Milky Way-like Galaxies in the TNG50 Simulation -- I: Halo Gas Properties and the Role of SMBH Feedback},
  doi           = {10.1093/mnras/stac3524},
  eprint        = {2211.00020},
  number        = {4},
  pages         = {5754--5777},
  volume        = {518},
  archiveprefix = {arXiv},
  copyright     = {arXiv.org perpetual, non-exclusive license},
  journal       = {MNRAS},
  month         = {dec},
  primaryclass  = {astro-ph.GA},
  publisher     = {Oxford University Press ({OUP})},
  ranking       = {rank3},
  year          = {2023},
}

@InProceedings{Plez2011,
  author    = {Plez, Bertrand},
  booktitle = {Journal of Physics Conference Series},
  title     = {Cool star model atmospheres for Gaia : ATLAS, MARCS, and PHOENIX},
  doi       = {10.1088/1742-6596/328/1/012005},
  pages     = {012005},
  publisher = {IOP},
  series    = {Journal of Physics Conference Series},
  url       = {https://ui.adsabs.harvard.edu/abs/2011JPhCS.328a2005P},
  volume    = {328},
  eid       = {012005},
  month     = dec,
  year      = {2011},
}

@Article{Tung2025,
  author        = {Tung, Pei-Cheng and Chen, Ke-Jung},
  title         = {Coevolution of Dwarf Galaxies and Their Circumgalactic Medium Across Cosmic Time},
  doi           = {10.3847/1538-4357/ade1d4},
  eid           = {127},
  eprint        = {2412.16440},
  number        = {1},
  pages         = {127},
  url           = {https://ui.adsabs.harvard.edu/abs/2025ApJ...988..127T},
  volume        = {988},
  archiveprefix = {arXiv},
  journal       = {ApJ},
  month         = jul,
  primaryclass  = {astro-ph.GA},
  ranking       = {rank3},
  year          = {2025},
}

@InCollection{Peimbert1994,
  author    = {Peimbert, M. and Colin, P. and Sarmiento, A.},
  booktitle = {Violent Star Formation, from 30 Doradus to QSOs},
  title     = {Abundances of {H} {II} regions and the chemical evolution of galaxies.},
  note      = {ADS Bibcode: 1994vsf..book...79P},
  pages     = {79},
  publisher = {Cambridge University Press},
  url       = {https://ui.adsabs.harvard.edu/abs/1994vsf..book...79P},
  urldate   = {2024-07-23},
  month     = jan,
  year      = {1994},
}

@Article{Steinwandel2025,
  author        = {Steinwandel, Ulrich P. and Rennehan, Douglas and Orr, Matthew E. and Fielding, Drummond B. and Kim, Chang-Goo},
  title         = {Pumping Iron: How Turbulent Metal Diffusion Impacts Multiphase Galactic Outflows},
  doi           = {10.3847/1538-4357/adf283},
  eid           = {16},
  eprint        = {2407.14599},
  number        = {1},
  pages         = {16},
  url           = {https://ui.adsabs.harvard.edu/abs/2025ApJ...991...16S},
  volume        = {991},
  archiveprefix = {arXiv},
  journal       = {ApJ},
  month         = sep,
  primaryclass  = {astro-ph.GA},
  year          = {2025},
}

@Article{Hidalgo2018,
  author    = {Sebastian L. Hidalgo and Adriano Pietrinferni and Santi Cassisi and Maurizio Salaris and Alessio Mucciarelli and Alessandro Savino and Antonio Aparicio and Victor Silva Aguirre and Kuldeep Verma},
  title     = {The Updated {BaSTI} Stellar Evolution Models and Isochrones. I. Solar-scaled Calculations},
  doi       = {10.3847/1538-4357/aab158},
  number    = {2},
  pages     = {125},
  url       = {https://ui.adsabs.harvard.edu/abs/2018ApJ...856..125H/abstract},
  volume    = {856},
  journal   = {ApJ},
  month     = {mar},
  publisher = {American Astronomical Society},
  ranking   = {rank4},
  year      = {2018},
}

@Article{Press1974,
  author   = {Press, W. H. and Schechter, P.},
  title    = {Formation of Galaxies and Clusters of Galaxies by Self-Similar Gravitational Condensation},
  doi      = {10.1086/152650},
  pages    = {425-438},
  url      = {https://ui.adsabs.harvard.edu/abs/1974ApJ...187..425P},
  volume   = {187},
  journal  = {ApJ},
  month    = feb,
  year     = {1974},
}

@Article{Gunasekera2023,
  author    = {Gunasekera, Chamani M. and van Hoof, Peter A. M. and Chatzikos, Marios and Ferland, Gary J.},
  title     = {The 23.01 {Release} of {Cloudy}},
  doi       = {10.3847/2515-5172/ad0e75},
  issn      = {2515-5172},
  note      = {ADS Bibcode: 2023RNAAS...7..246G},
  pages     = {246},
  url       = {https://ui.adsabs.harvard.edu/abs/2023RNAAS...7..246G},
  urldate   = {2025-04-02},
  volume    = {7},
  journal   = {RNAAS},
  month     = nov,
  publisher = {IOP},
  year      = {2023},
}

@Article{Chabrier2003,
  author   = {Chabrier, G.},
  title    = {Galactic Stellar and Substellar Initial Mass Function},
  doi      = {10.1086/376392},
  eprint   = {astro-ph/0304382},
  pages    = {763-795},
  url      = {http://adsabs.harvard.edu/abs/2003PASP..115..763C},
  volume   = {115},
  journal  = {PASP},
  month    = jul,
  year     = {2003},
}

@Article{Andrews2013,
  author    = {Andrews, Brett H. and Martini, Paul},
  title     = {The {Mass}-{Metallicity} {Relation} with the {Direct} {Method} on {Stacked} {Spectra} of {SDSS} {Galaxies}},
  doi       = {10.1088/0004-637X/765/2/140},
  issn      = {0004-637X},
  note      = {ADS Bibcode: 2013ApJ...765..140A},
  pages     = {140},
  url       = {https://ui.adsabs.harvard.edu/abs/2013ApJ...765..140A},
  urldate   = {2024-10-30},
  volume    = {765},
  comment   = {mzr},
  journal   = {ApJ},
  month     = mar,
  publisher = {IOP},
  year      = {2013},
}

@Article{Feldmann2023,
  author     = {Feldmann, Robert and Quataert, Eliot and Faucher-Giguère, Claude-André and Hopkins, Philip F. and Çatmabacak, Onur and Kereš, Dušan and Bassini, Luigi and Bernardini, Mauro and Bullock, James S. and Cenci, Elia and Gensior, Jindra and Liang, Lichen and Moreno, Jorge and Wetzel, Andrew},
  title      = {{FIREbox}: simulating galaxies at high dynamic range in a cosmological volume},
  doi        = {10.1093/mnras/stad1205},
  issn       = {0035-8711},
  note       = {ADS Bibcode: 2023MNRAS.522.3831F},
  pages      = {3831--3860},
  url        = {https://ui.adsabs.harvard.edu/abs/2023MNRAS.522.3831F},
  urldate    = {2025-01-30},
  volume     = {522},
  comment    = {roughly the same baryon resolution as TNG50 (which is 8e6)},
  journal    = {MNRAS},
  month      = jul,
  publisher  = {OUP},
  ranking    = {rank4},
  shorttitle = {{FIREbox}},
  year       = {2023},
}

@Article{Kirby2019,
  author      = {Evan N. Kirby and Justin L. Xie and Rachel Guo and Mithi A. C. de los Reyes and Maria Bergemann and Mikhail Kovalev and Ken J. Shen and Anthony L. Piro and Andrew McWilliam},
  date        = {2019-06-24},
  title       = {Evidence for Sub-Chandrasekhar Type Ia Supernovae from Stellar Abundances in Dwarf Galaxies},
  doi         = {10.3847/1538-4357/ab2c02},
  eprint      = {1906.10126v2},
  eprintclass = {astro-ph.SR},
  eprinttype  = {arXiv},
  issn        = {1538-4357},
  number      = {1},
  pages       = {45},
  volume      = {881},
  journal     = {ApJ},
  month       = aug,
  publisher   = {American Astronomical Society},
  year        = {2019},
}

@Article{Zu2016,
  author    = {Ying Zu and Rachel Mandelbaum},
  title     = {Mapping stellar content to dark matter haloes {\textendash} {II}. Halo mass is the main driver of galaxy quenching},
  doi       = {10.1093/mnras/stw221},
  number    = {4},
  pages     = {4360--4383},
  volume    = {457},
  journal   = {MNRAS},
  month     = {jan},
  publisher = {Oxford University Press ({OUP})},
  ranking   = {rank3},
  year      = {2016},
}

@Article{Garcia2024a,
  author        = {Garcia, Alex M. and Torrey, Paul and Grasha, Kathryn and Hernquist, Lars and Ellison, Sara and Zovaro, Henry R.~M. and Hemler, Z.~S. and Nelson, Erica J. and Kewley, Lisa J.},
  title         = {Interplay of stellar and gas-phase metallicities: unveiling insights for stellar feedback modelling with Illustris, IllustrisTNG, and EAGLE},
  doi           = {10.1093/mnras/stae737},
  eprint        = {2401.12310},
  number        = {4},
  pages         = {3342-3359},
  url           = {https://ui.adsabs.harvard.edu/abs/2024MNRAS.529.3342G},
  volume        = {529},
  archiveprefix = {arXiv},
  journal       = {MNRAS},
  month         = apr,
  primaryclass  = {astro-ph.GA},
  year          = {2024},
}

@Article{Bjorklund2021,
  author        = {Bj{\"o}rklund, R. and Sundqvist, J.~O. and Puls, J. and Najarro, F.},
  title         = {New predictions for radiation-driven, steady-state mass-loss and wind-momentum from hot, massive stars. II. A grid of O-type stars in the Galaxy and the Magellanic Clouds},
  doi           = {10.1051/0004-6361/202038384},
  eid           = {A36},
  eprint        = {2008.06066},
  pages         = {A36},
  url           = {https://ui.adsabs.harvard.edu/abs/2021A&A...648A..36B},
  volume        = {648},
  archiveprefix = {arXiv},
  journal       = {\aap},
  month         = apr,
  primaryclass  = {astro-ph.SR},
  year          = {2021},
}

@Article{Jiang2014,
  author    = {Fangzhou Jiang and Frank C. van den Bosch},
  title     = {Generating merger trees for dark matter haloes: a comparison of methods},
  doi       = {10.1093/mnras/stu280},
  number    = {1},
  pages     = {193--207},
  volume    = {440},
  journal   = {MNRAS},
  month     = {mar},
  publisher = {Oxford University Press ({OUP})},
  year      = {2014},
}

@Article{BernsteinCooper2014,
  author        = {Bernstein-Cooper, Elijah Z. and Cannon, John M. and Elson, Edward C. and Warren, Steven R. and Chengular, Jayaram and Skillman, Evan D. and Adams, Elizabeth A.~K. and Bolatto, Alberto D. and Giovanelli, Riccardo and Haynes, Martha P. and McQuinn, Kristen B.~W. and Pardy, Stephen A. and Rhode, Katherine L. and Salzer, John J.},
  title         = {ALFALFA Discovery of the Nearby Gas-rich Dwarf Galaxy Leo P. V. Neutral Gas Dynamics and Kinematics},
  doi           = {10.1088/0004-6256/148/2/35},
  eid           = {35},
  eprint        = {1404.5298},
  number        = {2},
  pages         = {35},
  url           = {https://ui.adsabs.harvard.edu/abs/2014AJ....148...35B},
  volume        = {148},
  archiveprefix = {arXiv},
  journal       = {AJ},
  month         = aug,
  primaryclass  = {astro-ph.GA},
  year          = {2014},
}

@Article{Kepley2007,
  author    = {Kepley, Amanda A. and Wilcots, Eric M. and Hunter, Deidre A. and Nordgren, Tyler},
  title     = {A {High}-{Resolution} {Study} of the {H} {I} {Content} of {Local} {Group} {Dwarf} {Irregular} {Galaxy} {WLM}},
  doi       = {10.1086/513716},
  issn      = {0004-6256},
  note      = {ADS Bibcode: 2007AJ....133.2242K},
  pages     = {2242--2257},
  url       = {https://ui.adsabs.harvard.edu/abs/2007AJ....133.2242K},
  urldate   = {2025-03-17},
  volume    = {133},
  journal   = {AJ},
  month     = may,
  publisher = {IOP},
  year      = {2007},
}

@Article{Zhang2008,
  author    = {Jun Zhang and Onsi Fakhouri and Chung-Pei Ma},
  title     = {How to grow a healthy merger tree},
  doi       = {10.1111/j.1365-2966.2008.13671.x},
  number    = {4},
  pages     = {1521--1538},
  volume    = {389},
  journal   = {MNRAS},
  month     = {aug},
  publisher = {Oxford University Press ({OUP})},
  year      = {2008},
}

@Software{DynamicHMC.jl,
  author    = {Tamas K. Papp and David Widmann and Dilum Aluthge and Yimin Yi and delehef and Julia TagBot and Morten Piibeleht and Pietro Monticone},
  title     = {tpapp/DynamicHMC.jl: v3.4.7},
  doi       = {10.5281/zenodo.3384417},
  copyright = {Creative Commons Attribution 4.0 International},
  publisher = {Zenodo},
  year      = {2023},
}

@Article{Feldmann2015,
  author        = {Feldmann, Robert},
  title         = {The equilibrium view on dust and metals in galaxies: Galactic outflows drive low dust-to-metal ratios in dwarf galaxies},
  doi           = {10.1093/mnras/stv552},
  eprint        = {1412.2755},
  number        = {3},
  pages         = {3274-3292},
  url           = {https://ui.adsabs.harvard.edu/abs/2015MNRAS.449.3274F},
  volume        = {449},
  archiveprefix = {arXiv},
  journal       = {MNRAS},
  month         = may,
  primaryclass  = {astro-ph.GA},
  year          = {2015},
}

@Article{Garcia2025,
  author     = {Garcia, Alex M. and Torrey, Paul and Bhagwat, Aniket and Wright, Ruby J. and Chen, Qian-hui and Grasha, Kathryn and Ridolfo, Sophia and Hemler, Z. S. and Sarkar, Arnab and Chakraborty, Priyanka and Nelson, Erica J. and Sanders, Ryan L. and Costa, Tiago and Vogelsberger, Mark and Kewley, Lisa J. and Ellison, Sara L. and Hernquist, Lars},
  title      = {Metallicity {Gradients} in {Modern} {Cosmological} {Simulations} {I}: {Tension} {Between} {Smooth} {Stellar} {Feedback} {Models} and {Observations}},
  doi        = {10.3847/1538-4357/adea51},
  issn       = {1538-4357},
  note       = {ADS Bibcode: 2025arXiv250303804G Type: article},
  number     = {2},
  pages      = {147},
  url        = {https://ui.adsabs.harvard.edu/abs/2025arXiv250303804G},
  urldate    = {2025-03-14},
  volume     = {989},
  journal    = {ApJ},
  month      = aug,
  publisher  = {American Astronomical Society},
  shorttitle = {Metallicity {Gradients} in {Modern} {Cosmological} {Simulations} {I}},
  year       = {2025},
}

@Article{Bassini2024,
  author        = {Bassini, Luigi and Feldmann, Robert and Gensior, Jindra and Faucher-Giguère, Claude-André and Cenci, Elia and Moreno, Jorge and Bernardini, Mauro and Liang, Lichen},
  title         = {Inflow and outflow properties, not total gas fractions, drive the evolution of the mass-metallicity relation},
  doi           = {10.1093/mnrasl/slae036},
  eprint        = {2401.13824},
  issn          = {1745-3933},
  number        = {1},
  pages         = {L14--L20},
  url           = {https://ui.adsabs.harvard.edu/abs/2024MNRAS.532L..14B/abstract},
  volume        = {532},
  archiveprefix = {arXiv},
  comment       = {Compare to Ma2016},
  copyright     = {arXiv.org perpetual, non-exclusive license},
  journal       = {MNRAS},
  month         = may,
  primaryclass  = {astro-ph.GA},
  publisher     = {Oxford University Press (OUP)},
  year          = {2024},
}

@Article{Garcia2024,
  author     = {Garcia, Alex M. and Torrey, Paul and Ellison, Sara and Grasha, Kathryn and Hernquist, Lars and Zovaro, Henry R. M. and Chen, Qian-Hui and Hemler, Z. S. and Kewley, Lisa J. and Nelson, Erica J. and Wright, Ruby J.},
  title      = {Does the fundamental metallicity relation evolve with redshift? {I}: the correlation between offsets from the mass-metallicity relation and star formation rate},
  doi        = {10.1093/mnras/stae1252},
  issn       = {0035-8711},
  note       = {ADS Bibcode: 2024MNRAS.531.1398G},
  pages      = {1398--1408},
  url        = {https://ui.adsabs.harvard.edu/abs/2024MNRAS.531.1398G},
  urldate    = {2025-03-20},
  volume     = {531},
  journal    = {MNRAS},
  month      = jun,
  publisher  = {OUP},
  shorttitle = {Does the fundamental metallicity relation evolve with redshift?},
  year       = {2024},
}

@Article{Narayanan2024,
  author    = {Narayanan, Desika and Lower, Sidney and Torrey, Paul and Brammer, Gabriel and Cui, Weiguang and Davé, Romeel and Iyer, Kartheik G. and Li, Qi and Lovell, Christopher C. and Sales, Laura V. and Stark, Daniel P. and Marinacci, Federico and Vogelsberger, Mark},
  title     = {Outshining by {Recent} {Star} {Formation} {Prevents} the {Accurate} {Measurement} of {High}-z {Galaxy} {Stellar} {Masses}},
  doi       = {10.3847/1538-4357/ad0966},
  issn      = {0004-637X},
  note      = {ADS Bibcode: 2024ApJ...961...73N},
  pages     = {73},
  url       = {https://ui.adsabs.harvard.edu/abs/2024ApJ...961...73N},
  urldate   = {2025-01-31},
  volume    = {961},
  journal   = {ApJ},
  month     = jan,
  publisher = {IOP},
  year      = {2024},
}

@Article{Optim.jl,
  author  = {Mogensen, Patrick Kofod and Riseth, Asbj{\o}rn Nilsen},
  title   = {Optim: A mathematical optimization package for {Julia}},
  doi     = {10.21105/joss.00615},
  number  = {24},
  pages   = {615},
  volume  = {3},
  journal = {Journal of Open Source Software},
  year    = {2018},
}

@Article{Lilly2013,
  author        = {Lilly, Simon J. and Carollo, C. Marcella and Pipino, Antonio and Renzini, Alvio and Peng, Yingjie},
  title         = {Gas Regulation of Galaxies: The Evolution of the Cosmic Specific Star Formation Rate, the Metallicity-Mass-Star-formation Rate Relation, and the Stellar Content of Halos},
  doi           = {10.1088/0004-637X/772/2/119},
  eid           = {119},
  eprint        = {1303.5059},
  number        = {2},
  pages         = {119},
  url           = {https://ui.adsabs.harvard.edu/abs/2013ApJ...772..119L},
  volume        = {772},
  archiveprefix = {arXiv},
  journal       = {ApJ},
  month         = aug,
  primaryclass  = {astro-ph.CO},
  year          = {2013},
}

@Article{Marigo2017,
  author    = {Marigo, Paola and Girardi, Léo and Bressan, Alessandro and Rosenfield, Philip and Aringer, Bernhard and Chen, Yang and Dussin, Marco and Nanni, Ambra and Pastorelli, Giada and Rodrigues, Thaíse S. and Trabucchi, Michele and Bladh, Sara and Dalcanton, Julianne and Groenewegen, Martin A. T. and Montalbán, Josefina and Wood, Peter R.},
  title     = {A {New} {Generation} of {PARSEC}-{COLIBRI} {Stellar} {Isochrones} {Including} the {TP}-{AGB} {Phase}},
  doi       = {10.3847/1538-4357/835/1/77},
  issn      = {0004-637X},
  note      = {ADS Bibcode: 2017ApJ...835...77M},
  pages     = {77},
  url       = {https://ui.adsabs.harvard.edu/abs/2017ApJ...835...77M},
  urldate   = {2024-08-21},
  volume    = {835},
  journal   = {ApJ},
  month     = jan,
  publisher = {IOP},
  year      = {2017},
}

@Article{Garcia2025a,
  author     = {Garcia, Alex M. and Torrey, Paul and Ellison, Sara L. and Grasha, Kathryn and Chen, Qian-Hui and Hemler, Z. S. and Zimmerman, Dhruv T. and Wright, Ruby J. and Zovaro, Henry R. M. and Nelson, Erica J. and Sanders, Ryan L. and Kewley, Lisa J. and Hernquist, Lars},
  title      = {Does the fundamental metallicity relation evolve with redshift? - {II}. {The} evolution in normalization of the mass-metallicity relation},
  doi        = {10.1093/mnras/stae2587},
  issn       = {0035-8711},
  note       = {ADS Bibcode: 2025MNRAS.536..119G},
  pages      = {119--144},
  url        = {https://ui.adsabs.harvard.edu/abs/2025MNRAS.536..119G},
  urldate    = {2025-03-20},
  volume     = {536},
  journal    = {MNRAS},
  month      = jan,
  publisher  = {OUP},
  shorttitle = {Does the fundamental metallicity relation evolve with redshift?},
  year       = {2025},
}

@Article{Pillepich2019,
  author        = {Pillepich, Annalisa and Nelson, Dylan and Springel, Volker and Pakmor, R{\"u}diger and Torrey, Paul and Weinberger, Rainer and Vogelsberger, Mark and Marinacci, Federico and Genel, Shy and van der Wel, Arjen and Hernquist, Lars},
  title         = {First results from the TNG50 simulation: the evolution of stellar and gaseous discs across cosmic time},
  doi           = {10.1093/mnras/stz2338},
  eprint        = {1902.05553},
  number        = {3},
  pages         = {3196-3233},
  url           = {https://ui.adsabs.harvard.edu/abs/2019MNRAS.490.3196P},
  volume        = {490},
  archiveprefix = {arXiv},
  journal       = {MNRAS},
  month         = dec,
  primaryclass  = {astro-ph.GA},
  year          = {2019},
}

@Article{Davies2019,
  author    = {L J M Davies and A S G Robotham and C del P Lagos and S P Driver and A R H Stevens and Y M Bah{\'{e}} and M Alpaslan and M N Bremer and M J I Brown and S Brough and J Bland-Hawthorn and L Cortese and P Elahi and M W Grootes and B W Holwerda and A D Ludlow and S McGee and M Owers and S Phillipps},
  title     = {Galaxy And Mass Assembly ({GAMA}): Environmental Quenching of Centrals and Satellites in Groups},
  doi       = {10.1093/mnras/sty3393},
  pages     = {5444–5458},
  volume    = {483},
  journal   = {MNRAS},
  month     = {jan},
  publisher = {Oxford University Press ({OUP})},
  ranking   = {rank2},
  year      = {2019},
}

@InProceedings{Castelli2003,
  author        = {Castelli, F. and Kurucz, R.~L.},
  booktitle     = {Modelling of Stellar Atmospheres},
  title         = {New Grids of ATLAS9 Model Atmospheres},
  doi           = {10.48550/arXiv.astro-ph/0405087},
  editor        = {{Piskunov}, N. and {Weiss}, W.~W. and {Gray}, D.~F.},
  eprint        = {astro-ph/0405087},
  pages         = {A20},
  series        = {IAU Symposium},
  url           = {https://ui.adsabs.harvard.edu/abs/2003IAUS..210P.A20C},
  volume        = {210},
  archiveprefix = {arXiv},
  month         = jan,
  primaryclass  = {astro-ph},
  year          = {2003},
}

@Article{Walcher2011,
  author    = {Walcher, Jakob and Groves, Brent and Budavári, Tamás and Dale, Daniel},
  title     = {Fitting the integrated spectral energy distributions of galaxies},
  doi       = {10.1007/s10509-010-0458-z},
  issn      = {1572-946X},
  number    = {1},
  pages     = {1--51},
  volume    = {331},
  journal   = {Astrophysics and Space Science},
  month     = jan,
  publisher = {Springer Science and Business Media LLC},
  year      = {2011},
}

@Article{Zee2006,
  author        = {van Zee, Liese and Skillman, Evan D. and Haynes, Martha P.},
  title         = {Oxygen and Nitrogen in Leo A and GR 8},
  doi           = {10.1086/498298},
  eprint        = {astro-ph/0509678},
  number        = {1},
  pages         = {269-282},
  url           = {https://ui.adsabs.harvard.edu/abs/2006ApJ...637..269V},
  volume        = {637},
  archiveprefix = {arXiv},
  journal       = {ApJ},
  month         = jan,
  primaryclass  = {astro-ph},
  year          = {2006},
}

@Article{Dolphin1997,
  author   = {Dolphin, Andrew},
  title    = {A new method to determine star formation histories of nearby galaxies},
  doi      = {10.1016/S1384-1076(97)00029-8},
  issn     = {1384-1076},
  note     = {ADS Bibcode: 1997NewA....2..397D},
  pages    = {397--409},
  url      = {https://ui.adsabs.harvard.edu/abs/1997NewA....2..397D},
  urldate  = {2024-04-25},
  volume   = {2},
  journal  = {New Astronomy},
  month    = nov,
  year     = {1997},
}

@Article{Cole2014,
  author        = {Cole, Andrew A. and Weisz, Daniel R. and Dolphin, Andrew E. and Skillman, Evan D. and McConnachie, Alan W. and Brooks, Alyson M. and Leaman, Ryan},
  title         = {Delayed Star Formation in Isolated Dwarf galaxies: Hubble Space Telescope Star Formation History of the Aquarius Dwarf Irregular},
  doi           = {10.1088/0004-637X/795/1/54},
  eid           = {54},
  eprint        = {1409.1630},
  number        = {1},
  pages         = {54},
  url           = {https://ui.adsabs.harvard.edu/abs/2014ApJ...795...54C},
  volume        = {795},
  archiveprefix = {arXiv},
  journal       = {ApJ},
  month         = nov,
  primaryclass  = {astro-ph.GA},
  year          = {2014},
}

@Article{Pillepich2018,
  author        = {Pillepich, A. and Springel, V. and Nelson, D. and Genel, S. and Naiman, J. and Pakmor, R. and Hernquist, L. and Torrey, P. and Vogelsberger, M. and Weinberger, R. and Marinacci, F.},
  title         = {Simulating galaxy formation with the IllustrisTNG model},
  doi           = {10.1093/mnras/stx2656},
  eprint        = {1703.02970},
  pages         = {4077-4106},
  url           = {http://adsabs.harvard.edu/abs/2018MNRAS.473.4077P},
  volume        = {473},
  archiveprefix = {arXiv},
  journal       = {MNRAS},
  month         = jan,
  year          = {2018},
}

@Article{Marszewski2024,
  author    = {Marszewski, Andrew and Sun, Guochao and Faucher-Giguère, Claude-André and Hayward, Christopher C. and Feldmann, Robert},
  title     = {The {High}-{Redshift} {Gas}-{Phase} {Mass}–{Metallicity} {Relation} in {FIRE}-2},
  doi       = {10.3847/2041-8213/ad4cee},
  issn      = {2041-8205},
  language  = {en},
  number    = {2},
  pages     = {L41},
  url       = {https://dx.doi.org/10.3847/2041-8213/ad4cee},
  urldate   = {2025-03-10},
  volume    = {967},
  journal   = {ApJL},
  month     = may,
  publisher = {The American Astronomical Society},
  year      = {2024},
}

@Article{Marszewski2025,
  author        = {Marszewski, Andrew and Faucher-Gigu{\`e}re, Claude-Andr{\'e} and Feldmann, Robert and Sun, Guochao},
  title         = {Explaining the Weak Evolution of the High-redshift Mass{\textendash}Metallicity Relation with Galaxy Burst Cycles},
  doi           = {10.3847/2041-8213/adf74b},
  eid           = {L4},
  eprint        = {2505.22712},
  number        = {1},
  pages         = {L4},
  url           = {https://ui.adsabs.harvard.edu/abs/2025ApJ...991L...4M},
  volume        = {991},
  archiveprefix = {arXiv},
  journal       = {ApJL},
  month         = sep,
  primaryclass  = {astro-ph.GA},
  year          = {2025},
}

@Article{Torrey2019,
  author    = {Torrey, Paul and Vogelsberger, Mark and Marinacci, Federico and Pakmor, Rüdiger and Springel, Volker and Nelson, Dylan and Naiman, Jill and Pillepich, Annalisa and Genel, Shy and Weinberger, Rainer and Hernquist, Lars},
  title     = {The evolution of the mass-metallicity relation and its scatter in {IllustrisTNG}},
  doi       = {10.1093/mnras/stz243},
  issn      = {0035-8711},
  note      = {ADS Bibcode: 2019MNRAS.484.5587T},
  pages     = {5587--5607},
  url       = {https://ui.adsabs.harvard.edu/abs/2019MNRAS.484.5587T},
  urldate   = {2024-10-29},
  volume    = {484},
  journal   = {MNRAS},
  month     = apr,
  publisher = {OUP},
  ranking   = {rank4},
  year      = {2019},
}

@Article{Dave2012,
  author        = {Dav{\'e}, Romeel and Finlator, Kristian and Oppenheimer, Benjamin D.},
  title         = {An analytic model for the evolution of the stellar, gas and metal content of galaxies},
  doi           = {10.1111/j.1365-2966.2011.20148.x},
  eprint        = {1108.0426},
  number        = {1},
  pages         = {98-107},
  url           = {https://ui.adsabs.harvard.edu/abs/2012MNRAS.421...98D},
  volume        = {421},
  archiveprefix = {arXiv},
  journal       = {MNRAS},
  month         = mar,
  primaryclass  = {astro-ph.CO},
  year          = {2012},
}

@Article{Lee2006,
  author   = {Lee, H. and Skillman, E. D. and Cannon, J. M. and Jackson, D. C. and Gehrz, R. D. and Polomski, E. F. and Woodward, C. E.},
  title    = {On Extending the Mass-Metallicity Relation of Galaxies by 2.5 Decades in Stellar Mass},
  doi      = {10.1086/505573},
  eprint   = {astro-ph/0605036},
  pages    = {970-983},
  url      = {http://adsabs.harvard.edu/abs/2006ApJ...647..970L},
  volume   = {647},
  journal  = {ApJ},
  month    = aug,
  ranking  = {rank5},
  year     = {2006},
}

@Article{Leaman2009,
  author     = {Leaman, Ryan and Cole, Andrew A. and Venn, Kim A. and Tolstoy, Eline and Irwin, Mike J. and Szeifert, Thomas and Skillman, Evan D. and McConnachie, Alan W.},
  title      = {Stellar {Metallicities} and {Kinematics} in a {Gas}-rich {Dwarf} {Galaxy}: {First} {Calcium} {Triplet} {Spectroscopy} of {Red} {Giant} {Branch} {Stars} in {WLM}},
  doi        = {10.1088/0004-637X/699/1/1},
  issn       = {0004-637X},
  note       = {ADS Bibcode: 2009ApJ...699....1L},
  pages      = {1--14},
  url        = {https://ui.adsabs.harvard.edu/abs/2009ApJ...699....1L},
  urldate    = {2024-06-25},
  volume     = {699},
  journal    = {ApJ},
  month      = jul,
  publisher  = {IOP},
  shorttitle = {Stellar {Metallicities} and {Kinematics} in a {Gas}-rich {Dwarf} {Galaxy}},
  year       = {2009},
}

@Article{Allard2012,
  author        = {Allard, F. and Homeier, D. and Freytag, B.},
  title         = {Models of very-low-mass stars, brown dwarfs and exoplanets},
  doi           = {10.1098/rsta.2011.0269},
  eprint        = {1112.3591},
  number        = {1968},
  pages         = {2765-2777},
  url           = {https://ui.adsabs.harvard.edu/abs/2012RSPTA.370.2765A},
  volume        = {370},
  archiveprefix = {arXiv},
  journal       = {Philosophical Transactions of the Royal Society of London Series A},
  month         = jun,
  primaryclass  = {astro-ph.SR},
  year          = {2012},
}

@Article{Cole2000,
  author    = {Shaun Cole and Cedric G. Lacey and Carlton M. Baugh and Carlos S. Frenk},
  title     = {Hierarchical galaxy formation},
  doi       = {10.1046/j.1365-8711.2000.03879.x},
  number    = {1},
  pages     = {168--204},
  url       = {https://ui.adsabs.harvard.edu/abs/2000MNRAS.319..168C/abstract},
  volume    = {319},
  journal   = {MNRAS},
  month     = {apr},
  publisher = {Oxford University Press ({OUP})},
  year      = {2000},
}

@Article{Savino2025,
  author    = {Savino, Alessandro and Weisz, Daniel R. and Dolphin, Andrew E. and Durbin, Meredith J. and Kallivayalil, Nitya and Wetzel, Andrew and Anderson, Jay and Besla, Gurtina and Boylan-Kolchin, Michael and Brown, Thomas M. and Bullock, James S. and Cole, Andrew A. and Collins, Michelle L. M. and Cooper, M. C. and Deason, Alis J. and Dotter, Aaron L. and Fardal, Mark and Ferguson, Annette M. N. and Fritz, Tobias K. and Geha, Marla C. and Gilbert, Karoline M. and Guhathakurta, Puragra and Ibata, Rodrigo and Irwin, Michael J. and Jeon, Myoungwon and Kirby, Evan N. and Lewis, Geraint F. and Mackey, Dougal and Majewski, Steven R. and Martin, Nicolas and McConnachie, Alan and Patel, Ekta and Rich, R. Michael and Skillman, Evan D. and Simon, Joshua D. and Sohn, Sangmo Tony and Tollerud, Erik J. and van der Marel, Roeland P.},
  title     = {The {Hubble} {Space} {Telescope} {Survey} of {M31} {Satellite} {Galaxies}. {IV}. {Survey} {Overview} and {Lifetime} {Star} {Formation} {Histories}},
  doi       = {10.3847/1538-4357/ada24f},
  issn      = {0004-637X},
  note      = {ADS Bibcode: 2025ApJ...979..205S},
  pages     = {205},
  url       = {https://ui.adsabs.harvard.edu/abs/2025ApJ...979..205S},
  urldate   = {2025-04-10},
  volume    = {979},
  journal   = {ApJ},
  month     = feb,
  publisher = {IOP},
  year      = {2025},
}

@Article{Dave2019,
  author        = {Dav{\'e}, R. and Angl{\'e}s-Alc{\'a}zar, D. and Narayanan, D. and Li, Q. and Rafieferantsoa, M. H. and Appleby, S.},
  title         = {SIMBA: Cosmological simulations with black hole growth and feedback},
  doi           = {10.1093/mnras/stz937},
  eprint        = {1901.10203},
  pages         = {2827-2849},
  url           = {https://ui.adsabs.harvard.edu/abs/2019MNRAS.486.2827D},
  volume        = {486},
  archiveprefix = {arXiv},
  journal       = {MNRAS},
  month         = jun,
  year          = {2019},
}

@Article{Savino2023,
  author        = {Savino, Alessandro and Weisz, Daniel R. and Skillman, Evan D. and Dolphin, Andrew and Cole, Andrew A. and Kallivayalil, Nitya and Wetzel, Andrew and Anderson, Jay and Besla, Gurtina and Boylan-Kolchin, Michael and Brown, Thomas M. and Bullock, James S. and Collins, Michelle L.~M. and Cooper, M.~C. and Deason, Alis J. and Dotter, Aaron L. and Fardal, Mark and Ferguson, Annette M.~N. and Fritz, Tobias K. and Geha, Marla C. and Gilbert, Karoline M. and Guhathakurta, Puragra and Ibata, Rodrigo and Irwin, Michael J. and Jeon, Myoungwon and Kirby, Evan N. and Lewis, Geraint F. and Mackey, Dougal and Majewski, Steven R. and Martin, Nicolas and McConnachie, Alan and Patel, Ekta and Rich, R. Michael and Simon, Joshua D. and Sohn, Sangmo Tony and Tollerud, Erik J. and van der Marel, Roeland P.},
  title         = {The Hubble Space Telescope Survey of M31 Satellite Galaxies. II. The Star Formation Histories of Ultrafaint Dwarf Galaxies},
  doi           = {10.3847/1538-4357/acf46f},
  eid           = {86},
  eprint        = {2305.13360},
  number        = {2},
  pages         = {86},
  url           = {https://ui.adsabs.harvard.edu/abs/2023ApJ...956...86S},
  volume        = {956},
  archiveprefix = {arXiv},
  journal       = {ApJ},
  month         = oct,
  primaryclass  = {astro-ph.GA},
  ranking       = {rank3},
  year          = {2023},
}

@Article{Cole2007,
  author        = {Cole, Andrew A. and Skillman, Evan D. and Tolstoy, Eline and Gallagher, III, John S. and Aparicio, Antonio and Dolphin, Andrew E. and Gallart, Carme and Hidalgo, Sebastian L. and Saha, Abhijit and Stetson, Peter B. and Weisz, Daniel R.},
  title         = {Leo A: A Late-blooming Survivor of the Epoch of Reionization in the Local Group},
  doi           = {10.1086/516711},
  eprint        = {astro-ph/0702646},
  number        = {1},
  pages         = {L17-L20},
  url           = {https://ui.adsabs.harvard.edu/abs/2007ApJ...659L..17C},
  volume        = {659},
  archiveprefix = {arXiv},
  journal       = {\apjl},
  month         = apr,
  primaryclass  = {astro-ph},
  year          = {2007},
}

@Article{Cole2008,
  author        = {Cole, N. and Newberg, H. J. and Magdon-Ismail, M. and Desell, T. and Dawsey, K. and Hayashi, W. and Liu, X. F. and Purnell, J. and Szymanski, B. and Varela, C. and Willett, B. and Wisniewski, J.},
  title         = {Maximum Likelihood Fitting of Tidal Streams with Application to the Sagittarius Dwarf Tidal Tails},
  doi           = {10.1086/589681},
  eprint        = {0805.2121},
  pages         = {750-766},
  url           = {https://ui.adsabs.harvard.edu/abs/2008ApJ...683..750C},
  volume        = {683},
  archiveprefix = {arXiv},
  journal       = {ApJ},
  month         = aug,
  year          = {2008},
}

@Article{Wetzel2023,
  author   = {Wetzel, Andrew and Hayward, Christopher C. and Sanderson, Robyn E. and Ma, Xiangcheng and Anglés-Alcázar, Daniel and Feldmann, Robert and Chan, T. K. and El-Badry, Kareem and Wheeler, Coral and Garrison-Kimmel, Shea and Nikakhtar, Farnik and Panithanpaisal, Nondh and Arora, Arpit and Gurvich, Alexander B. and Samuel, Jenna and Sameie, Omid and Pandya, Viraj and Hafen, Zachary and Hummels, Cameron and Loebman, Sarah and Boylan-Kolchin, Michael and Bullock, James S. and Faucher-Giguère, Claude-André and Kereš, Dušan and Quataert, Eliot and Hopkins, Philip F.},
  title    = {Public {Data} {Release} of the {FIRE}-2 {Cosmological} {Zoom}-in {Simulations} of {Galaxy} {Formation}},
  doi      = {10.3847/1538-4365/acb99a},
  issn     = {0067-0049},
  note     = {ADS Bibcode: 2023ApJS..265...44W},
  pages    = {44},
  url      = {https://ui.adsabs.harvard.edu/abs/2023ApJS..265...44W},
  urldate  = {2024-03-08},
  volume   = {265},
  journal  = {\apjs},
  month    = apr,
  ranking  = {rank5},
  year     = {2023},
}

@Article{Durbin2025,
  author        = {Durbin, Meredith J. and Choi, Yumi and Savino, Alessandro and Weisz, Daniel and Dolphin, Andrew E. and Dalcanton, Julianne J. and Jeon, Myoungwon and Kallivayalil, Nitya and Li, Ting S. and Pace, Andrew B. and Patel, Ekta and Sacchi, Elena and Skillman, Evan D. and Sohn, Sangmo Tony and van der Marel, Roeland P. and Wetzel, Andrew and Williams, Benjamin F.},
  title         = {The HST Legacy Archival Uniform Reduction of Local Group Imaging (LAURELIN). I. Photometry and Star Formation Histories for 36 Ultra-faint Dwarf Galaxies},
  doi           = {10.3847/1538-4357/ae00c8},
  eid           = {arXiv:2505.18252},
  eprint        = {2505.18252},
  issn          = {1538-4357},
  number        = {1},
  pages         = {arXiv:2505.18252},
  url           = {https://ui.adsabs.harvard.edu/abs/2025arXiv250518252D},
  volume        = {992},
  archiveprefix = {arXiv},
  journal       = {ApJ},
  month         = oct,
  primaryclass  = {astro-ph.GA},
  publisher     = {American Astronomical Society},
  ranking       = {rank5},
  year          = {2025},
}

@Article{Harvey2025,
  author        = {Harvey, Thomas and Conselice, Christopher J. and Adams, Nathan J. and Austin, Duncan and Li, Qiong and Rusakov, Vadim and Westcott, Lewi and Goolsby, Caio M. and Lovell, Christopher C. and Cochrane, Rachel K. and Vijayan, Aswin P. and Trussler, James},
  title         = {Behind the spotlight: a systematic assessment of outshining using NIRCam medium bands in the JADES Origins Field},
  doi           = {10.1093/mnras/staf1396},
  eprint        = {2504.05244},
  number        = {4},
  pages         = {2998-3027},
  url           = {https://ui.adsabs.harvard.edu/abs/2025MNRAS.542.2998H},
  volume        = {542},
  archiveprefix = {arXiv},
  journal       = {MNRAS},
  month         = oct,
  primaryclass  = {astro-ph.GA},
  year          = {2025},
}

@Article{Benson2017,
  author        = {Benson, Andrew J.},
  title         = {The mass function of unprocessed dark matter haloes and merger tree branching rates},
  doi           = {10.1093/mnras/stx343},
  eprint        = {1610.01057},
  number        = {3},
  pages         = {3454-3466},
  url           = {https://ui.adsabs.harvard.edu/abs/2017MNRAS.467.3454B},
  volume        = {467},
  archiveprefix = {arXiv},
  journal       = {MNRAS},
  month         = may,
  primaryclass  = {astro-ph.GA},
  year          = {2017},
}

@Article{Benson2012,
  author    = {Andrew J. Benson},
  title     = {Galacticus: A semi-analytic model of galaxy formation},
  doi       = {10.1016/j.newast.2011.07.004},
  number    = {2},
  pages     = {175--197},
  volume    = {17},
  comment   = {New Astronomy},
  journal   = {New Astronomy},
  month     = {feb},
  publisher = {Elsevier {BV}},
  ranking   = {rank4},
  year      = {2012},
}

@Article{Blanc2019,
  author   = {Blanc, Guillermo A. and Lu, Yu and Benson, Andrew and Katsianis, Antonios and Barraza, Marcelo},
  title    = {A {Characteristic} {Mass} {Scale} in the {Mass}-{Metallicity} {Relation} of {Galaxies}},
  doi      = {10.3847/1538-4357/ab16ec},
  issn     = {0004-637X},
  note     = {ADS Bibcode: 2019ApJ...877....6B},
  pages    = {6},
  url      = {https://ui.adsabs.harvard.edu/abs/2019ApJ...877....6B},
  urldate  = {2024-03-04},
  volume   = {877},
  journal  = {ApJ},
  month    = may,
  year     = {2019},
}

@Article{KadoFong2024,
  author       = {Kado-Fong, Erin and Geha, Marla and Mao, Yao-Yuan and Reyes, Mithi A. C. de los and Wechsler, Risa H. and Asali, Yasmeen and Kallivayalil, Nitya and Nadler, Ethan O. and Tollerud, Erik J. and Weiner, Benjamin},
  title        = {SAGAbg. I. A Near-unity Mass-loading Factor in Low-mass Galaxies via Their Low-redshift Evolution in Stellar Mass, Oxygen Abundance, and Star Formation Rate},
  doi          = {10.3847/1538-4357/ad3042},
  eprint       = {2401.16469},
  issn         = {1538-4357},
  number       = {1},
  pages        = {129},
  volume       = {966},
  copyright    = {Creative Commons Attribution 4.0 International},
  journal      = {ApJ},
  month        = may,
  primaryclass = {astro-ph.GA},
  publisher    = {American Astronomical Society},
  year         = {2024},
}

@Article{RocaFabrega2024,
  author     = {Roca-Fàbrega, Santi and Kim, Ji-hoon and Primack, Joel R. and Jung, Minyong and Genina, Anna and Hausammann, Loic and Kim, Hyeonyong and Lupi, Alessandro and Nagamine, Kentaro and Powell, Johnny W. and Revaz, Yves and Shimizu, Ikkoh and Strawn, Clayton and Velázquez, Héctor and Abel, Tom and Ceverino, Daniel and Dong, Bili and Quinn, Thomas R. and Shin, Eun-jin and Segovia-Otero, Alvaro and Agertz, Oscar and Barrow, Kirk S. S. and Cadiou, Corentin and Dekel, Avishai and Hummels, Cameron and Oh, Boon Kiat and Teyssier, Romain},
  title      = {The {AGORA} {High}-resolution {Galaxy} {Simulations} {Comparison} {Project} {IV}: {Halo} and {Galaxy} {Mass} {Assembly} in a {Cosmological} {Zoom}-in {Simulation} at \$z{\textbackslash}le2\$},
  doi        = {10.3847/1538-4357/ad43de},
  issn       = {1538-4357},
  note       = {ADS Bibcode: 2024arXiv240206202R Type: article},
  number     = {2},
  pages      = {125},
  url        = {https://ui.adsabs.harvard.edu/abs/2024arXiv240206202R},
  urldate    = {2024-03-21},
  volume     = {968},
  journal    = {ApJ},
  month      = jun,
  publisher  = {American Astronomical Society},
  shorttitle = {The {AGORA} {High}-resolution {Galaxy} {Simulations} {Comparison} {Project} {IV}},
  year       = {2024},
}

@Article{Cohen2025,
  author    = {Cohen, Roger E. and McQuinn, Kristen B. W. and Savino, Alessandro and Newman, Max J. B. and Weisz, Daniel R. and Dolphin, Andrew E. and Boyer, Martha L. and Correnti, Matteo and Geha, Marla C. and Gennaro, Mario and Gilbert, Karoline M. and Kallivayalil, Nitya and Warfield, Jack T. and Williams, Benjamin F. and Brooks, Alyson M. and Cole, Andrew A. and Skillman, Evan D. and Garling, Christopher T. and Kalirai, Jason S. and Anderson, Jay},
  title     = {The {JWST} {Resolved} {Stellar} {Populations} {Early} {Release} {Science} {Program}. {VIII}. {The} {Spatially} {Resolved} {Star} {Formation} {History} of {WLM}},
  doi       = {10.3847/1538-4357/adb61d},
  issn      = {0004-637X},
  note      = {ADS Bibcode: 2025ApJ...981..153C},
  pages     = {153},
  url       = {https://ui.adsabs.harvard.edu/abs/2025ApJ...981..153C},
  urldate   = {2025-03-17},
  volume    = {981},
  journal   = {ApJ},
  month     = mar,
  publisher = {IOP},
  year      = {2025},
}

@Article{Escala2018,
  author     = {Escala, Ivanna and Wetzel, Andrew and Kirby, Evan N. and Hopkins, Philip F. and Ma, Xiangcheng and Wheeler, Coral and Kereš, Dušan and Faucher-Giguère, Claude-André and Quataert, Eliot},
  title      = {Modelling chemical abundance distributions for dwarf galaxies in the {Local} {Group}: the impact of turbulent metal diffusion},
  doi        = {10.1093/mnras/stx2858},
  issn       = {0035-8711},
  note       = {ADS Bibcode: 2018MNRAS.474.2194E},
  pages      = {2194--2211},
  url        = {https://ui.adsabs.harvard.edu/abs/2018MNRAS.474.2194E},
  urldate    = {2025-03-11},
  volume     = {474},
  journal    = {MNRAS},
  month      = feb,
  publisher  = {OUP},
  shorttitle = {Modelling chemical abundance distributions for dwarf galaxies in the {Local} {Group}},
  year       = {2018},
}

@Article{Rosenfield2016,
  author    = {Rosenfield, Philip and Marigo, Paola and Girardi, Léo and Dalcanton, Julianne J. and Bressan, Alessandro and Williams, Benjamin F. and Dolphin, Andrew},
  title     = {Evolution of {Thermally} {Pulsing} {Asymptotic} {Giant} {Branch} {Stars}. {V}. {Constraining} the {Mass} {Loss} and {Lifetimes} of {Intermediate}-mass, {Low}-metallicity {AGB} {Stars}},
  doi       = {10.3847/0004-637X/822/2/73},
  issn      = {0004-637X},
  note      = {ADS Bibcode: 2016ApJ...822...73R},
  pages     = {73},
  url       = {https://ui.adsabs.harvard.edu/abs/2016ApJ...822...73R},
  urldate   = {2024-08-24},
  volume    = {822},
  comment   = {https://github.com/philrosenfield/padova_tracks},
  journal   = {ApJ},
  month     = may,
  publisher = {IOP},
  year      = {2016},
}

@Article{Zentner2007,
  author       = {Andrew R. Zentner},
  date         = {2006-11-14},
  journaltitle = {Int.J.Mod.Phys.D16:763-816,2007},
  title        = {The Excursion Set Theory of Halo Mass Functions, Halo Clustering, and Halo Growth},
  doi          = {10.1142/S0218271807010511},
  eprint       = {astro-ph/0611454v1},
  eprintclass  = {astro-ph},
  eprinttype   = {arXiv},
  pages        = {763-815},
  volume       = {16},
  journal      = {International Journal of Modern Physics D},
  year         = {2007},
}

@Article{Schlafly2011,
  author        = {Schlafly, Edward F. and Finkbeiner, Douglas P.},
  title         = {{Measuring Reddening with SDSS Stellar Spectra and Recalibrating SFD}},
  doi           = {10.1088/0004-637X/737/2/103},
  eprint        = {1012.4804},
  pages         = {103--116},
  volume        = {737},
  archiveprefix = {arXiv},
  arxivid       = {1012.4804},
  journal       = {ApJ},
  year          = {2011},
}

@Article{Salaris2022,
  author    = {Salaris, Maurizio and Cassisi, Santi and Pietrinferni, Adriano and Hidalgo, Sebastian},
  title     = {The updated {BASTI} stellar evolution models and isochrones - {III}. {White} dwarfs},
  doi       = {10.1093/mnras/stab3359},
  issn      = {0035-8711},
  note      = {ADS Bibcode: 2022MNRAS.509.5197S},
  pages     = {5197--5208},
  url       = {https://ui.adsabs.harvard.edu/abs/2022MNRAS.509.5197S},
  urldate   = {2024-04-28},
  volume    = {509},
  journal   = {MNRAS},
  month     = feb,
  publisher = {OUP},
  year      = {2022},
}

@Article{Nelson2020,
  author    = {Nelson, Dylan and Sharma, Prateek and Pillepich, Annalisa and Springel, Volker and Pakmor, Rüdiger and Weinberger, Rainer and Vogelsberger, Mark and Marinacci, Federico and Hernquist, Lars},
  title     = {Resolving small-scale cold circumgalactic gas in {TNG50}},
  doi       = {10.1093/mnras/staa2419},
  issn      = {0035-8711},
  note      = {ADS Bibcode: 2020MNRAS.498.2391N},
  pages     = {2391--2414},
  url       = {https://ui.adsabs.harvard.edu/abs/2020MNRAS.498.2391N},
  urldate   = {2025-04-08},
  volume    = {498},
  journal   = {MNRAS},
  month     = oct,
  publisher = {OUP},
  year      = {2020},
}

@Article{Benson2005,
  author   = {Benson, A. J. and Kamionkowski, M. and Hassani, S. H.},
  title    = {Self-consistent theory of halo mergers},
  doi      = {10.1111/j.1365-2966.2005.08679.x},
  eprint   = {astro-ph/0407136},
  pages    = {847-858},
  url      = {https://ui.adsabs.harvard.edu/abs/2005MNRAS.357..847B},
  volume   = {357},
  journal  = {MNRAS},
  month    = mar,
  year     = {2005},
}

@Article{HermosaMunoz2020,
  author        = {Hermosa Mu{\~n}oz, L. and Taibi, S. and Battaglia, G. and Iorio, G. and Rejkuba, M. and Leaman, R. and Cole, A.~A. and Irwin, M. and Jablonka, P. and Kacharov, N. and McConnachie, A. and Starkenburg, E. and Tolstoy, E.},
  title         = {Kinematic and metallicity properties of the Aquarius dwarf galaxy from FORS2 MXU spectroscopy},
  doi           = {10.1051/0004-6361/201936136},
  eid           = {A10},
  eprint        = {1911.13081},
  pages         = {A10},
  url           = {https://ui.adsabs.harvard.edu/abs/2020A&A...634A..10H},
  volume        = {634},
  archiveprefix = {arXiv},
  journal       = {\aap},
  month         = feb,
  primaryclass  = {astro-ph.GA},
  year          = {2020},
}

@Article{Urbaneja2023,
  author        = {Urbaneja, Miguel A. and Bresolin, Fabio and Kudritzki, Rolf-Peter},
  title         = {The Metallicity and Distance of Leo A from Blue Supergiants},
  doi           = {10.3847/1538-4357/acfc3d},
  eid           = {52},
  eprint        = {2309.11952},
  number        = {1},
  pages         = {52},
  url           = {https://ui.adsabs.harvard.edu/abs/2023ApJ...959...52U},
  volume        = {959},
  archiveprefix = {arXiv},
  journal       = {ApJ},
  month         = dec,
  primaryclass  = {astro-ph.GA},
  year          = {2023},
}

@Article{Pandya2020,
  author    = {Viraj Pandya and Rachel S. Somerville and Daniel Angl{\'{e}}s-Alc{\'{a}}zar and Christopher C. Hayward and Greg L. Bryan and Drummond B. Fielding and John C. Forbes and Blakesley Burkhart and Shy Genel and Lars Hernquist and Chang-Goo Kim and Stephanie Tonnesen and Tjitske Starkenburg},
  title     = {First Results from {SMAUG}: The Need for Preventative Stellar Feedback and Improved Baryon Cycling in Semianalytic Models of Galaxy Formation},
  doi       = {10.3847/1538-4357/abc3c1},
  number    = {1},
  pages     = {4},
  volume    = {905},
  journal   = {ApJ},
  month     = {dec},
  publisher = {American Astronomical Society},
  ranking   = {rank4},
  year      = {2020},
}

@Article{Gallart2005,
  author   = {Gallart, C. and Zoccali, M. and Aparicio, A.},
  title    = {The {Adequacy} of {Stellar} {Evolution} {Models} for the {Interpretation} of the {Color}-{Magnitude} {Diagrams} of {Resolved} {Stellar} {Populations}},
  doi      = {10.1146/annurev.astro.43.072103.150608},
  issn     = {0066-4146},
  note     = {ADS Bibcode: 2005ARA\&A..43..387G},
  pages    = {387--434},
  url      = {https://ui.adsabs.harvard.edu/abs/2005ARA&A..43..387G},
  urldate  = {2024-07-27},
  volume   = {43},
  comment  = {Annual Review of Astronomy and Astrophysics},
  journal  = {\araa},
  month    = sep,
  ranking  = {rank4},
  year     = {2005},
}

@Article{Gordon2016,
  author        = {Gordon, Karl D. and Fouesneau, Morgan and Arab, Heddy and Tchernyshyov, Kirill and Weisz, Daniel R. and Dalcanton, Julianne J. and Williams, Benjamin F. and Bell, Eric F. and Bianchi, Luciana and Boyer, Martha and Choi, Yumi and Dolphin, Andrew and Girardi, L{\'e}o and Hogg, David W. and Kalirai, Jason S. and Kapala, Maria and Lewis, Alexia R. and Rix, Hans-Walter and Sandstrom, Karin and Skillman, Evan D.},
  title         = {The Panchromatic Hubble Andromeda Treasury. XV. The BEAST: Bayesian Extinction and Stellar Tool},
  doi           = {10.3847/0004-637X/826/2/104},
  eid           = {104},
  eprint        = {1606.06182},
  number        = {2},
  pages         = {104},
  url           = {https://ui.adsabs.harvard.edu/abs/2016ApJ...826..104G},
  volume        = {826},
  archiveprefix = {arXiv},
  journal       = {\apj},
  month         = aug,
  primaryclass  = {astro-ph.GA},
  ranking       = {rank4},
  year          = {2016},
}

@Article{Leitherer1999,
  author   = {Leitherer, C. and Schaerer, D. and Goldader, J. D. and Delgado, R. M. G. and Robert, C. and Kune, D. F. and de Mello, D. F. and Devost, D. and Heckman, T. M.},
  title    = {Starburst99: Synthesis Models for Galaxies with Active Star Formation},
  doi      = {10.1086/313233},
  eprint   = {astro-ph/9902334},
  pages    = {3-40},
  url      = {http://adsabs.harvard.edu/abs/1999ApJS..123....3L},
  volume   = {123},
  journal  = {ApJS},
  month    = jul,
  year     = {1999},
}

@Article{Kobayashi2006,
  author        = {Kobayashi, Chiaki and Umeda, Hideyuki and Nomoto, Ken'ichi and Tominaga, Nozomu and Ohkubo, Takuya},
  title         = {Galactic Chemical Evolution: Carbon through Zinc},
  doi           = {10.1086/508914},
  eprint        = {astro-ph/0608688},
  number        = {2},
  pages         = {1145-1171},
  url           = {https://ui.adsabs.harvard.edu/abs/2006ApJ...653.1145K},
  volume        = {653},
  archiveprefix = {arXiv},
  journal       = {ApJ},
  month         = dec,
  primaryclass  = {astro-ph},
  year          = {2006},
}

@Article{Pandya2021,
  author        = {Pandya, Viraj and Fielding, Drummond B. and Angl{\'e}s-Alc{\'a}zar, Daniel and Somerville, Rachel S. and Bryan, Greg L. and Hayward, Christopher C. and Stern, Jonathan and Kim, Chang-Goo and Quataert, Eliot and Forbes, John C. and Faucher-Gigu{\`e}re, Claude-Andr{\'e} and Feldmann, Robert and Hafen, Zachary and Hopkins, Philip F. and Kere{\v{s}}, Du{\v{s}}an and Murray, Norman and Wetzel, Andrew},
  title         = {Characterizing mass, momentum, energy, and metal outflow rates of multiphase galactic winds in the FIRE-2 cosmological simulations},
  doi           = {10.1093/mnras/stab2714},
  eprint        = {2103.06891},
  number        = {2},
  pages         = {2979-3008},
  url           = {https://ui.adsabs.harvard.edu/abs/2021MNRAS.508.2979P},
  volume        = {508},
  archiveprefix = {arXiv},
  journal       = {MNRAS},
  month         = dec,
  primaryclass  = {astro-ph.GA},
  ranking       = {rank3},
  year          = {2021},
}

@Article{Nelson2019,
  author        = {Nelson, Dylan and Pillepich, Annalisa and Springel, Volker and Pakmor, R{\"u}diger and Weinberger, Rainer and Genel, Shy and Torrey, Paul and Vogelsberger, Mark and Marinacci, Federico and Hernquist, Lars},
  title         = {First results from the TNG50 simulation: galactic outflows driven by supernovae and black hole feedback},
  doi           = {10.1093/mnras/stz2306},
  eprint        = {1902.05554},
  number        = {3},
  pages         = {3234-3261},
  url           = {https://ui.adsabs.harvard.edu/abs/2019MNRAS.490.3234N},
  volume        = {490},
  archiveprefix = {arXiv},
  journal       = {MNRAS},
  month         = dec,
  primaryclass  = {astro-ph.GA},
  ranking       = {rank4},
  year          = {2019},
}

@Article{Curti2020,
  author    = {Curti, Mirko and Mannucci, Filippo and Cresci, Giovanni and Maiolino, Roberto},
  title     = {The mass-metallicity and the fundamental metallicity relation revisited on a fully {Te}-based abundance scale for galaxies},
  doi       = {10.1093/mnras/stz2910},
  issn      = {0035-8711},
  note      = {ADS Bibcode: 2020MNRAS.491..944C},
  pages     = {944--964},
  url       = {https://ui.adsabs.harvard.edu/abs/2020MNRAS.491..944C},
  urldate   = {2024-11-06},
  volume    = {491},
  journal   = {MNRAS},
  month     = jan,
  publisher = {OUP},
  year      = {2020},
}

@Article{Curti2024,
  author        = {Curti, Mirko and Maiolino, Roberto and Curtis-Lake, Emma and Chevallard, Jacopo and Carniani, Stefano and D'Eugenio, Francesco and Looser, Tobias J. and Scholtz, Jan and Charlot, Stephane and Cameron, Alex and {\"U}bler, Hannah and Witstok, Joris and Boyett, Kristian and Laseter, Isaac and Sandles, Lester and Arribas, Santiago and Bunker, Andrew and Giardino, Giovanna and Maseda, Michael V. and Rawle, Tim and Rodr{\'\i}guez Del Pino, Bruno and Smit, Renske and Willott, Chris J. and Eisenstein, Daniel J. and Hausen, Ryan and Johnson, Benjamin and Rieke, Marcia and Robertson, Brant and Tacchella, Sandro and Williams, Christina C. and Willmer, Christopher and Baker, William M. and Bhatawdekar, Rachana and Egami, Eiichi and Helton, Jakob M. and Ji, Zhiyuan and Kumari, Nimisha and Perna, Michele and Shivaei, Irene and Sun, Fengwu},
  title         = {JADES: Insights into the low-mass end of the mass-metallicity-SFR relation at 3 < z < 10 from deep JWST/NIRSpec spectroscopy},
  doi           = {10.1051/0004-6361/202346698},
  eid           = {A75},
  eprint        = {2304.08516},
  pages         = {A75},
  url           = {https://ui.adsabs.harvard.edu/abs/2024A&A...684A..75C},
  volume        = {684},
  archiveprefix = {arXiv},
  journal       = {\aap},
  month         = apr,
  primaryclass  = {astro-ph.GA},
  year          = {2024},
}

@Article{VanNest2023,
  author    = {Van Nest, Jordan and Munshi, Ferah and Christensen, Charlotte and Brooks, Alyson M. and Tremmel, Michael and Quinn, Thomas R.},
  title     = {The {Role} of {Mass} and {Environment} on {Satellite} {Distributions} around {Milky} {Way} {Analogs} in the {ROMULUS25} {Simulation}},
  doi       = {10.3847/1538-4357/acf861},
  issn      = {0004-637X},
  note      = {ADS Bibcode: 2023ApJ...956...96V},
  pages     = {96},
  url       = {https://ui.adsabs.harvard.edu/abs/2023ApJ...956...96V},
  urldate   = {2024-08-17},
  volume    = {956},
  journal   = {ApJ},
  month     = oct,
  publisher = {IOP},
  year      = {2023},
}

@Article{Segers2016,
  author        = {Segers, Marijke C. and Crain, Robert A. and Schaye, Joop and Bower, Richard G. and Furlong, Michelle and Schaller, Matthieu and Theuns, Tom},
  title         = {Recycled stellar ejecta as fuel for star formation and implications for the origin of the galaxy mass-metallicity relation},
  doi           = {10.1093/mnras/stv2562},
  eprint        = {1507.08281},
  number        = {2},
  pages         = {1235-1258},
  url           = {https://ui.adsabs.harvard.edu/abs/2016MNRAS.456.1235S},
  volume        = {456},
  archiveprefix = {arXiv},
  journal       = {MNRAS},
  month         = feb,
  primaryclass  = {astro-ph.GA},
  year          = {2016},
}

@Article{Dolphin2012,
  author    = {Andrew E. Dolphin},
  title     = {On the Estimation of Systematic Uncertainties of Star Formation Histories},
  doi       = {10.1088/0004-637x/751/1/60},
  number    = {1},
  pages     = {60},
  url       = {https://ui.adsabs.harvard.edu/abs/2012ApJ...751...60D/abstract},
  volume    = {751},
  comment   = {SFH},
  journal   = {ApJ},
  month     = {may},
  publisher = {{IOP} Publishing},
  ranking   = {rank3},
  year      = {2012},
}

@Article{Conroy2018,
  author        = {Conroy, Charlie and Villaume, Alexa and van Dokkum, Pieter G. and Lind, Karin},
  title         = {Metal-rich, Metal-poor: Updated Stellar Population Models for Old Stellar Systems},
  doi           = {10.3847/1538-4357/aaab49},
  eid           = {139},
  eprint        = {1801.10185},
  number        = {2},
  pages         = {139},
  url           = {https://ui.adsabs.harvard.edu/abs/2018ApJ...854..139C},
  volume        = {854},
  archiveprefix = {arXiv},
  journal       = {ApJ},
  month         = feb,
  primaryclass  = {astro-ph.GA},
  ranking       = {rank4},
  year          = {2018},
}

@Article{Tammann2011,
  author        = {Tammann, G.~A. and Reindl, B. and Sandage, A.},
  title         = {New period-luminosity and period-color relations of classical Cepheids. IV. The low-metallicity galaxies IC 1613, WLM, Pegasus, Sextans A and B, and Leo A in comparison to SMC},
  doi           = {10.1051/0004-6361/201016382},
  eid           = {A134},
  eprint        = {1012.4940},
  pages         = {A134},
  url           = {https://ui.adsabs.harvard.edu/abs/2011A&A...531A.134T},
  volume        = {531},
  archiveprefix = {arXiv},
  journal       = {\aap},
  month         = jul,
  primaryclass  = {astro-ph.CO},
  year          = {2011},
}

@Article{Conroy2013,
  author        = {Conroy, Charlie},
  title         = {Modeling the Panchromatic Spectral Energy Distributions of Galaxies},
  doi           = {10.1146/annurev-astro-082812-141017},
  eprint        = {1301.7095},
  number        = {1},
  pages         = {393-455},
  url           = {https://ui.adsabs.harvard.edu/abs/2013ARA&A..51..393C},
  volume        = {51},
  archiveprefix = {arXiv},
  comment       = {Annual Reviews of Astronomy & Astrophysics},
  journal       = {\araa},
  month         = aug,
  primaryclass  = {astro-ph.CO},
  ranking       = {rank3},
  year          = {2013},
}

@Article{Vink2001,
  author        = {Vink, Jorick S. and de Koter, A. and Lamers, H.~J.~G.~L.~M.},
  title         = {Mass-loss predictions for O and B stars as a function of metallicity},
  doi           = {10.1051/0004-6361:20010127},
  eprint        = {astro-ph/0101509},
  pages         = {574-588},
  url           = {https://ui.adsabs.harvard.edu/abs/2001A&A...369..574V},
  volume        = {369},
  archiveprefix = {arXiv},
  comment       = {Telford2024 prefers the Bjorklund2021mass loss rates},
  journal       = {\aap},
  month         = apr,
  primaryclass  = {astro-ph},
  year          = {2001},
}

@Article{Kurucz2005,
  author   = {Kurucz, Robert L.},
  title    = {ATLAS12, SYNTHE, ATLAS9, WIDTH9, et cetera},
  pages    = {14},
  url      = {https://ui.adsabs.harvard.edu/abs/2005MSAIS...8...14K},
  volume   = {8},
  journal  = {Memorie della Societa Astronomica Italiana Supplementi},
  month    = jan,
  year     = {2005},
}

@Article{Knollmann2009,
  author        = {Knollmann, Steffen R. and Knebe, Alexander},
  title         = {AHF: Amiga's Halo Finder},
  doi           = {10.1088/0067-0049/182/2/608},
  eprint        = {0904.3662},
  number        = {2},
  pages         = {608-624},
  url           = {https://ui.adsabs.harvard.edu/abs/2009ApJS..182..608K},
  volume        = {182},
  archiveprefix = {arXiv},
  journal       = {\apjs},
  month         = jun,
  primaryclass  = {astro-ph.CO},
  year          = {2009},
}

@Article{Kirby2011a,
  author        = {Kirby, Evan N. and Martin, Crystal L. and Finlator, Kristian},
  title         = {Metals Removed by Outflows from Milky Way Dwarf Spheroidal Galaxies},
  doi           = {10.1088/2041-8205/742/2/L25},
  eid           = {L25},
  eprint        = {1110.5624},
  number        = {2},
  pages         = {L25},
  url           = {https://ui.adsabs.harvard.edu/abs/2011ApJ...742L..25K},
  volume        = {742},
  archiveprefix = {arXiv},
  journal       = {ApJL},
  month         = dec,
  primaryclass  = {astro-ph.GA},
  year          = {2011},
}

@Article{Dolphin2016a,
  author        = {Dolphin, Andrew E.},
  title         = {{On the Incorporation of Metallicity Data Into Measurements of Star Formation History From Resolved Stellar Populations}},
  doi           = {10.3847/0004-637X/825/2/153},
  issn          = {1538-4357},
  pages         = {153--158},
  url           = {http://stacks.iop.org/0004-637X/825/i=2/a=153?key=crossref.94ba9d0955365dd957a050f374ac98ca},
  volume        = {825},
  comment       = {SFH},
  journal       = {ApJ},
  publisher     = {IOP Publishing},
  year          = {2016},
}

@Article{Springel2001,
  author        = {Springel, Volker and White, Simon D.~M. and Tormen, Giuseppe and Kauffmann, Guinevere},
  title         = {Populating a cluster of galaxies - I. Results at z=0},
  doi           = {10.1046/j.1365-8711.2001.04912.x},
  eprint        = {astro-ph/0012055},
  number        = {3},
  pages         = {726-750},
  url           = {https://ui.adsabs.harvard.edu/abs/2001MNRAS.328..726S},
  volume        = {328},
  archiveprefix = {arXiv},
  comment       = {SUBFIND},
  journal       = {MNRAS},
  month         = dec,
  primaryclass  = {astro-ph},
  year          = {2001},
}

@Article{Springel2003,
  author   = {Springel, V. and Hernquist, L.},
  title    = {Cosmological smoothed particle hydrodynamics simulations: a hybrid multiphase model for star formation},
  doi      = {10.1046/j.1365-8711.2003.06206.x},
  eprint   = {astro-ph/0206393},
  pages    = {289-311},
  url      = {https://ui.adsabs.harvard.edu/abs/2003MNRAS.339..289S},
  volume   = {339},
  journal  = {MNRAS},
  month    = feb,
  year     = {2003},
}

@InCollection{Kurucz2014,
  author    = {Kurucz, Robert L.},
  booktitle = {Determination of Atmospheric Parameters of B, A, F, and G-Type Stars},
  title     = {Model Atmosphere Codes: ATLAS12 and ATLAS9},
  doi       = {10.1007/978-3-319-06956-2_4},
  editor    = {{Niemczura}, Ewa and {Smalley}, Barry and {Pych}, Wojtek},
  isbn      = {978-3-319-06955-5},
  pages     = {39-51},
  publisher = {Springer International Publishing},
  url       = {https://ui.adsabs.harvard.edu/abs/2014dapb.book...39K},
  year      = {2014},
}

@Software{Distributions.jl,
  author    = {Dahua Lin and David Widmann and Simon Byrne and John Myles White and Andreas Noack and Mathieu Besançon and Douglas Bates and John Pearson and John Zito and Alex Arslan and Moritz Schauer and Kevin Squire and David Anthoff and Theodore Papamarkou and Jan Drugowitsch and Benjamin Deonovic and Seth Axen and Avik Sengupta and Viral B. Shah and Giuseppe Ragusa and Glenn Moynihan and Brian J Smith and Christoph Dann and Mike J Innes and Martin O'Leary and Tamas K. Papp and Mohamed Tarek and Michael},
  title     = {JuliaStats/Distributions.jl: v0.25.118},
  doi       = {10.5281/ZENODO.2647458},
  copyright = {Creative Commons Attribution 4.0 International},
  publisher = {Zenodo},
  year      = {2025},
}

@Article{Agertz2020a,
  author     = {Agertz, Oscar and Pontzen, Andrew and Read, Justin I. and Rey, Martin P. and Orkney, Matthew and Rosdahl, Joakim and Teyssier, Romain and Verbeke, Robbert and Kretschmer, Michael and Nickerson, Sarah},
  title      = {{EDGE}: the mass-metallicity relation as a critical test of galaxy formation physics},
  doi        = {10.1093/mnras/stz3053},
  issn       = {0035-8711},
  note       = {ADS Bibcode: 2020MNRAS.491.1656A},
  pages      = {1656--1672},
  url        = {https://ui.adsabs.harvard.edu/abs/2020MNRAS.491.1656A},
  urldate    = {2024-01-08},
  volume     = {491},
  journal    = {MNRAS},
  month      = jan,
  shorttitle = {{EDGE}},
  year       = {2020},
}

@Article{Julia,
  author    = {Bezanson, Jeff and Edelman, Alan and Karpinski, Stefan and Shah, Viral B.},
  title     = {Julia: A Fresh Approach to Numerical Computing},
  doi       = {10.1137/141000671},
  issn      = {1095-7200},
  number    = {1},
  pages     = {65--98},
  volume    = {59},
  journal   = {SIAM Review},
  month     = jan,
  publisher = {Society for Industrial & Applied Mathematics (SIAM)},
  year      = {2017},
}

@Article{Weinberger2017,
  author    = {Rainer Weinberger and Volker Springel and Lars Hernquist and Annalisa Pillepich and Federico Marinacci and Rüdiger Pakmor and Dylan Nelson and Shy Genel and Mark Vogelsberger and Jill Naiman and Paul Torrey},
  title     = {Simulating galaxy formation with black hole driven thermal and kinetic feedback},
  doi       = {10.1093/mnras/stw2944},
  number    = {3},
  pages     = {3291--3308},
  volume    = {465},
  journal   = {MNRAS},
  month     = {nov},
  publisher = {Oxford University Press ({OUP})},
  year      = {2017},
}

@Article{Aparicio2004,
  author     = {Aparicio, Antonio and Gallart, Carme},
  title      = {{IAC}-{STAR}: {A} {Code} for {Synthetic} {Color}-{Magnitude} {Diagram} {Computation}},
  doi        = {10.1086/382836},
  issn       = {0004-6256},
  note       = {ADS Bibcode: 2004AJ....128.1465A},
  pages      = {1465--1477},
  url        = {https://ui.adsabs.harvard.edu/abs/2004AJ....128.1465A},
  urldate    = {2024-04-25},
  volume     = {128},
  comment    = {IAC-star does CMD / Hertzsprung-Russel diagram simulation},
  journal    = {AJ},
  month      = sep,
  publisher  = {IOP},
  ranking    = {rank3},
  shorttitle = {{IAC}-{STAR}},
  year       = {2004},
}

@Article{Johnson2021,
  author    = {Johnson, Benjamin D. and Leja, Joel and Conroy, Charlie and Speagle, Joshua S.},
  title     = {Stellar {Population} {Inference} with {Prospector}},
  doi       = {10.3847/1538-4365/abef67},
  issn      = {0067-0049},
  note      = {ADS Bibcode: 2021ApJS..254...22J},
  pages     = {22},
  url       = {https://ui.adsabs.harvard.edu/abs/2021ApJS..254...22J},
  urldate   = {2024-04-23},
  volume    = {254},
  comment   = {FSPS update},
  journal   = {\apjs},
  month     = jun,
  publisher = {IOP},
  year      = {2021},
}

@Article{Nakajima2023,
  author        = {Nakajima, Kimihiko and Ouchi, Masami and Isobe, Yuki and Harikane, Yuichi and Zhang, Yechi and Ono, Yoshiaki and Umeda, Hiroya and Oguri, Masamune},
  title         = {JWST Census for the Mass-Metallicity Star Formation Relations at z = 4-10 with Self-consistent Flux Calibration and Proper Metallicity Calibrators},
  doi           = {10.3847/1538-4365/acd556},
  eid           = {33},
  eprint        = {2301.12825},
  number        = {2},
  pages         = {33},
  url           = {https://ui.adsabs.harvard.edu/abs/2023ApJS..269...33N},
  volume        = {269},
  archiveprefix = {arXiv},
  journal       = {\apjs},
  month         = dec,
  primaryclass  = {astro-ph.GA},
  year          = {2023},
}

@Article{Dotter2016,
  author    = {Aaron Dotter},
  title     = {{MESA} {ISOCHRONES} {AND} {STELLAR} {TRACKS} ({MIST}) 0: {METHODS} {FOR} {THE} {CONSTRUCTION} {OF} {STELLAR} {ISOCHRONES}},
  doi       = {10.3847/0067-0049/222/1/8},
  number    = {1},
  pages     = {8},
  url       = {https://ui.adsabs.harvard.edu/abs/2016ApJS..222....8D/abstract},
  volume    = {222},
  comment   = {See Choi2016 for next paper in series},
  journal   = {\apjs},
  month     = {jan},
  publisher = {American Astronomical Society},
  year      = {2016},
}

@Article{RuizEscobedo2018,
  author        = {Ruiz-Escobedo, Francisco and Pe{\~n}a, Miriam and Hern{\'a}ndez-Mart{\'\i}nez, Liliana and Garc{\'\i}a-Rojas, Jorge},
  title         = {Chemistry in the dIrr galaxy Leo A},
  doi           = {10.1093/mnras/sty2265},
  eprint        = {1808.06027},
  number        = {1},
  pages         = {396-404},
  url           = {https://ui.adsabs.harvard.edu/abs/2018MNRAS.481..396R},
  volume        = {481},
  archiveprefix = {arXiv},
  journal       = {MNRAS},
  month         = nov,
  primaryclass  = {astro-ph.GA},
  year          = {2018},
}

@Article{Pietrinferni2004,
  author    = {Adriano Pietrinferni and Santi Cassisi and Maurizio Salaris and Fiorella Castelli},
  title     = {A Large Stellar Evolution Database for Population Synthesis Studies. I. Scaled Solar Models and Isochrones},
  doi       = {10.1086/422498},
  number    = {1},
  pages     = {168--190},
  volume    = {612},
  journal   = {ApJ},
  month     = {sep},
  publisher = {{IOP} Publishing},
  year      = {2004},
}

@Article{Pietrinferni2006,
  author    = {Adriano Pietrinferni and Santi Cassisi and Maurizio Salaris and Fiorella Castelli},
  title     = {A Large Stellar Evolution Database for Population Synthesis Studies. {II}. Stellar Models and Isochrones for an alpha-enhanced Metal Distribution},
  doi       = {10.1086/501344},
  number    = {2},
  pages     = {797--812},
  volume    = {642},
  journal   = {ApJ},
  month     = {may},
  publisher = {{IOP} Publishing},
  year      = {2006},
}

@Article{Tollet2019,
  author        = {Tollet, {\'E}. and Cattaneo, A. and Macci{\`o}, A. V. and Dutton, A. A. and Kang, X.},
  title         = {NIHAO XIX: how supernova feedback shapes the galaxy baryon cycle},
  doi           = {10.1093/mnras/stz545},
  eprint        = {1902.03888},
  pages         = {2511-2531},
  url           = {http://adsabs.harvard.edu/abs/2019MNRAS.485.2511T},
  volume        = {485},
  archiveprefix = {arXiv},
  journal       = {MNRAS},
  month         = may,
  year          = {2019},
}

@Article{Bond1991,
  author    = {J. R. Bond and S. Cole and G. Efstathiou and N. Kaiser},
  title     = {Excursion set mass functions for hierarchical Gaussian fluctuations},
  doi       = {10.1086/170520},
  pages     = {440},
  volume    = {379},
  journal   = {ApJ},
  month     = {oct},
  publisher = {{IOP} Publishing},
  year      = {1991},
}

@Article{Muratov2017,
  author    = {Alexander L. Muratov and Du{\v{s}}an Kere{\v{s}} and Claude-Andr{\'{e}} Faucher-Gigu{\`{e}}re and Philip F. Hopkins and Xiangcheng Ma and Daniel Angl{\'{e}}s-Alc{\'{a}}zar and T. K. Chan and Paul Torrey and Zachary H. Hafen and Eliot Quataert and Norman Murray},
  title     = {Metal flows of the circumgalactic medium, and the metal budget in galactic haloes},
  doi       = {10.1093/mnras/stx667},
  number    = {4},
  pages     = {4170--4188},
  volume    = {468},
  journal   = {MNRAS},
  month     = {mar},
  publisher = {Oxford University Press ({OUP})},
  year      = {2017},
}

@Article{Giovanelli2013,
  author        = {Giovanelli, Riccardo and Haynes, Martha P. and Adams, Elizabeth A.~K. and Cannon, John M. and Rhode, Katherine L. and Salzer, John J. and Skillman, Evan D. and Bernstein-Cooper, Elijah Z. and McQuinn, Kristen B.~W.},
  title         = {ALFALFA Discovery of the Nearby Gas-rich Dwarf Galaxy Leo P. I. H I Observations},
  doi           = {10.1088/0004-6256/146/1/15},
  eid           = {15},
  eprint        = {1305.0272},
  number        = {1},
  pages         = {15},
  url           = {https://ui.adsabs.harvard.edu/abs/2013AJ....146...15G},
  volume        = {146},
  archiveprefix = {arXiv},
  journal       = {\aj},
  month         = jul,
  primaryclass  = {astro-ph.CO},
  year          = {2013},
}

@Article{Li2023c,
  author        = {Li, Mingyu and Cai, Zheng and Bian, Fuyan and Lin, Xiaojing and Li, Zihao and Wu, Yunjing and Sun, Fengwu and Zhang, Shiwu and Golden-Marx, Emmet and Sun, Zechang and Zou, Siwei and Fan, Xiaohui and Egami, Eiichi and Charlot, Stephane and Bruzual, Gustavo and Chevallard, Jacopo},
  title         = {The Mass–Metallicity Relation of Dwarf Galaxies at Cosmic Noon from JWST Observations},
  doi           = {10.3847/2041-8213/acf470},
  eprint        = {2211.01382},
  issn          = {2041-8213},
  number        = {1},
  pages         = {L18},
  url           = {https://ui.adsabs.harvard.edu/abs/2023ApJ...955L..18L/abstract},
  volume        = {955},
  archiveprefix = {arXiv},
  copyright     = {Creative Commons Attribution 4.0 International},
  journal       = {ApJL},
  month         = sep,
  primaryclass  = {astro-ph.GA},
  publisher     = {American Astronomical Society},
  year          = {2023},
}

@Article{McQuinn2015,
  author    = {Kristen B. W. McQuinn and Evan D. Skillman and Andrew Dolphin and John M. Cannon and John J. Salzer and Katherine L. Rhode and Elizabeth A. K. Adams and Danielle Berg and Riccardo Giovanelli and L{\'{e}}o Girardi and Martha P. Haynes},
  title     = {{LEO} P: {AN} {UNQUENCHED} {VERY} {LOW}-{MASS} {GALAXY}},
  doi       = {10.1088/0004-637x/812/2/158},
  number    = {2},
  pages     = {158},
  volume    = {812},
  journal   = {ApJ},
  month     = {oct},
  publisher = {{IOP} Publishing},
  year      = {2015},
}

@Article{Blitz2006,
  author     = {Blitz, Leo and Rosolowsky, Erik},
  title      = {The {Role} of {Pressure} in {GMC} {Formation} {II}: {The} {H2}-{Pressure} {Relation}},
  doi        = {10.1086/505417},
  issn       = {0004-637X},
  note       = {ADS Bibcode: 2006ApJ...650..933B},
  pages      = {933--944},
  url        = {https://ui.adsabs.harvard.edu/abs/2006ApJ...650..933B},
  urldate    = {2025-04-02},
  volume     = {650},
  journal    = {ApJ},
  month      = oct,
  publisher  = {IOP},
  shorttitle = {The {Role} of {Pressure} in {GMC} {Formation} {II}},
  year       = {2006},
}

@Article{Kirby2017a,
  author        = {Kirby, Evan N. and Rizzi, Luca and Held, Enrico V. and Cohen, Judith G. and Cole, Andrew A. and Manning, Ellen M. and Skillman, Evan D. and Weisz, Daniel R.},
  title         = {Chemistry and Kinematics of the Late-forming Dwarf Irregular Galaxies Leo A, Aquarius, and Sagittarius DIG},
  doi           = {10.3847/1538-4357/834/1/9},
  eid           = {9},
  eprint        = {1610.08505},
  number        = {1},
  pages         = {9},
  url           = {https://ui.adsabs.harvard.edu/abs/2017ApJ...834....9K},
  volume        = {834},
  archiveprefix = {arXiv},
  journal       = {ApJ},
  month         = jan,
  primaryclass  = {astro-ph.GA},
  year          = {2017},
}

@Article{Schlegel1998,
  author        = {Schlegel, David J. and Finkbeiner, Douglas P. and Davis, Marc},
  title         = {{Maps of Dust Infrared Emission for Use in Estimation of Reddening and Cosmic Microwave Background Radiation Foregrounds}},
  doi           = {10.1086/305772},
  eprint        = {9710327},
  pages         = {525--553},
  volume        = {500},
  archiveprefix = {arXiv},
  arxivid       = {astro-ph/9710327},
  isbn          = {0004-637X},
  journal       = {ApJ},
  primaryclass  = {astro-ph},
  year          = {1998},
}

@Article{Wang2025,
  author        = {Wang, Bingjie and Leja, Joel and Atek, Hakim and Bezanson, Rachel and Burnham, Emilie and Dayal, Pratika and Feldmann, Robert and Greene, Jenny E. and Johnson, Benjamin D. and Labb{\'e}, Ivo and Maseda, Michael V. and Nanayakkara, Themiya and Price, Sedona H. and Suess, Katherine A. and Weaver, John R. and Whitaker, Katherine E.},
  title         = {Population Models for Star Formation Timescales in Early Galaxies: The First Step toward Solving Outshining in Star Formation History Inference},
  doi           = {10.3847/1538-4357/adddb8},
  eid           = {184},
  eprint        = {2504.15255},
  number        = {2},
  pages         = {184},
  url           = {https://ui.adsabs.harvard.edu/abs/2025ApJ...987..184W},
  volume        = {987},
  archiveprefix = {arXiv},
  journal       = {ApJ},
  month         = jul,
  primaryclass  = {astro-ph.GA},
  year          = {2025},
}

@Article{Muratov2015,
  author        = {Muratov, A. L. and Kere{\v s}, D. and Faucher-Gigu{\`e}re, C.-A. and Hopkins, P. F. and Quataert, E. and Murray, N.},
  title         = {Gusty, gaseous flows of FIRE: galactic winds in cosmological simulations with explicit stellar feedback},
  doi           = {10.1093/mnras/stv2126},
  eprint        = {1501.03155},
  pages         = {2691-2713},
  url           = {http://adsabs.harvard.edu/abs/2015MNRAS.454.2691M},
  volume        = {454},
  archiveprefix = {arXiv},
  journal       = {MNRAS},
  month         = dec,
  year          = {2015},
}

@Article{Pietrinferni2013,
  author    = {Adriano Pietrinferni and Santi Cassisi and Maurizio Salaris and Sebastian Hidalgo},
  title     = {The {BaSTI} Stellar Evolution Database: models for extremely metal-poor and super-metal-rich stellar populations},
  doi       = {10.1051/0004-6361/201321950},
  pages     = {A46},
  volume    = {558},
  journal   = {A{\&}A},
  month     = {oct},
  publisher = {{EDP} Sciences},
  ranking   = {rank2},
  year      = {2013},
}

@Article{TorresRios2024,
  author    = {Torres-Ríos, G. and Pérez, I. and Verley, S. and Domínguez-Gómez, J. and Argudo-Fernández, M. and Duarte Puertas, S. and Jiménez, A. and Ruiz-Lara, T. and Zurita, A. and Bidaran, B. and Conrado, A. and Espada, D. and García-Benito, R. and González Delgado, R. M. and Falcón-Barroso, J. and Florido, E. and Sánchez-Blázquez, P. and Sánchez-Menguiano, L.},
  title     = {Effect of the local and large-scale environment on the star formation histories of galaxies},
  doi       = {10.1051/0004-6361/202450675},
  issn      = {0004-6361},
  note      = {ADS Bibcode: 2024A\&A...691A.341T},
  pages     = {A341},
  url       = {https://ui.adsabs.harvard.edu/abs/2024A&A...691A.341T},
  urldate   = {2025-01-30},
  volume    = {691},
  journal   = {\aap},
  month     = nov,
  publisher = {EDP},
  year      = {2024},
}

@Article{Matplotlib,
  author    = {Hunter, J. D.},
  title     = {Matplotlib: A 2D graphics environment},
  doi       = {10.1109/MCSE.2007.55},
  number    = {3},
  pages     = {90--95},
  volume    = {9},
  journal   = {Computing In Science \& Engineering},
  publisher = {IEEE COMPUTER SOC},
  year      = {2007},
}

@Article{McQuinn2024,
  author   = {McQuinn, Kristen. B. W. and B. Newman, Max J. and Savino, Alessandro and Dolphin, Andrew E. and Weisz, Daniel R. and Williams, Benjamin F. and Boyer, Martha L. and Cohen, Roger E. and Correnti, Matteo and Cole, Andrew A. and Geha, Marla C. and Gennaro, Mario and Kallivayalil, Nitya and Sandstrom, Karin M. and Skillman, Evan D. and Anderson, Jay and Bolatto, Alberto and Boylan-Kolchin, Michael and Garling, Christopher T. and Gilbert, Karoline M. and Girardi, Léo and Kalirai, Jason S. and Mazzi, Alessandro and Pastorelli, Giada and Richstein, Hannah and Warfield, Jack T.},
  title    = {The {JWST} {Resolved} {Stellar} {Populations} {Early} {Release} {Science} {Program}. {IV}. {The} {Star} {Formation} {History} of the {Local} {Group} {Galaxy} {WLM}},
  doi      = {10.3847/1538-4357/ad1105},
  issn     = {0004-637X},
  note     = {ADS Bibcode: 2024ApJ...961...16M},
  pages    = {16},
  url      = {https://ui.adsabs.harvard.edu/abs/2024ApJ...961...16M},
  urldate  = {2024-01-29},
  volume   = {961},
  journal  = {ApJ},
  month    = jan,
  year     = {2024},
}

@Article{Dolphin2002,
  author   = {Dolphin, A. E.},
  title    = {Numerical methods of star formation history measurement and applications to seven dwarf spheroidals},
  doi      = {10.1046/j.1365-8711.2002.05271.x},
  eprint   = {astro-ph/0112331},
  pages    = {91-108},
  url      = {http://adsabs.harvard.edu/abs/2002MNRAS.332...91D},
  volume   = {332},
  comment  = {SFH},
  journal  = {MNRAS},
  month    = may,
  ranking  = {rank5},
  year     = {2002},
}

@Article{McQuinn2024b,
  author     = {McQuinn, Kristen B. W. and Newman, Max J. B. and Skillman, Evan D. and Telford, O. Grace and Brooks, Alyson and Adams, Elizabeth A. K. and Berg, Danielle A. and Boyer, Martha L. and Cannon, John M. and Dolphin, Andrew E. and Pahl, Anthony and Rhode, Katherine L. and Salzer, John J. and Cohen, Roger E. and Goldman, Steve R.},
  title      = {The {Ancient} {Star} {Formation} {History} of the {Extremely} {Low}-{Mass} {Galaxy} {Leo} {P}: {An} {Emerging} {Trend} of a {Post}-{Reionization} {Pause} in {Star} {Formation}},
  doi        = {10.3847/1538-4357/ad8158},
  eprint     = {2409.19050},
  issn       = {1538-4357},
  note       = {ADS Bibcode: 2024arXiv240919050M Type: article},
  number     = {1},
  pages      = {60},
  url        = {https://ui.adsabs.harvard.edu/abs/2024arXiv240919050M},
  urldate    = {2024-10-03},
  volume     = {976},
  journal    = {ApJ},
  month      = nov,
  publisher  = {American Astronomical Society},
  shorttitle = {The {Ancient} {Star} {Formation} {History} of the {Extremely} {Low}-{Mass} {Galaxy} {Leo} {P}},
  year       = {2024},
}

@Article{Dolphin2003,
  author    = {Dolphin, Andrew E. and Saha, A. and Skillman, Evan D. and Dohm-Palmer, R. C. and Tolstoy, Eline and Cole, A. A. and Gallagher, J. S. and Hoessel, J. G. and Mateo, Mario},
  title     = {Deep {Hubble} {Space} {Telescope} {Imaging} of {Sextans} {A}. {III}. {The} {Star} {Formation} {History}},
  doi       = {10.1086/375761},
  issn      = {0004-6256},
  note      = {ADS Bibcode: 2003AJ....126..187D},
  pages     = {187--196},
  url       = {https://ui.adsabs.harvard.edu/abs/2003AJ....126..187D},
  urldate   = {2025-01-30},
  volume    = {126},
  journal   = {AJ},
  month     = jul,
  publisher = {IOP},
  year      = {2003},
}

@Article{Ianjamasimanana2020,
  author    = {Ianjamasimanana, Roger and Namumba, Brenda and Ramaila, Athanaseus J. T. and Saburova, Anna S. and Józsa, Gyula I. G. and Myburgh, Talon and Thorat, Kshitij and Carignan, Claude and Maina, Eric and de Blok, W. J. G. and Andati, Lexy A. L. and Hugo, Benjamin V. and Kleiner, Dane and Kamphuis, Peter and Serra, Paolo and Smirnov, Oleg M. and Maccagni, Filippo M. and Makhathini, Sphesihle and Molnár, Dániel Cs and Perkins, Simon and Ramatsoku, Mpati and White, Sarah V.},
  title     = {{MeerKAT}-16 {H} {I} observation of the {dIrr} galaxy {WLM}},
  doi       = {10.1093/mnras/staa1967},
  issn      = {0035-8711},
  note      = {ADS Bibcode: 2020MNRAS.497.4795I},
  pages     = {4795--4813},
  url       = {https://ui.adsabs.harvard.edu/abs/2020MNRAS.497.4795I},
  urldate   = {2025-03-17},
  volume    = {497},
  journal   = {MNRAS},
  month     = oct,
  publisher = {OUP},
  year      = {2020},
}

@Article{Buck2019,
  author        = {Buck, T. and Dutton, A. A. and Macci{\`o}, A. V.},
  title         = {An observational test for star formation prescriptions in cosmological hydrodynamical simulations},
  doi           = {10.1093/mnras/stz969},
  eprint        = {1812.05613},
  pages         = {1481-1487},
  url           = {https://ui.adsabs.harvard.edu/abs/2019MNRAS.486.1481B},
  volume        = {486},
  archiveprefix = {arXiv},
  journal       = {MNRAS},
  month         = jun,
  ranking       = {rank3},
  year          = {2019},
}

@Article{Telford2024,
  author        = {Telford, O. Grace and Chisholm, John and Sander, Andreas A.~C. and Ramachandran, Varsha and McQuinn, Kristen B.~W. and Berg, Danielle A.},
  title         = {Observations of Extremely Metal-poor O Stars: Weak Winds and Constraints for Evolution Models},
  doi           = {10.3847/1538-4357/ad697e},
  eid           = {85},
  eprint        = {2407.20313},
  number        = {1},
  pages         = {85},
  url           = {https://ui.adsabs.harvard.edu/abs/2024ApJ...974...85T},
  volume        = {974},
  archiveprefix = {arXiv},
  journal       = {ApJ},
  month         = oct,
  primaryclass  = {astro-ph.SR},
  year          = {2024},
}

@Article{Karakas2010,
  author        = {Karakas, A.~I.},
  title         = {Updated stellar yields from asymptotic giant branch models},
  doi           = {10.1111/j.1365-2966.2009.16198.x},
  eprint        = {0912.2142},
  number        = {3},
  pages         = {1413-1425},
  url           = {https://ui.adsabs.harvard.edu/abs/2010MNRAS.403.1413K},
  volume        = {403},
  archiveprefix = {arXiv},
  journal       = {MNRAS},
  month         = apr,
  primaryclass  = {astro-ph.SR},
  year          = {2010},
}

@Article{Hunter2012,
  author        = {Hunter, D. A. and Ficut-Vicas, D. and Ashley, T. and Brinks, E. and Cigan, P. and Elmegreen, B. G. and Heesen, V. and Herrmann, K. A. and Johnson, M. and Oh, S.-H. and Rupen, M. P. and Schruba, A. and Simpson, C. E. and Walter, F. and Westpfahl, D. J. and Young, L. M. and Zhang, H.-X.},
  title         = {Little Things},
  doi           = {10.1088/0004-6256/144/5/134},
  eid           = {134},
  eprint        = {1208.5834},
  pages         = {134},
  url           = {http://adsabs.harvard.edu/abs/2012AJ....144..134H},
  volume        = {144},
  archiveprefix = {arXiv},
  journal       = {AJ},
  month         = nov,
  ranking       = {rank3},
  year          = {2012},
}

@Article{Mitchell2020a,
  author   = {Mitchell, Peter D. and Schaye, Joop and Bower, Richard G. and Crain, Robert A.},
  title    = {Galactic outflow rates in the {EAGLE} simulations},
  doi      = {10.1093/mnras/staa938},
  issn     = {0035-8711},
  note     = {ADS Bibcode: 2020MNRAS.494.3971M},
  pages    = {3971--3997},
  url      = {https://ui.adsabs.harvard.edu/abs/2020MNRAS.494.3971M},
  urldate  = {2024-03-06},
  volume   = {494},
  comment  = {Appendix B for parametric fits},
  journal  = {MNRAS},
  month    = may,
  year     = {2020},
}

@Article{Chen2019a,
  author        = {Chen, Yang and Girardi, L{\'e}o and Fu, Xiaoting and Bressan, Alessandro and Aringer, Bernhard and Dal Tio, Piero and Pastorelli, Giada and Marigo, Paola and Costa, Guglielmo and Zhang, Xing},
  title         = {YBC: a stellar bolometric corrections database with variable extinction coefficients. Application to PARSEC isochrones},
  doi           = {10.1051/0004-6361/201936612},
  eid           = {A105},
  eprint        = {1910.09037},
  pages         = {A105},
  url           = {https://ui.adsabs.harvard.edu/abs/2019A&A...632A.105C},
  volume        = {632},
  archiveprefix = {arXiv},
  journal       = {\aap},
  month         = dec,
  primaryclass  = {astro-ph.SR},
  ranking       = {rank4},
  year          = {2019},
}

@Article{Nadler2023a,
  author        = {Nadler, Ethan O. and Benson, Andrew and Driskell, Trey and Du, Xiaolong and Gluscevic, Vera},
  title         = {Growing the first galaxies' merger trees},
  doi           = {10.1093/mnras/stad666},
  eprint        = {2212.08584},
  number        = {3},
  pages         = {3201-3220},
  url           = {https://ui.adsabs.harvard.edu/abs/2023MNRAS.521.3201N},
  volume        = {521},
  archiveprefix = {arXiv},
  journal       = {MNRAS},
  month         = may,
  primaryclass  = {astro-ph.GA},
  ranking       = {rank3},
  year          = {2023},
}

@Article{Ramesh2023a,
  author   = {Ramesh, Rahul and Nelson, Dylan and Pillepich, Annalisa},
  title    = {The circumgalactic medium of {Milky} {Way}-like galaxies in the {TNG50} simulation - {II}. {Cold}, dense gas clouds and high-velocity cloud analogs},
  doi      = {10.1093/mnras/stad951},
  issn     = {0035-8711},
  note     = {ADS Bibcode: 2023MNRAS.522.1535R},
  pages    = {1535--1555},
  url      = {https://ui.adsabs.harvard.edu/abs/2023MNRAS.522.1535R},
  urldate  = {2024-01-09},
  volume   = {522},
  journal  = {MNRAS},
  month    = jun,
  ranking  = {rank3},
  year     = {2023},
}

@Article{Maoz2012,
  author        = {Maoz, Dan and Mannucci, Filippo and Brandt, Timothy D.},
  title         = {The delay-time distribution of Type Ia supernovae from Sloan II},
  doi           = {10.1111/j.1365-2966.2012.21871.x},
  eprint        = {1206.0465},
  number        = {4},
  pages         = {3282-3294},
  url           = {https://ui.adsabs.harvard.edu/abs/2012MNRAS.426.3282M},
  volume        = {426},
  archiveprefix = {arXiv},
  journal       = {MNRAS},
  month         = nov,
  primaryclass  = {astro-ph.CO},
  year          = {2012},
}

@Article{Li2021b,
  author   = {Li, Fei and Rahman, Mubdi and Murray, Norman and Hafen, Zachary and Faucher-Giguère, Claude-André and Stern, Jonathan and Hummels, Cameron B. and Hopkins, Philip F. and El-Badry, Kareem and Kereš, Dušan},
  title    = {Probing the {CGM} of low-redshift dwarf galaxies using {FIRE} simulations},
  doi      = {10.1093/mnras/staa3322},
  issn     = {0035-8711},
  note     = {ADS Bibcode: 2021MNRAS.500.1038L},
  pages    = {1038--1053},
  url      = {https://ui.adsabs.harvard.edu/abs/2021MNRAS.500.1038L},
  urldate  = {2024-02-21},
  volume   = {500},
  comment  = {COS-DWARFS comparison study (Bordoloi2014)},
  journal  = {MNRAS},
  month    = jan,
  ranking  = {rank3},
  year     = {2021},
}

@Article{Portinari1998,
  author       = {L. Portinari and C. Chiosi and A. Bressan},
  date         = {1997-11-27},
  journaltitle = {Astron.Astrophys.334:505-539,1998},
  title        = {Galactic chemical enrichment with new metallicity dependent yields},
  eprint       = {http://arxiv.org/abs/astro-ph/9711337v1},
  eprintclass  = {astro-ph},
  eprinttype   = {arXiv},
  pages        = {505-539},
  volume       = {334},
  journal      = {A{\&}A},
  year         = {1998},
}

@Article{Hopkins2014,
  author        = {Hopkins, P. F. and Kere{\v s}, D. and O{\~n}orbe, J. and Faucher-Gigu{\`e}re, C.-A. and Quataert, E. and Murray, N. and Bullock, J. S.},
  title         = {Galaxies on FIRE (Feedback In Realistic Environments): stellar feedback explains cosmologically inefficient star formation},
  doi           = {10.1093/mnras/stu1738},
  eprint        = {1311.2073},
  pages         = {581-603},
  volume        = {445},
  archiveprefix = {arXiv},
  journal       = {MNRAS},
  month         = nov,
  year          = {2014},
}

@Article{Chisholm2018,
  author    = {Chisholm, J and Tremonti, C and Leitherer, C},
  title     = {Metal-enriched galactic outflows shape the mass–metallicity relationship},
  doi       = {10.1093/mnras/sty2380},
  issn      = {1365-2966},
  number    = {2},
  pages     = {1690--1706},
  volume    = {481},
  journal   = {MNRAS},
  month     = sep,
  publisher = {Oxford University Press (OUP)},
  ranking   = {rank4},
  year      = {2018},
}

@Article{Ahvazi2024,
  author    = {Ahvazi, Niusha and Benson, Andrew and Sales, Laura V. and Nadler, Ethan O. and Weerasooriya, Sachi and Du, Xiaolong and Bovill, Mia Sauda},
  title     = {A comprehensive model for the formation and evolution of the faintest {Milky} {Way} dwarf satellites},
  doi       = {10.1093/mnras/stae761},
  issn      = {0035-8711},
  pages     = {3387--3407},
  url       = {https://ui.adsabs.harvard.edu/abs/2024MNRAS.529.3387A},
  urldate   = {2024-04-22},
  volume    = {529},
  journal   = {MNRAS},
  month     = apr,
  publisher = {OUP},
  year      = {2024},
}

@Article{Knebe2018,
  author        = {Knebe, A. and Stoppacher, D. and Prada, F. and Behrens, C. and Benson, A. and Cora, S. A. and Croton, D. J. and Padilla, N. D. and Ruiz, A. N. and Sinha, M. and Stevens, A. R. H. and Vega-Mart{\'{\i}}nez, C. A. and Behroozi, P. and Gonzalez-Perez, V. and Gottl{\"o}ber, S. and Klypin, A. A. and Yepes, G. and Enke, H. and Libeskind, N. I. and Riebe, K. and Steinmetz, M.},
  title         = {MULTIDARK-GALAXIES: data release and first results},
  doi           = {10.1093/mnras/stx2662},
  eprint        = {1710.08150},
  pages         = {5206-5231},
  url           = {http://adsabs.harvard.edu/abs/2018MNRAS.474.5206K},
  volume        = {474},
  archiveprefix = {arXiv},
  journal       = {MNRAS},
  month         = mar,
  year          = {2018},
}

@Article{Gill2004,
  author        = {Gill, Stuart P.~D. and Knebe, Alexander and Gibson, Brad K.},
  title         = {The evolution of substructure - I. A new identification method},
  doi           = {10.1111/j.1365-2966.2004.07786.x},
  eprint        = {astro-ph/0404258},
  number        = {2},
  pages         = {399-409},
  url           = {https://ui.adsabs.harvard.edu/abs/2004MNRAS.351..399G},
  volume        = {351},
  archiveprefix = {arXiv},
  comment       = {MHF AHF},
  journal       = {MNRAS},
  month         = jun,
  primaryclass  = {astro-ph},
  year          = {2004},
}

@Article{Moutard2018,
  author    = {Thibaud Moutard and Marcin Sawicki and St{\'{e}}phane Arnouts and Anneya Golob and Nicola Malavasi and Christophe Adami and Jean Coupon and Olivier Ilbert},
  title     = {On the fast quenching of young low-mass galaxies up to z $\sim$ 0.6: new spotlight on the lead role of environment},
  doi       = {10.1093/mnras/sty1543},
  number    = {2},
  pages     = {2147--2160},
  volume    = {479},
  journal   = {MNRAS},
  month     = {jun},
  publisher = {Oxford University Press ({OUP})},
  year      = {2018},
}

@Article{Hopkins2018,
  author        = {Hopkins, P. F. and Wetzel, A. and Kere{\v s}, D. and Faucher-Gigu{\`e}re, C.-A. and Quataert, E. and Boylan-Kolchin, M. and Murray, N. and Hayward, C. C. and Garrison-Kimmel, S. and Hummels, C. and Feldmann, R. and Torrey, P. and Ma, X. and Angl{\'e}s-Alc{\'a}zar, D. and Su, K.-Y. and Orr, M. and Schmitz, D. and Escala, I. and Sanderson, R. and Grudi{\'c}, M. Y. and Hafen, Z. and Kim, J.-H. and Fitts, A. and Bullock, J. S. and Wheeler, C. and Chan, T. K. and Elbert, O. D. and Narayanan, D.},
  title         = {FIRE-2 simulations: physics versus numerics in galaxy formation},
  doi           = {10.1093/mnras/sty1690},
  eprint        = {1702.06148},
  pages         = {800-863},
  url           = {http://adsabs.harvard.edu/abs/2018MNRAS.480..800H},
  volume        = {480},
  archiveprefix = {arXiv},
  journal       = {MNRAS},
  month         = oct,
  ranking       = {rank5},
  year          = {2018},
}

@Article{Wright2024,
  author        = {Wright, Ruby J. and Somerville, Rachel S. and Lagos, Claudia del P. and Schaller, Matthieu and Dav{\'e}, Romeel and Angl{\'e}s-Alc{\'a}zar, Daniel and Genel, Shy},
  title         = {The baryon cycle in modern cosmological hydrodynamical simulations},
  doi           = {10.1093/mnras/stae1688},
  eprint        = {2402.08408},
  number        = {3},
  pages         = {3417-3440},
  url           = {https://ui.adsabs.harvard.edu/abs/2024MNRAS.532.3417W},
  volume        = {532},
  archiveprefix = {arXiv},
  journal       = {MNRAS},
  month         = aug,
  primaryclass  = {astro-ph.GA},
  year          = {2024},
}

@Article{Pietrinferni2021,
  author    = {Pietrinferni, Adriano and Hidalgo, Sebastian and Cassisi, Santi and Salaris, Maurizio and Savino, Alessandro and Mucciarelli, Alessio and Verma, Kuldeep and Silva Aguirre, Victor and Aparicio, Antonio and Ferguson, Jason W.},
  title     = {Updated {BaSTI} {Stellar} {Evolution} {Models} and {Isochrones}. {II}. α-enhanced {Calculations}},
  doi       = {10.3847/1538-4357/abd4d5},
  issn      = {0004-637X},
  note      = {ADS Bibcode: 2021ApJ...908..102P},
  pages     = {102},
  url       = {https://ui.adsabs.harvard.edu/abs/2021ApJ...908..102P},
  urldate   = {2024-04-28},
  volume    = {908},
  journal   = {ApJ},
  month     = feb,
  publisher = {IOP},
  year      = {2021},
}

@Article{AnglesAlcazar2017,
  author   = {Anglés-Alcázar, Daniel and Faucher-Giguère, Claude-André and Kereš, Dušan and Hopkins, Philip F. and Quataert, Eliot and Murray, Norman},
  title    = {The cosmic baryon cycle and galaxy mass assembly in the {FIRE} simulations},
  doi      = {10.1093/mnras/stx1517},
  issn     = {0035-8711},
  note     = {ADS Bibcode: 2017MNRAS.470.4698A},
  pages    = {4698--4719},
  url      = {https://ui.adsabs.harvard.edu/abs/2017MNRAS.470.4698A},
  urldate  = {2024-03-04},
  volume   = {470},
  journal  = {MNRAS},
  month    = oct,
  year     = {2017},
}

@Article{Pietrinferni2024,
  author    = {Pietrinferni, Adriano and Salaris, Maurizio and Cassisi, Santi and Savino, Alessandro and Mucciarelli, Alessio and Hyder, David and Hidalgo, Sebastian},
  title     = {The updated {BaSTI} stellar evolution models and isochrones - {IV}. α-{Depleted} calculations},
  doi       = {10.1093/mnras/stad3267},
  issn      = {0035-8711},
  note      = {ADS Bibcode: 2024MNRAS.527.2065P},
  pages     = {2065--2070},
  url       = {https://ui.adsabs.harvard.edu/abs/2024MNRAS.527.2065P},
  urldate   = {2024-04-28},
  volume    = {527},
  journal   = {MNRAS},
  month     = jan,
  publisher = {OUP},
  year      = {2024},
}

@Article{Papovich2018,
  author    = {Casey Papovich and Lalitwadee Kawinwanichakij and Ryan F. Quadri and Karl Glazebrook and Ivo Labb{\'{e}} and Kim-Vy H. Tran and Ben Forrest and Glenn G. Kacprzak and Lee R. Spitler and Caroline M. S. Straatman and Adam R. Tomczak},
  title     = {The Effects of Environment on the Evolution of the Galaxy Stellar Mass Function},
  doi       = {10.3847/1538-4357/aaa766},
  number    = {1},
  pages     = {30},
  volume    = {854},
  journal   = {ApJ},
  month     = {feb},
  publisher = {American Astronomical Society},
  ranking   = {rank3},
  year      = {2018},
}

@Article{Jain2026,
  author        = {Jain, Shweta and Sanders, Ryan L. and Khostovan, Ali Ahmad and Jones, Tucker and Shapley, Alice E. and Reddy, Naveen A. and Garcia, Alex M. and Torrey, Paul and Coil, Alison},
  title         = {A Uniform Analysis of Gas-phase Metallicity Evolution with 1–3 Gyr Time Sampling over the Past 12 Gyr},
  doi           = {10.3847/1538-4357/ae4590},
  eprint        = {2508.18369},
  issn          = {1538-4357},
  number        = {1},
  pages         = {109},
  volume        = {1000},
  archiveprefix = {arXiv},
  comment       = {MZR},
  copyright     = {arXiv.org perpetual, non-exclusive license},
  journal       = {ApJ},
  month         = mar,
  primaryclass  = {astro-ph.GA},
  publisher     = {American Astronomical Society},
  year          = {2026},
}

\end{document}